\documentclass{article}
\usepackage{epubtk}    
\usepackage{booktabs}  
\usepackage{graphicx,subfigure}  
\usepackage{amsmath,amssymb}

\newcommand{\vecx}{\ensuremath{\boldsymbol{x}}}

\newcommand{\veck}{\ensuremath{\boldsymbol{k}}}
\newcommand{\vecf}{\ensuremath{\boldsymbol{f}}}
\newcommand{\vecg}{\ensuremath{\boldsymbol{g}}}
\newcommand{\vecu}{\ensuremath{\boldsymbol{u}}}
\newcommand{\vecU}{\ensuremath{\boldsymbol{U}}}

\newcommand{\vecnab}{\ensuremath{\boldsymbol{\nabla}}}

\newcommand{\dd}{\ensuremath{\mathrm{d}}}
\newcommand{\DD}{\ensuremath{\mathrm{D}}}

\newcommand{\FT}[1]{\ensuremath{\widehat{#1}}}

\newcommand{\vect}[1]{\ensuremath{\boldsymbol{#1}}}
\newcommand{\tens}[1]{\ensuremath{\boldsymbol{\mathsf{#1}}}}

\newcommand{\frho}{\ensuremath{\langle\rho\rangle}}
\newcommand{\fU}{\ensuremath{\tilde{U}}}
\newcommand{\fE}{\ensuremath{\tilde{E}}}
\newcommand{\fe}{\ensuremath{\tilde{e}}}
\newcommand{\fP}{\ensuremath{\langle P\rangle}}

\newcommand{\fvecu}{\ensuremath{\tilde{\vecu}}}
\newcommand{\fvecU}{\ensuremath{\tilde{\vecU}}}

\newcommand{\fh}{\ensuremath{\tilde{h}}}
\newcommand{\Fconv}{\ensuremath{\boldsymbol{\mathfrak{F}}^{\mathrm{(conv)}}}}
\newcommand{\Fdiff}{\ensuremath{\boldsymbol{\mathfrak{F}}^{\mathrm{(kin)}}}}
\newcommand{\Fpress}{\ensuremath{\boldsymbol{\mathfrak{F}}^{\mathrm{(press)}}}}

\newcommand{\Masgs}{\ensuremath{\mathcal{M}_{\rm sgs}}}
\newcommand{\Reyn}{\ensuremath{\mathrm{Re}}}

\newcommand{\adot}{\ensuremath{\dot{a}}}

\begin{document}

\title{Large Eddy Simulations in Astrophysics}

\author{\epubtkAuthorData{Wolfram Schmidt}{%
Institut f\"ur Astrophysik \\
Universit\"at G\"ottingen\\
Friedrich-Hund-Platz 1\\ 
D-37077 G\"ottingen, Germany}{%
schmidt@astro.physik.uni-goettingen.de}{%
http://www.uni-goettingen.de/en/354642.html}%
}

\date{April 2014}
\maketitle

\begin{abstract}
In this review, the methodology of large eddy simulations (LES) is introduced and
applications in astrophysics are discussed. As theoretical framework,
the scale decomposition of the dynamical equations for neutral fluids 
by means of spatial filtering is explained. For cosmological applications,
the filtered equations in comoving coordinates are also presented.
To obtain a closed set of equations that can be evolved in LES, 
several subgrid scale models for the interactions between numerically
resolved and unresolved scales are discussed, in particular the 
subgrid scale turbulence energy equation model. It is then shown
how model coefficients can be calculated, either by
dynamical procedures or, a priori, from high-resolution data.
For astrophysical applications, adaptive mesh refinement is
often indispensable. It is shown that the subgrid scale turbulence energy
model allows for a particularly elegant and physically well motivated 
way of preserving momentum and energy conservation in AMR simulations.
Moreover, the notion of shear-improved models for inhomogeneous
and non-stationary turbulence is introduced. Finally, applications of
LES to turbulent combustion in thermonuclear supernovae, star formation
and feedback in galaxies, and cosmological structure formation
are reviewed.

\end{abstract}


\newpage


\section{Introduction}
\label{sec:intro}

Turbulent flows with high Reynolds numbers are often encountered in computational astrophysics.
Examples are the solar wind, stellar convection zones, star-forming clouds, and probably
the gas in galaxy clusters. This review concentrates on computational methods that
treat turbulence in the limit of high Reynolds numbers by explicitly
solving the compressible Euler equations for the large-scale dynamics of the flow, 
while incorporating small-scale effects such as viscous dissipation into 
a subgrid-scale model. Since the non-linear turbulent interactions between different
scales are at least partially resolved, this type of simulation is called large eddy simulation (LES). 

The relative importance of non-linear interactions 
and viscous damping is specified by the Reynolds number. 
It is determined by the characteristic velocity $V$ of the flow,
its integral length scale $L$, and the microscopic viscosity $\nu$:
\begin{equation}
	\label{eq:Re}
	\Reyn = \frac{VL}{\nu} 
\end{equation}
The flow becomes turbulent if the non-linear interactions are much stronger than viscous 
damping. Generally, this happens if $\Reyn$ reaches values greater than a few $10^3$,
but $\Reyn$ can become much greater than that.
For instance, an estimate for the turbulent convection zone of the Sun is $\Reyn\sim 10^{14}$
\cite{Canuto94}.

In principle, we can also define a scale-dependent Reynolds number $\Reyn(\ell) = v'(\ell)\ell/\nu$, 
where $v'(\ell)$ is the typical magnitude of velocity fluctuations on the
length scale $\ell$. The length sale of strong viscous damping
is then given by $\Reyn(\ell_{\rm K})\sim 1$. For incompressible turbulence,
substitution of the Kolmogorov-Obukhov scaling law $v'(\ell)\sim(\epsilon\ell)^{1/3}$
yields \cite{Frisch}
\[
	\frac{\epsilon^{1/3}\ell_{\rm K}^{4/3}}{\nu}\sim 1\,.
\]
Since the mean dissipation rate $\epsilon\sim V^3/L$, it follows that
\begin{equation}
	\frac{L}{\ell_{\rm K}}\sim\Reyn^{3/4}.
\end{equation}
The problem of high \Reyn\ is thus a problem of largely different length scales
or, equivalently, a high number of degrees of freedom.

In a numerical simulation of turbulence, the range of length scales is
limited by the grid scale $\Delta$, which is simply the linear size of the
grid cells. Only if $\Delta\lesssim\ell_{\rm K}$, turbulence
can be fully resolved by a so-called \emph{direct numerical simulation} (DNS).
However, DNS become infeasible for very large \Reyn\ 
because the total amount of floating point operations (FLOPs) increases with
$(L/\Delta)^4\gtrsim (L/\ell_{\rm K})^4\sim \Reyn^{\,3}$. 
The scaling may differ for highly compressible turbulence, but the basic problem 
remains the same. 
For a DNS of solar convection over one dynamical time scale, it would be necessary to
perform very roughly $10^{42}$ FLOP, which would take far longer than the current age
of the Universe on the fastest existing computer. 

In practice, however, it is neither feasible nor useful to account for 
all degrees of freedom in a simulation of high-\Reyn\ turbulence.
To reproduce statistical properties, a much coarser sampling of the
degrees of freedom can be quite sufficient. This is why LES encompass
only the energy-containing scales and structures dominated by 
non-linear interactions, which are part of the turbulent cascade down to
a cutoff scale much greater than the microscopic dissipation scale. 
The cutoff scale is given by grid scale $\Delta$. 
The defining criterion for LES is thus $L\gg\Delta\gg \ell_{\rm K}$ or, equivalently,
\[
	\Reyn\gg\Reyn(\Delta)\gg 1\,.
\]
Here, $\Reyn(\Delta)\sim v'(\Delta)\Delta/\nu$ is the Reynolds number of subgrid-scale turbulence.
The product $v'(\Delta)\Delta$ can be interpreted as turbulent viscosity of the numerically
unresolved eddies of size $\ell\lesssim \Delta$.
The effective Reynolds number of the numerically computed flow is therefore given by
\begin{equation}
	\label{eq:Re_eff}
	\Reyn_{\rm eff} = \frac{\Reyn}{\Reyn(\Delta)}\sim \frac{VL}{v'(\Delta)\Delta}\sim 
	\left(\frac{L}{\Delta}\right)^{4/3}\,.
\end{equation}
This means that LES reduces the number of degrees of freedom by replacing 
the microscopic viscosity $\nu$ by a turbulent viscosity
of the order $v'(\Delta)\Delta\gg \nu$. As a result, the purely non-linear turbulent dynamics of the ``large eddies"
is separated from microscopic dissipation.\epubtkFootnote{
	For many applications, particularly in astrophysics, the
	definition used here is appropriate.
	In a broader sense, LES may include the case where microscopic dissipation is partially resolved.
	DNS can then be considered as limiting case of LES for $\Reyn(\Delta)\sim 1$. 
}
The biggest challenge when implementing this concept is
to find an appropriate model for the coupling between the small- and large-scale dynamics.

A mathematical framework for LES is based on the notion of a filter, which separates large-scale
($\ell\gtrsim\Delta$) from small-scale ($\ell\lesssim\Delta$) fluctuations. 
Filters can be used to decompose the equations of fluid dynamics into equations for smoothed
variables, which have a very similar mathematical structure as the unfiltered equations, and equations for
second-order moments of the fluctuations. 
The latter are interpreted as subgrid-scale variables. 
In Section~\ref{sec:separation}, we will carry out the decomposition of the compressible
Navier-Stokes equation by applying the filter formalism of Germano \cite{Germano92}.
This formalism comprises the so-called Reynolds-averaged Navier-Stokes (RANS)
equations as limiting case if the filter length is comparable to the integral length scale of the flow.
Simulations based on the RANS equations work with low $\Reyn_{\rm eff}$, while
LES have high $\Reyn_{\rm eff}$. In principle, second-order moments can be expressed in terms
of higher-order moments. Since this would entail an infinite hierarchy of moments, 
the set of variables is limited by introducing closures. Usually, one attempts to find
closures for the second-order moments by expressing them in terms of the filtered variables.
This is what is called a subgrid-scale (SGS) model.\epubtkFootnote{In astrophysics,
	the term subgrid-scale model may comprise models
	that capture sub-resolution physics other than turbulence.	
	A typical example are star-formation models in galaxy simulations. }
For example, a complete second-order closure model for turbulent convection is formulated in 
\cite{Canuto94}.
Much simpler, yet often employed is the one-equation model for the
SGS turbulence energy $K$, i.~e., the local kinetic energy of 
numerically unresolved turbulent eddies.
For this reason, it is sometimes called the $K$-equation model. 
Closures for the transport and source terms in the SGS turbulence energy equation are presented 
in some detail in Section~\ref{sec:subgrid}, followed by a discussion of how the closure coefficients 
can be determined (Section~\ref{sec:closure}). Of particular importance is the prediction of the
local turbulent viscosity, which is is given by $\Delta\sqrt{K}$ times a dimension-less coefficient.
The turbulent viscosity is required to calculate the turbulent stresses, which enter the 
equations for the filtered variables analogous to the viscous stresses in the unfiltered
Navier-Stokes equations (see Section~\ref{sec:turb_stress}).
 
Filtering the dynamical equations is usually considered to be equivalent to numerical discretization.
The filter length can then be identified with the grid scale $\Delta$. Since the
numerical truncation errors of stable finite difference or finite volume schemes are
more or less diffusion-like terms, they produce a numerical viscosity that
effectively reduces the Reynolds number to a value comparable to equation~(\ref{eq:Re_eff}). 
It is actually a common assumption that numerical viscosity approximates the
turbulent viscosity on the grid scale. This leads to the notion of an \emph{implicit} large
eddy simulation (ILES) \cite{Garnier}, 
which is widely used for simulating turbulent flows in astrophysics. 
Numerous numerical studies demonstrated that ILES is a very robust method, which reliably predicts 
scaling laws of compressible turbulence at sufficiently high resolution 
\cite{SyPort00,KritNor07,BenzBif08,SchmFeder08,FederRom10,KritWag13}. This is a consequence of
the independence of inertial-range scaling from the dissipation mechanism, be it microscopic, turbulent
or numerical viscosity, provided that the dynamical range of the simulation is large enough.
In simulations of statistically stationary isotropic turbulence, however, the inertial subrange
is very narrow for computationally feasible resolutions because the bottleneck
effect distorts the spectrum over a large range of high wave numbers
below the Nyquist wavenumber \cite{Falko94,DobHaug03,SchmHille06}.
It appears that LES with an explicit SGS model, such as the $K$-equation model, 
can reduce the bottleneck effect 
to some degree and reproduce scalings from ILES or DNS at lower resolution 
\cite{HauBrand06,WoodPort06,Schm09b}. 
However, more systematic studies covering the parameters space of forced compressible turbulence
are necessary to confirm this effect.

There are, of course, alternative methods of scale separation and a large variety of
SGS models (for a comprehensive overview, see the monographs \cite{Sagaut,Garnier}). 
An example are the Camassa-Holm equations, 
which follow from the incompressible Navier-Stokes equations 
by decomposing the trajectories of fluid elements into mean and fluctuating parts in
the Lagrangian framework \cite{ChenFoi98}. Since the filtered component of the
velocity is defined by an inverse Helmholtz operator of the form $(1-\alpha^2\nabla^2)^{-1}$, which
is explicitly applied to determine the turbulent stresses in the filtered velocity equation, 
the resulting model is called Lagrangian-averaged Navier-Stokes $\alpha$-model
(LANS-$\alpha$). Depending on the choice of $\alpha$, the variables computed in LES based
on LANS-$\alpha$ are typically smoothed over length scales somewhat larger than the grid resolution. 
In other words, this type of simulation partially resolves the sub-filter scales, which
improves the controllability of the model. While there is no handle on the
competition between the SGS model and numerical truncation errors on the grid scale in convectional LES, 
LANS-$\alpha$ can, in principle, alleviate this problem by adjusting the balance between
truncation and model errors \cite{GraHolm07}. Although the idea is very elegant, the numerical
studies discussed in \cite{GraHolm07,GraHolm08} show that the applicability of LANS-$\alpha$ and similar models 
is limited, particularly for very high $\Reyn$. Moreover, the generalization to 
compressible turbulence is not straightforward. Models such as LANS-$\alpha$ are not further covered 
by this review, but they might be an option for magnetohydrodynamical LES \cite{GraMin09}.

Currently, LES are mainly applied to complex astrophysical systems. In simulations of cosmological structure formation, 
which are discussed in Section~\ref{sec:clusters}, the length scales on which turbulence is driven by gravity
are varying.
Although adaptive mesh refinement is applied to track down collapsing structures, it is difficult to
to resolve a wide range of length scales between the smallest driving scale and the
the grid scale at the highest refinement level. In this situation, SGS effects can become fairly large.
However, the variable grid scale complicates the scale separation in AMR simulations because 
energy has to be transferred between the resolved and SGS energy variables if a region is refined or de-refined.
Section~\ref{sec:consrv} describes how to combine LES and AMR. This method, 
for which the acronym FEARLESS (Fluid mEchanics for Adaptively Refined Large Eddy SimulationS) 
was coined in \cite{MaierIap09}, has been applied to galaxy clusters, the intergalactic medium, 
and primordial atomic cooling halos. The results from these simulations indicate that 
the contribution of the numerically unresolved turbulent pressure to the support against
gravity is non-negligible and the turbulent viscosity tends to stabilize disk-like structures 
around collapsed gas clouds. Moreover,
the SGS model provides indicators of turbulence production and dissipation and
allows for the computation of the turbulent velocity dispersion. A difficulty is
that turbulence production by cosmological structure formation is highly inhomogeneous. 
This entails the problem that the SGS model should dynamically adapt to conditions
ranging from laminar flow to developed turbulence. Inhomogeneous and non-stationary
turbulence can be treated by dynamical
procedures for the calculation of closure coefficients or shear-improved SGS models, which
decompose the numerically resolved flow into mean and fluctuating components. These
techniques are outlined in Sections~\ref{sec:dyn_proc} and~\ref{sec:kalman}.

Furthermore, SGS models offer unique possibilities for modeling physical processes 
that are influenced by turbulence. An example is turbulent deflagration, where the
turbulent diffusivity predicted by the SGS model dominates the effective flame propagation 
speed in underresolved numerical simulations. 
Turbulent deflagration plays a role at least in the initial phase of
thermonuclear explosions of white dwarfs (see Section~\ref{sec:SN_Ia}), which is one of the
scenarios that are thought to produce type Ia supernovae.
A recent application along similar lines are LES of isolated disk galaxies, where
the SGS turbulence energy is a crucial parameter for calculating the star formation rate
and the feedback due to supernova blast wave (see Section~\ref{sec:galaxy}).
Since the impact of feedback processes on the formation of galaxies and their evolution
leaves many questions unanswered, galaxies are a particularly promising field of application.

While great progress has been made for compressible hydrodynamics, magnetohydrodynamical
LES are still in their infancy.  Several SGS models have been proposed in the
context of terrestrial plasma physics \cite{MuellCar02a,MuellCar02b,HauBrand06,CherKar07,GraMin09,SonOber12}, 
but their applicability to astrophysical plasmas is unclear.
Astrophysical MHD turbulence, particularly in the interstellar medium, 
extends to the supersonic and super-Alfv\'{e}nic regimes. Moreover, plasmas become
collisionless for high temperatures and low densities. A typical example is the
solar corona. It is also likely to be the case in the intracluster medium.
Since the fluid-dynamical description is not applicable in this case, kinetic methods have to
be employed. Nevertheless, MHD-LES could provide a reasonable approximation on length
scales that are sufficiently large compared to the characteristic scales of
kinetic processes. In any case, SGS models for MHD turbulence will be
a very challenging problem because of the local anisotropy of turbulent fluctuations,
the potentially strong back-reaction from smaller to larger scales, and
complicated dissipative processes such as turbulent
reconnection \cite{BrandSub05,Buech07,ZweiYa09}. In this area, extensive fundamental studies will be necessary.

\newpage


\section{Scale Separation}
\label{sec:separation}

Large eddy simulations are based on the notion of scale separation. Although
turbulence is a multi-scale phenomenon, with interactions among different 
length scales, a separation into smoothed and fluctuating components can
be rigorously defined by means of filter operators. Of course, the filtering
of non-linear terms gives rise to interactions between these components.
Filter operators were originally applied in the context of mean-field theories, 
but can be generalized to LES. For incompressible hydrodynamical turbulence, 
Germano \cite{Germano92} introduced a general framework that encompasses 
mean field theories as limiting case.

The smoothed component of a generic field variable $q(\vecx,t)$ is defined 
by means of a spatial low-pass filter,
which is a convolution of $q$ with an appropriate filter kernel $G$ (see Chapter 2 in \cite{Sagaut}):
\begin{equation}
  \label{eq:q_flt}
  \langle q\rangle_G(\vecx) =
  \int G(\vecx-\vecx')q(\vecx',t)\,\dd^{3}x'.
\end{equation}
A homogeneous isotropic low-pass filter has the following properties:
\begin{itemize}
\item The filter kernel is independent of direction:
	\[
		G(\vecx-\vecx') = G(r),\quad\mbox{where\ }r=|\vecx-\vecx'|\,.
	\]
\item Filtering smoothes out fluctuations on length scales smaller
  than the filter length $\Delta_G$. Length scales that are large in comparison to
  $\Delta_G$ are not affected. This implies
  \[
    G(\vecx-\vecx') \sim \left\{\begin{array}{ll}
				1/\Delta_G^3 &\mbox{if}\ |\vecx-\vecx'|\ll \Delta_G,\\
				0 &\mbox{if}\ |\vecx-\vecx'|\gg \Delta_G.
				\end{array}\right.
  \]
\item The filter operator is linear, conserves constants, and commutes with spatial derivatives: 
	\[
		\langle\vecnab q\rangle_G = \vecnab\langle q\rangle_G.
	\]
\end{itemize}
The simplest low-pass filter is the box or top-hat filter. For Cartesian coordinates $x_i$, the kernel
of the box filter is defined by
\begin{equation}
  \label{eq:box_filter}
  G_{\rm box}(\vecx-\vecx') = \prod_{i=1}^3 G_i(x_i-x_i'),\quad\mbox{where}\quad
  G_i(x_i-x_i')=\left\{\begin{array}{ll}
				1/\Delta_i &\mbox{if}\ |x_i-x_i'|\le\Delta_i/2,\\
				0 &\mbox{otherwise}.
				\end{array}\right.
\end{equation}
Usually, $\Delta_i$ is assumed to be equal for all spatial dimensions.
The mean value of $q$ in a rectangular domain with periodic boundary conditions 
follows in the limit that $\Delta_i$ is the linear size of the domain in each dimension. 

The construction of a filter is particularly simple in Fourier space. For a low-pass filter,
the Fourier transform of the filter kernel, the so-called transfer function $\FT{G}(\veck)$, 
drops rapidly to zero for wavenumbers
$k\gtrsim k_{\mathrm{c}}=\pi/\Delta_{G}$. Since
\begin{equation}
	\langle\hat{q}\rangle_G(\veck,t)=\hat{q}(\veck,t)\FT{G}(\veck),
\end{equation}
only the Fourier modes
$\hat{q}(\veck,t)$ with $k\lesssim k_{\mathrm{c}}$ contribute significantly to 
the corresponding filtered field $\langle q\rangle_G(\vecx)$ in physical space.
The simplest case is the sharp cutoff filter, for which
\begin{equation}
  \label{eq:spect_filter}
  \FT{G}_{\rm sharp}(\veck) = \left\{\begin{array}{ll}
				1 &\mbox{if}\ k\le k_{\rm c},\\
				0 &\mbox{otherwise}.
				\end{array}\right.
\end{equation}
The sharp cutoff filter, however, is not equivalent to the box filter, which has
the Fourier representation
\begin{equation}
  \label{eq:sharp_filter}
  \FT{G}_{\rm box}(\veck) = \prod_{i=1}^3\frac{\sin(k\Delta_i/2)}{k\Delta_i/2}.
\end{equation}
A filter that is intermediate between these two cases is the Gaussian filter.

\subsection{Decomposition of the compressible Navier-Stokes equations}
\label{sc:decomp_navier}

The compressible Navier-Stokes equations for the mass density $\rho$, the momentum density $\rho\vecu$, and
the energy density $\rho E$ of a neutral fluid subject to gravitational and mechanical accelerations $\vecg$ and $\vecf$,
respectively, are
\begin{align}
 \label{eq:navier_rho}
  \frac{\partial}{\partial t}\rho + \vecnab\cdot(\vecu\rho)\, &= 0\,, \\
 \label{eq:navier_momt}
  \frac{\partial}{\partial t}\rho\vecu + 
  \vecnab\cdot\left(\rho\vecu\otimes\vecu\right)\, &=
   \rho(\vect{g} + \vect{f}) -\vecnab P + \vecnab\cdot\tens{\sigma}\,, \\
 \label{eq:navier_energy}
   \frac{\partial}{\partial t} \rho E + \vecnab\cdot(\rho\vecu E)\, &=
   \rho\vect{u}\cdot(\vect{g} + \vect{f}) -\vecnab\cdot(\vecu P) + \vecnab\cdot(\vecu\cdot\tens{\sigma})\,.
\end{align}
Thermal conduction is neglected here. The energy per unit mass can be expressed as
\begin{equation}
  \label{eq:e_tot}
    E = e + \frac{1}{2}u^{2}\,,
\end{equation}
where $e$ is the internal or thermal gas energy. For a perfect gas, 
$e$ is related to the gas pressure $P$ and the temperature $T$ via the ideal gas law:
\begin{equation}
  \label{eq:e_int}
    e = \frac{P}{(\gamma-1)\rho} = \frac{k_{\mathrm{B}}T}{(\gamma-1)\mu m_{\mathrm{H}}}\,,
\end{equation}
where $\gamma$ is the adiabatic exponent, $k_{\mathrm{B}}$ the Boltzmann constant, 
$\mu$ the mean molecular weight, and $m_{\mathrm{H}}$ the mass of the hydrogen atom.
The viscous stress tensor $\tens{\sigma}$ is defined by
\begin{equation}
  \label{eq:visc_stress}
  \sigma_{ij} =
    2\eta S_{ij}^{\,\ast} + \zeta d\delta_{ij} = 
    2\eta\left(S_{ij} - \frac{1}{3}d\delta_{ij}\right) + \zeta d\delta_{ij}\,
\end{equation}
where the two coefficients $\eta$ and $\zeta$ are the dynamic and bulk
viscosities of the fluid,
\begin{equation}
    \label{eq:strain}
	S_{ij}=\frac{1}{2}\left(\frac{\partial u_i}{\partial x_j}+\frac{\partial u_j}{\partial x_i}\right)
\end{equation}
is the rate-of-strain tensor, and the trace $S_{ii}$ is equal to the divergence $d=\vecnab\cdot\vecu$.
The gravitational acceleration is given by $\vecg=-\vecnab\phi$, where the gravitational potential
$\phi$ is determined by the Poisson equation
\begin{equation}
  \label{eq:poisson}
	\nabla^2\phi = 4\pi G(\rho-\rho_0)
\end{equation}
for a constant background density $\rho_0$ ($G$ is Newton's constant).

Mean-field equations for compressible turbulence are derived in \cite{Canuto97}.
Much in the same way, a general low-pass filter $\langle\;\rangle_G$
can be applied to the system of PDEs~(\ref{eq:navier_rho})--(\ref{eq:navier_energy}).
Alternative formulations can be found in \cite{Garnier}, Section 2.4.
For brevity, we omit the subscript $G$ in the following. 
Since $\langle\;\rangle$ commutes with differential operators, 
the smoothed mass density $\langle\rho\rangle$ obeys an equation of exactly
the same form as the continuity equation,
\begin{equation}
  \label{eq:dens_flt}
  \frac{\partial}{\partial t}\frho + \vect{\nabla}\cdot\frho\fvecu = 0\,,
\end{equation}
if we set $\langle\rho\vecu\rangle=\frho\fvecu$. This identiy implies the
definition of the \emph{Favre-filtered} velocity
\begin{equation}
  \label{eq:favre_flt}
  \fvecu = \frac{\langle\rho\vecu\rangle}{\frho}\,.
\end{equation}

Filtering the momentum equation results in
\[
  \frac{\partial}{\partial t}\frho\fvecu + 
  \vecnab\cdot\left\langle\rho\vecu\otimes\vecu\right\rangle =
   \left\langle\rho(-\vecnab\phi + \vect{f})\right\rangle -\vecnab \fP + \vecnab\cdot\langle\tens{\sigma}\rangle
\]
Owing to the non-linearities, however, we are facing some difficulties here. 
To obtain a PDE with the same basic structure as the unfiltered momentum equation,
the advection term on the left-hand side should read $\vecnab\cdot\left[\langle\rho\rangle\fvecu\otimes\fvecu\right]$.
The solution is to split the filtered non-linear terms:
\begin{equation}
	\label{eq:turb_stress}
	\left\langle\rho\vecu\otimes\vecu\right\rangle
	= \langle\rho\rangle\fvecu\otimes\fvecu - \tens{\tau}(\rho\vecu,\vecu)\quad
	\mbox{where}\quad
	\tens{\tau} := 
	-\langle\rho\vecu\otimes\vecu\rangle + 
	\frac{\langle\rho\vecu\rangle \otimes \langle\rho\vecu\rangle}{\langle\rho\rangle}\,.
\end{equation} 
Since the Poisson equation~(\ref{eq:poisson}) is linear, 
the smoothed potential $\langle\phi\rangle$ is solely determined by $\langle\rho\rangle$. The self-gravity
term $\langle\rho\vecnab\phi\rangle$, however, has to be split by defining
\begin{equation}
	\label{eq:grav_sgs}
	\vect{\gamma} := 
	-\langle\rho\vecnab\phi\rangle + \langle\rho\rangle\vecnab\langle\phi\rangle.
\end{equation}
The specific force $\vect{f}$, on the other hand, usually varies only over the
largest scales of the system. If the filter length is small compared to these scales, 
$\langle\rho\vect{f}\rangle\simeq \langle\rho\rangle\vect{f}$ is a good approximation.
Thus, the filtered momentum equation can be casted into the following form 
\cite{Garnier,MoinSqui91,Yoshi91,Germano92,Canuto97,SchmNie06b}:
\begin{equation}
  \label{eq:momt_flt}
  \frac{\partial}{\partial t}\frho\fvecu + 
  \vecnab\cdot\left[\langle\rho\rangle\fvecu\otimes\fvecu\right] =
   \langle\rho\rangle(-\vecnab\langle\phi\rangle + \vect{f}) -\vecnab \fP + 
   \vecnab\cdot\left[\langle\tens{\sigma}\rangle+\tens{\tau}\right]
   + \vect{\gamma}\,.
\end{equation}

Now, what is the physical interpretation of the terms $\tens{\tau}$ and $\vect{\gamma}$?
Let us first consider the weakly compressible limit. By assuming that $\rho$ varies only little over the filter length,
density factors can be pulled out of brackets. In this case, $\fvecu\simeq\langle\vecu\rangle$. 
By defining the fluctuation of the velocity as $\vect{u}'=\vecu-\fvecu$, it follows that
\[
    \tens{\tau} \simeq
    \rho\left[\langle\vecu\rangle \otimes \langle\vecu\rangle
    -\langle\langle\vecu\rangle \otimes \langle\vecu\rangle\rangle 
    - 2\langle\langle\vecu\rangle \otimes \vecu'\rangle
    - \langle\vecu'\otimes\vecu'\rangle\right]\,.
\]
If we futher assume that $\langle\;\rangle$ is a Reynolds operator (see Section 3.3 in \cite{Sagaut}), 
which is not generally true for filters but applies, for example, to global averages, filtered quantities 
can be pulled out of brackets and the above expression simplifies to
\[
    \tens{\tau} \simeq -\rho\langle\vecu'\otimes\vecu'\rangle\,.
\]
Although this simple relation holds only for a Reynolds operator in the weakly compressible limit,
$\tens{\tau}$ is generally interpreted as the stress tensor associated with
the turbulent velocity fluctuations below the filter length. 
For this reason, $\tens{\tau}$ is called the subgrid-scale turbulence stress tensor
in the context of LES. The non-linear interactions
of the filtered flow (the ``large eddies'') with small-scale fluctuations below the grid scale $\Delta$
are given by $\vecnab\cdot\tens{\tau}$ in equation~(\ref{eq:momt_flt}). Likewise, 
the term $\vect{\gamma}$ defined by equation~(\ref{eq:grav_sgs}) accounts for the momentum transfer due self-gravitating
fluctuations in the density. The trace of $\tens{\tau}$ defines the fraction of kinetic energy 
on length scales smaller than the filter length:
\begin{equation}
  \label{eq:sgs_energy}
  \langle\rho\rangle K: = -\frac{1}{2}\tau_{ii} = \frac{1}{2}\langle\rho u^2\rangle - \frac{1}{2}\langle\rho\rangle\tilde{u}^2\,.
\end{equation}
If the filter length is the grid scale, $\rho K$ is called the subgrid-scale turbulence energy. The first term
on the right-hand side of equation~(\ref{eq:sgs_energy}) is the total kinetic energy, the second term the kinetic energy 
on length scales greater than the filter length (i.~e., the numerically resolved kinetic energy in LES).

In the limit of high Reynolds numbers, 
the viscous dissipation scale (also known as Kolmogorov scale)
is typically much smaller than the filter length. In this case, scaling arguments for incompressible turbulence 
imply $\langle\tens{\sigma}\rangle\ll\tens{\tau}$, i.~e., the filtered viscous stresses are negligible
compared to the stresses associated with the turbulent velocity fluctuations \cite{RoepSchm09}.
Since the scaling of compressible turbulence tends to be stiffer than for incompressible turbulence 
\cite{KritNor07,SchmFeder08,SchmFeder09}, 
one can reasonably assume that this conclusion is generally applicable. 
The filtered momentum equation~(\ref{eq:momt_flt}) thus can be written as
\begin{equation}
  \label{eq:momt_flt_high_Re}
  \frac{\partial}{\partial t}\frho\fvecu + 
  \vecnab\cdot\left[\langle\rho\rangle\fvecu\otimes\fvecu\right] =\,
   \langle\rho\rangle(-\vecnab\langle\phi\rangle + \vect{f}) -\vecnab\!\left(\fP+\frac{2}{3}\rho K\right)\\
   + \vecnab\cdot\tens{\tau}^\ast + \vect{\gamma}\,.
\end{equation}
where $\tens{\tau}^\ast$ is the trace-free part of $\tens{\tau}$:
\begin{equation}
  \tau_{ij}^\ast=\tau_{ij}-\frac{1}{3}\tau_{kk}\delta_{ij}=\tau_{ij}+\frac{2}{3}\rho K\delta_{ij}.
\end{equation}
As one can see from equation~(\ref{eq:momt_flt_high_Re}), the trace of $\tens{\tau}$ is associated
with the turbulent pressure $\frac{2}{3}\rho K$ at the filter length scale. 

In contrast to the filtered momentum density, which can be expressed as $\langle\rho\vecu\rangle=\langle\rho\rangle\fvecu$, 
the energy density on length scales greater than the filter length is given by
\begin{equation}
	\langle\rho\rangle\fE:=
	\langle\rho\rangle\left(\tilde{e} + \frac{1}{2}\tilde{u}^2\right)
	= \langle\rho E\rangle - \langle\rho\rangle K,
\end{equation}
where the second equality follows from equations~(\ref{eq:e_tot}) and~(\ref{eq:sgs_energy}). 
Consequently, $\langle\rho\rangle\fE\ne \langle\rho E\rangle$.\epubtkFootnote{
	In \cite{Garnier}, $\langle\rho\rangle\fE$ is identified with $\langle\rho E\rangle$
	and a different symbol is used for $\tilde{e} + \frac{1}{2}\tilde{u}^2$.
	However, we do not follow this nomenclature here.
} A PDE for $\langle\rho\rangle\fE$
follows form the contraction of equation~(\ref{eq:momt_flt_high_Re}) with $\fvecu$
plus the filtered internal energy equation. The subtraction of this PDE from the 
filtered equation for the total energy yields the PDE for $\rho K$
(see Section 3.3 in \cite{Sagaut} and \cite{MoinSqui91,Yoshi91,Germano92,Canuto97,SchmNie06b}).
In the limit of high Reynolds numbers, the resulting equations are:
\begin{align}
	\label{eq:pde_energy_res}
	\begin{split}
	\frac{\partial}{\partial t}\langle\rho\rangle\fE + \vecnab\cdot\langle\rho\rangle\fvecu\fE =\,
    	&\langle\rho\rangle\fvecu\cdot(-\vecnab\langle\phi\rangle + \vecf) 
    	 + \vecnab\cdot\left[-\fvecu\left(\fP+\frac{2}{3}\rho K\right) + \fvecu\cdot\tens{\tau}^\ast + \Fconv\right]\\
	&- \Sigma + \langle\rho\rangle(\epsilon + \lambda) + \fvecu\cdot\vect{\gamma}\,,
	\end{split}\\
	\label{eq:pde_energy_sgs}
	\frac{\partial}{\partial t}\langle\rho\rangle K + \vecnab\cdot\langle\rho\rangle\fvecu K =\,
		&\Gamma + \Sigma - \langle\rho\rangle(\epsilon + \lambda) + \vecnab\cdot\left[\Fdiff+\Fpress\right].
\end{align}
The additional source and transport terms resulting form the scale separation of the energy are defined as follows.
\begin{itemize}
\item Gravitational energy injection on subgrid scales:
	\begin{equation}
		\Gamma = 
		-\langle\rho\fvecu\cdot\vecnab\phi\rangle + \fvecu\cdot\langle\rho\vecnab\phi\rangle =\
		-\langle\rho\fvecu\cdot\vecnab\phi\rangle + \langle\rho\rangle\fvecu\cdot\vecnab\langle\phi\rangle
		-\fvecu\cdot\vect{\gamma}\,. 
	\end{equation} 
\item Rate of subgrid-scale turbulence energy production:\epubtkFootnote{Also called turbulence energy flux,
	although this is not a transport term. 
	In the incompressible limit, $\Sigma$ corresponds to the energy transfer in spectral space.}
	\begin{equation}
		\Sigma = \tau_{ij}\tilde{S}_{ij}\,,
		\label{eq:sgs_prod}
	\end{equation} 
	where $\tau_{ij}$ is defined by equations~(\ref{eq:turb_stress})
	and $\tilde{S}_{ij}$ is the rate-of-strain tensor associated with the Favre-filtered velocity:\epubtkFootnote{
		The definition of $\tilde{S}_{ij}$ is a consequence of integration by parts of
		$\tilde{u}_i\partial_j\tau_{ij}$. The symbol $\tilde{S}_{ij}$ is used for convenience. 
		It is important to keep in mind that $\tilde{S}_{ij}\ne
	  \langle\rho S_{ij}\rangle/\langle\rho\rangle$ because 
	  $\partial_j\tilde{u}_i =\partial_j[\langle\rho u_i\rangle/\langle\rho\rangle]\ne 
	  \langle\rho\partial_j u_i\rangle/\langle\rho\rangle$.}
	\begin{equation}
	  \label{eq:strain_flt}
	  \tilde{S}_{ij}:=
	  \frac{1}{2}\left(\frac{\partial\tilde{u}_i}{\partial x_j}+\frac{\partial\tilde{u}_j}{\partial x_i}\right)\,.
	\end{equation}
\item Rate of viscous energy dissipation in the limit of high Reynolds numbers:\epubtkFootnote{
	  	Since $S_{ij}$ is a velocity derivative, it is of the order of the velocity fluctuation at the
	  	smallest length scales. For incompressible turbulence, Kolmogorov scaling implies
	  	$\langle\sigma\rangle\langle S\rangle\sim \rho\epsilon (\Delta/\ell_{\rm K})^{-4/3}\sim \rho\epsilon/\Reyn(\Delta)$,
	  	where $\ell_{\rm K}$ is the Kolmogorov length. For high Reynolds numbers, 
	  	the ratio $\Delta/\ell_{\rm K}$ is typically very large. As a result,
		$\langle\sigma\rangle\langle S\rangle$ is negligible compared to $\rho\epsilon\simeq\langle\sigma S\rangle$.
	  	From the same estimates follows
	  	$\tau \sim (\Delta/\ell_{\rm K})^{4/3}\langle\sigma\rangle\sim \Reyn(\Delta)\langle\sigma\rangle$ 
		\cite{RoepSchm09}, which is applied to obtain
	  	equation~(\ref{eq:momt_flt_high_Re}) for the filtered momentum.}
	\begin{equation}
		\langle\rho\rangle\epsilon = 
		\langle\sigma_{ij}S_{ij}\rangle - \langle\sigma_{ij}\rangle\tilde{S}_{ij}
		\simeq \langle\sigma_{ij}S_{ij}\rangle =
		\langle\eta|S^{\ast}|^{2}+\zeta d^2\rangle\,,
		\label{eq:sgs_diss}
	\end{equation}
	where $S_{ij}$ is defined by equation~(\ref{eq:strain}), $|S^{\ast}|^{2} = 2S_{ij}^{\ast}S_{ij}^{\ast}$ 
	is the squared norm of the trace-free rate-of-strain tensor $S_{ij}^{\ast}=S_{ij}-\frac{1}{3}d\delta_{ij}$ 
	and $d=S_{ii}$. Although the viscous stresses can be neglected in the filtered momentum equation,
	viscous dissipation is crucial for the energy balance of turbulent flows. 
\item Rate of subgrid-scale pressure dilatation:
	\begin{equation}
		\langle\rho\rangle\lambda = -\langle d P\rangle + \tilde{d}\langle P\rangle, \\
		\label{eq:sgs_press_dilt}
	\end{equation}
	where $\tilde{d}=\tilde{S}_{ii}=\partial\tilde{u}_i/\partial x_i$.
\item Convective internal energy flux on sub-grid scales:\epubtkFootnote{In \cite{SchmNie06b}, the
      convective flux in equation~(\ref{eq:pde_energy_res}) is erroneously defined in terms of the enthalpy.}
	\begin{equation}
  		\label{eq:flux_conv}
  		\Fconv = 
		-\langle\rho\vecu e\rangle + \rho\fvecu\fe\,.
	\end{equation}
\item The flux associated with pressure fluctuations:
	\begin{equation}
		\label{eq:flux_press}
		\Fpress = -\langle\vecu P\rangle+\fvecu\langle P\rangle.
	\end{equation}
      For ideal gas with adiabatic exponent $\gamma$, $\Fpress=(\gamma-1)\Fconv$.
\item The flux of turbulent energy diffusion on sub-grid scales:
	\begin{equation}
	  \label{eq:flux_diff}
	  \Fdiff =
	  -\frac{1}{2}\langle\rho u^2 \vecu\rangle +
	  \frac{1}{2}\langle\rho u^2\rangle \fvecu - \fvecu\cdot\tens{\tau}\,.
	\end{equation}
      For Reynolds operators in the weakly compressible limit, \Fdiff can be expressed as
      a third-order moment of the velocity fluctuation: 
      $2\mathfrak{F}^{(\mathrm{kin})}_j\simeq -\rho\langle u_i^\prime u_i^\prime u_j^\prime\rangle$
      \cite{Germano92}.
\item There is also a viscous flux, which can be neglected relative to other flux terms if the
      Reynolds number is sufficiently high.
\end{itemize}

By adding the equations~(\ref{eq:pde_energy_res}) and~(\ref{eq:pde_energy_sgs}), we obtain an equation
for the filtered total energy
\begin{equation}
	\begin{split}
	\label{eq:pde_energy_tot}
	\frac{\partial}{\partial t}\langle\rho\rangle(\fE+K) +& \vecnab\cdot\langle\rho\rangle\fvecu(\fE+K) =
    	\langle\rho\rangle\fvecu(-\vecnab\langle\phi\rangle + \vecf) + \Gamma + \fvecu\cdot\vect{\gamma}\\
	&+ \vecnab\cdot\left[-\fvecu\left(\fP+\frac{2}{3}\rho K\right) + \fvecu\cdot\tens{\tau}^\ast + 
	   \Fdiff + \Fconv + \Fpress\right]\,.
	\end{split}
\end{equation}
Except for the gravitational source terms, production and dissipation rates cancel out. The fluxes on the very right
are related to turbulent transport processes below filter length. 
In particular, the sum of $\Fconv$ and $\Fpress$ can be expressed as
convective enthalpy flux,
\begin{equation}
 	\Fconv + \Fpress= 
	-\langle\rho\vecu h\rangle + \rho\fvecu\fh\,,
\end{equation}
where $\rho h=\rho e+P$, which corresponds to $-\rho\langle\vect{u}'h'\rangle$ in the weakly compressible limit.
For a closed system of PDEs, it is necessary to compute all terms defined above in terms of known
quantities. A rigorous calculation requires further PDEs, which involve higher-order moments and so on ad infinitum.
This is known as the closure problem. A subgrid-scale model truncates the closure problem by approximating
moments above a given order by lower-order moments.
  
\subsection{Cosmological fluid dynamics}
\label{sc:cosmo}

In cosmological simulations, the equations of fluid dynamics are solved in a comoving coordinate system. Coordinates of observers that are stationary relative to the Hubble expansion of the Universe are constant in this system. The expansion
is characterized by the scale factor $a(t)$, which is determined by the Friedmann equations for a homogeneous
and isotropic cosmology \cite{Peacock}. If the proper coordinates, which include changes of position
due to the expansion of the Universe, are denoted by $\vecx_{\rm proper}$ and $t_{\rm proper}$, the corresponding 
comoving coordinates are $\vecx=\vecx_{\rm proper}/a$ and $t=t_{\rm proper}$. Derivative operators transform as
\[
	\left.\frac{\partial}{\partial t}\right|_{\rm proper}=\frac{\partial}{\partial t}-\frac{\adot}{a}\vecx\cdot\vecnab
	\quad\mbox{and}\quad
	\vecnab_{\rm proper}=\frac{1}{a}\vecnab\,.
\]
Furthermore, the invariance of mass implies that the comoving baryonic density $\rho$ is related to 
the proper density by $\rho = a^3 \rho_{\rm proper}$. It can then be shown that 
continuity equation for $\rho$ in comoving coordinates assumes exactly the same form as 
equation~(\ref{eq:navier_rho}):
\[
	\frac{\partial \rho}{\partial t} + \nabla \cdot (\rho \vecu) = 0\, , 	
\]
Here, $\vecu$ is the so-called peculiar velocity, which is defined as
\begin{equation}
	\vecu = \dot{\vecx} = \frac{1}{a}\vecu_{\rm proper}-H\vecx,
\end{equation}
where $\vecu_{\rm proper}=\dot{\vecx}_{\rm proper}$ is the proper velocity and $H=\adot/a$ the Hubble constant.
This means that, in the comoving coordinate system, matter moves with velocity $\vecu$ relative to the Hubble flow $H\vecx$.
With some algebra, also the momentum and energy equations can be transformed to comoving coordinates. The resulting
equations do not have the same form as equations~(\ref{eq:navier_momt}) and~(\ref{eq:navier_energy}), but include additional terms with prefactors $H$. However, a particularly simple representation of the momentum and energy equations is obtained if the \emph{proper} peculiar velocity
\begin{equation}
	\vecU = a\vecu = \vecu_{\rm proper}-\adot\vecx
\end{equation}
is used in place of $\vecu$.

Filtered dynamical equations for cosmological fluids were first derived in \cite{MaierPhD,MaierIap09}
and presented in an alternative formulation in \cite{SchmAlm13}. The applied filter kernel is static
in comoving coordinates, i.~e., the filter length increases proportional to the cosmological
scale factor $a$. Consequently, commutation of the filter with time derivatives is unaffected by the
cosmological expansion and equations for filtered dynamical variables follow completely analogous to 
Section~\ref{sc:decomp_navier}. By neglecting gravitational terms associated with fluctuations below the
filter length, the following equations for the filtered mass density 
$\langle\rho\rangle$, the filtered momentum density $\langle\rho\vecU\rangle=\langle\rho\rangle\fvecU$, and 
energy density $\langle\rho\rangle\fE$, where $\fE = \fe + \frac{1}{2}\fU^2$, are obtained:
\begin{align}
	\label{eq:cosmo_mass_les}
	\frac{\partial \langle\rho\rangle}{\partial t} 
		+ \frac{1}{a} \vecnab \cdot [\langle\rho\rangle\fvecU] =&\, 0\, , \\
	\label{eq:cosmo_momt_les}
	\frac{\partial}{\partial t}a \langle\rho\rangle \fvecU  
	 + \vecnab \cdot [\langle\rho\rangle \fvecU\otimes\fvecU] =&\,
	 - \langle\rho\rangle\vecnab\langle\phi\rangle
	 - \vecnab \fP + \vecnab\cdot\tens{\tau} + \vect{\gamma}\, , \\
	\label{eq:cosmo_energy_les}
	\begin{split}
	\frac{\partial}{\partial t}a^2\langle\rho\rangle\fE + a \vecnab\cdot[\langle\rho\rangle\fvecU\fE] =\,
    &- a\langle\rho\rangle\fvecU\cdot\vecnab\langle\phi\rangle 
     + a \vecnab\cdot\left[-\fvecU\fP + \fvecU\cdot\tens{\tau} + \Fconv\right]\\
	&- a \dot{a} [2 - 3 (\gamma - 1)] \langle\rho\rangle\fe 
     - a \left[\Sigma + \langle\rho\rangle(\epsilon + \lambda)\right]
     + a \fvecU\cdot\vect{\gamma}\,,
	\end{split}
\end{align}
Here, the filtered internal energy density is $\langle\rho\rangle\fe=\langle\rho e\rangle = \fP/(\gamma - 1)$, 
where $P=a^3 P_{\rm proper}$, 
and the gravitational potential $\langle\phi\rangle$ of baryonic and dark matter density fluctuations
is given by the cosmological Poisson equation
\begin{equation}
	\nabla^2\langle\phi\rangle = \frac{3H_0^2\Omega_{\rm m}(t_0)}{2a}\langle\delta_{\rm m}\rangle,
\end{equation}
where $H_0=\adot(t_0)$ is the Hubble constant and $\Omega_{\rm m}(t_0)$ the density parameter of matter
at redshift zero. 
Since the mean matter density $\rho_{\rm m, 0}$ is constant in comoving coordinates, 
the source term of the Poisson equation can expressed in terms of the
density fluctuation $\delta_{\rm m}=(\langle\rho_{\rm dm}+\rho\rangle-\rho_{\rm m, 0})/\rho_{\rm m, 0}$ for 
the local dark matter density $\rho_{\rm dm}$ and baryonic mass density $\rho$. 
The density parameter is defined by
$\Omega_{\rm m}(t_0)=\rho_{\rm m, 0}/\rho_{\rm crit,0}$, where $\rho_{\rm crit,0}=3H_0^2/(8\pi G)$ is the
critical density at time $t=t_0$.

The kinetic energy associated with peculiar velocity fluctuations below the filter length,
\[
  \langle\rho\rangle K = 
  \frac{1}{2}\langle\rho U^2\rangle - \frac{1}{2}\langle\rho\rangle\fU^2\,,
\]
is given by the dynamical equation
\begin{equation}
	\label{eq:cosmo_k_les}
	\frac{\partial}{\partial t}a^2\langle\rho\rangle K + a\vecnab\cdot[\langle\rho\rangle\fvecU K] =\,
	a\left[\Gamma + \Sigma - \langle\rho\rangle(\epsilon + \lambda)\right] + a\vecnab\cdot\left[\Fdiff+\Fpress\right]\,.
\end{equation}
Since the prefactors of all terms except for the time derivatives in the momentum and energy equations are unity and $a$, respectively, the definitions of all source and transport terms in equations~(\ref{eq:cosmo_momt_les}), (\ref{eq:cosmo_energy_les}) and (\ref{eq:cosmo_k_les}) are analogous to the definitions given in Section~\ref{sc:decomp_navier}, with $u_i$ being replaced by $U_i$. This suggests that closures for turbulence in a
static space are applicable to cosmological fluids as well. Although cosmological expansion, in principle, causes
a dampening of the kinetic energy \cite{SchmAlm13}, this effect is subdominant for 
turbulent eddies even on the largest scales in galaxy clusters because turbulence is driven in gravitationally
bound gas on time scales shorter than the current Hubble time $1/H_0$. 

\newpage


\section{Subgrid-Scale Models}
\label{sec:subgrid}

There is a beautiful correspondence between finite-volume discretization and filtering. Finite-volume
methods solve an equation for the cell averages of some dynamical variable q(\vect{x}):
\[
	Q_{ijk}=
	\int_{z_k-\Delta/2}^{z_k+\Delta/2}\int_{y_j-\Delta/2}^{y_j+\Delta/2}\int_{x_i-\Delta/2}^{x_i+\Delta/2}
	 q(x,y,z)\,\dd x\,\dd y\,\dd z
\]
Here, $(x_i,y_j,z_k)$ are the cell-centered coordinates and $\Delta$ is the linear size of a grid cell. It is not
difficult to see that $Q_{ijk}$ equals the box-filtered variable $\langle q\rangle_G$ for a box filter $G$ with
filter length $\Delta$ at the discrete points $(x_i,y_j,z_k)$.\epubtkFootnote{
	An \emph{effective} filter length for isotropic turbulence simulations can be calculated from the
	second moment of the compensated energy spectrum as shown in \cite{SchmHille06}.}
Also finite differences correspond to low-pass filters. Thus, numerical
discretization can be interpreted as \emph{implicit filtering}.
The numerical errors in the approximations to $Q_{ijk}$
can be characterized by the truncation error of the finite-volume method. For stable schemes with
flux limiters, these errors are usually associated with terms of the diffusion type, i.~e.,
proportional to $\nabla^2 q$.\epubtkFootnote{
	For higher-order methods such as PPM \cite{ColWood84}, the leading-order truncation errors
	correspond to hyper-viscosity terms proportional to $\nabla^4 q$.}
In ILES, this is what causes the dissipation of kinetic energy
into heat (a detailed account of ILES is given in \cite{Garnier}). 
A similar expression follows if the Boussinesq expression for the turbulent
stresses is used as explicit SGS model for the interaction between numerically resolved and
unresolved turbulent eddies. An important difference, however, is that the
turbulent viscosity in the resulting diffusion terms in the momentum and energy
equations is controlled by the dynamical variable $K$, which is the kinetic energy associated with turbulent 
velocity fluctuations below the grid scale. In this section, we mainly discuss the computation of $K$ in LES.
The approach we follow here is known as \emph{functional} modeling. The aim is to model 
only statistical effects of SGS turbulence on the dynamics of the filtered fields,
which are identified with the numerical solution. An alternative strategy is \emph{structural} modeling 
(see Chapter 5 in \cite{Garnier}), which is not covered here.

\subsection{Closures for the turbulence stress tensor}
\label{sec:turb_stress}

The SGS turbulence stress tensor, which is associated with the non-linear energy transfer between
large and small scales, is the central quantity that has to be modeled in LES. The most commonly
used closure is the eddy-viscosity closure. The underlying assumption is that the form of the
trace-free part $\tens{\tau}^\ast$ is analogous to the anisotropic viscous stress tensor $\tens{\sigma}^{\ast}$, with the correspondence\epubtkFootnote{
	This idea was originally proposed by Boussinesq in the 19th century \cite{Boussin77}.}
\begin{align}
	S_{ij} &\longleftrightarrow \tilde{S}_{ij},\\
	\eta   &\longleftrightarrow \langle\rho\rangle\nu_{\rm sgs}\,,
\end{align}
where $S_{ij}$ and $\tilde{S}_{ij}$ are defined by equations~(\ref{eq:strain}) and~(\ref{eq:strain_flt}), respectively. 
The turbulent viscosity $\nu_{\rm sgs}$ is assumed to depend on the grid scale and the unresolved turbulent velocity fluctuation (see, for example, Section 4.3 in \cite{Sagaut}):
\begin{equation}
	\label{eq:visc_sgs}
	\nu_{\rm sgs} = C_{\nu}\Delta \sqrt{K}\,.
\end{equation}
Hence,
\begin{equation}
	\label{eq:tau_eddy_flt}
	\tau_{ij}^{(\rm eddy)} = 2\langle\rho\rangle\left(\nu_{\rm sgs}\tilde{S}_{ij}^\ast-\frac{1}{3}K\delta_{ij}\right)\,.
\end{equation}
For brevity, we drop brackets and tildes indicating filtered and Favre-filtered quantities from now
onwards, so that the turbulent stresses can be written as
\begin{equation}
	\label{eq:tau_eddy}
	\tau_{ij}^{(\rm eddy)} = 2\rho\left(\nu_{\rm sgs}S_{ij}^{\,\ast}-\frac{1}{3}K\delta_{ij}\right)\,.
\end{equation}
In the following, it is understood that all quantities are either numerically resolved variables or 
modeled in terms of these variables. The production rate (energy flux) corresponding to the eddy-viscosity
closure is
\begin{equation}
	\label{eq:flux_eddy}
	\Sigma^{(\rm eddy)} = C_{\nu}\rho\Delta K^{1/2}|S^\ast|^2-\frac{2}{3}\rho K d\,,
\end{equation}
where $d=S_{ii}$ is the divergence. The eddy-viscosity coefficient $C_{\nu}$ is typically in the range
from $0.05$ and $0.1$ \cite{Sagaut,SchmNie06b,SchmFeder11}.

In the incompressible case $(d = 0)$, the eddy-viscosity closure admits only positive energy flux.
However, direct numerical simulation data show that there is a certain amount of backscattering
from smaller to larger scales, corresponding to a negative energy flux \cite{SchmNie06b,SchmFeder11}. 
This motivated a closure for the turbulent viscosity that is constructed from the determinant of the trace-free rate-of-strain
tensor \cite{PortWood01}:\epubtkFootnote{The expression for the turbulent viscosity is determined by the physical dimension 
	of viscosity, the positivity in the incompressible limit, and the requirement that it must be a scalar, independent from 
	the frame of reference. Scalars associated with the rate-of-strain tensor $\tens{S}$ are $d$, $|S^\ast|$, and 
	$\det\tens{S}^\ast$.}
\begin{equation}
	\label{eq:visc_sgs_det}
	\nu_{\rm sgs} = -\frac{C_{1}\Delta^2\det\tens{S}^\ast}{|S^\ast|^2}\,.
\end{equation}
By substituting the above expression into equation~(\ref{eq:tau_eddy}) for the turbulent stresses, it follows that
the production rate is given by
\begin{equation}
	\label{eq:flux_det}
	\Sigma^{(\rm det)} = -C_{1}\rho\Delta^2\det\tens{S}^\ast-\frac{2}{3}\rho K d\,,
\end{equation}
Since the determinant can be positive under certain flow conditions, in principle, this closure
accounts for backscattering (also known as inverse cascade). This phenomenon can be explained
by the so-called the "tornado" topology, i. e., the alignment of vortices along a single stretching
direction \cite{PortWood01}. Then the flow is contracting in one dimension and expanding in the other two, 
which results in a positive determinant. A negative determinant, on the other hand, corresponds to the
standard situation of a forward cascade transporting energy from larger to smaller eddies.

While the determinant closure modifies only the turbulent viscosity, a different expression for
the SGS turbulence stress tensor is proposed for compressible turbulence in \cite{WoodPort06}. Based on Taylor
series expansions of the velocity around grid cell centers, an appropriate normalization leads to the
non-linear closure
\begin{equation}
	\label{eq:tau_nonlin}
	\tau_{ij}^{(\rm nonlin)} = 4\rho K\frac{u_{i,k}u_{j,k}}{|\vecnab\otimes\vecu|^2}\,,
\end{equation}
where $|\vecnab\otimes\vecu|=(2u_{i,k}u_{i,k})^{1/2}$ is the norm of the velocity derivative. The above expression 
satisfies the identity $\tau_{ii} = -2\rho K$. However, it is generally not adequate as a model for the turbulence stress
tensor in LES \cite{SchmFeder11}. In contrast to the eddy-viscosity closure, rotation invariance is violated because
of the antisymmetric part of $\vecnab\otimes\vecu$. This would cause spurious production of $K$ in a uniformly
rotating fluid. A further problem is that $K = 0$ would be a fixed point of equation~(\ref{eq:pde_energy_sgs}) if all
other sources of turbulence energy are zero. This results in unphysical behavior. With the eddy-viscosity closure, 
on the other hand, $K$ can grow sufficiently fast from arbitrarily small initial values because $\nu_{\rm sgs}$ 
is proportional to $\sqrt{K}$ rather than $K$. For this reason, a linear combination of $\tau_{ij}^{(\rm nonlin)}$
and $2\nu_{\rm sgs}S_{ij}^{\,\ast}$ is used in \cite{PortWood01}, where $\nu_{\rm sgs}$ is given by equation~(\ref{eq:visc_sgs_det}).
The additional determinant term, with a small coefficient $C_1$, has the function of a seed term that triggers the production 
of turbulence energy, while the production rate vanishes for a uniformly rotating fluid.

With the standard turbulent viscosity defined by equation~(\ref{eq:visc_sgs}), the same idea leads to the
following generalized two-coefficient closure \cite{SchmFeder11}:
\begin{equation}
	\label{eq:tau_mixed}
	\tau_{ij} = 2\rho\left[C_1\Delta (2K)^{1/2}S_{ij}^{\,\ast} -
	2C_2\rho K\frac{u_{i,k}u_{j,k}}{|\vecnab\otimes\vecu|^2} - \frac{1}{3}(1-C_2)K\delta_{ij}\right]\,.
\end{equation}
The coefficient $C_2$ determines the relative contributions from the non-linear and divergence terms
to the trace $\tau_{ii}$. The purely non-linear closure corresponds to $C_1 = 0$ and $C_2 = 1$. Equation~(\ref{eq:tau_eddy}),
on the other hand, is obtained if $C_1 = C_{\nu}/\sqrt{2}$ and $C_2 = 0$. For the application in LES, it is
necessary to calibrate the closure coefficients $C_1$ and $C_2$. For supersonic turbulence, $C_1 = 0.02$
and $C_2 = 0.7$ appear to be robust values (see Section~\ref{sec:closure}). The rate of production following from
the generalized closure is
\begin{equation}
	\label{eq:flux_mixed}
	\Sigma = C_{1}\rho\Delta (2K)^{1/2}|S^\ast|^2
	-4C_2\rho K\frac{u_{i,k}u_{j,k}S_{ij}^{\,\ast}}{|\vecnab\otimes\vecu|^2} - \frac{2}{3}\rho K d\,.
\end{equation}
The first term dominates if $K^{1/2}$ is small compared to $\Delta|S^\ast|$. For strong turbulence intensity, 
i. e., $K^{1/2}\gtrsim\Delta|\vecnab\otimes\vecu|$,
the second term contributes significantly. The transition is further influenced by the ratio $C_2/C_1$.

\subsection{The Sarkar-Smagorinsky model for weakly compressible turbulence}
\label{sec:smag_sarkar}

In the case of isotropic incompressible turbulence, the mean SGS turbulence energy $\overline{K}$ 
for a sharp cutoff at the length scale $\Delta$ is obtained by integrating the Kolmogorov spectrum 
$E(k)$ over wavenumbers $k\ge\pi/\Delta$:
\begin{equation}
	\label{eq:ksgs_kolmogorov}
    \overline{K} = 
    \int_{\pi/\Delta}E(k)\dd k =
    \frac{3}{2}C\bar{\epsilon}^{2/3}
    \left(\frac{\pi}{\Delta}\right)^{-2/3}.
\end{equation}
The mean dissipation rate is therefore given by
\begin{equation}
    \bar{\epsilon} =
    \pi\left(\frac{3C}{2}\right)^{-3/2}
    \frac{\overline{K}^{3/2}}{\Delta} \approx
    0.81\frac{\overline{K}^{3/2}}{\Delta}
\end{equation}
for the Kolmogorov constant $C\approx 1.65$ \cite{Pope}. It is commonly assumed that an
expression of this form also holds for the \emph{local} dissipation rate in LES
(see, for example, \cite{Sagaut}):
\begin{equation}
    \label{eq:diss_close}
    \epsilon = 
    C_{\epsilon}\frac{K^{3/2}}{\Delta}\,,
\end{equation}
with $C_{\epsilon}\sim 1$. The above dimensional closure for the dissipation rate 
basically means that the time scale of energy dissipation is given by
$\tau_{\epsilon}\sim\Delta/\sqrt{K}$.

For subsonic compressible turbulence, closures for the dissipation rate and
pressure dilatation are obtained by separating the pressure fluctuations into
a rapid osciallatory and a slow component \cite{Sarkar92}. The resulting combined
expression for $\epsilon+\lambda$ reads
\begin{equation}
  \label{eq:sarkar}
  \rho(\epsilon+\lambda) =
  C_{\epsilon}\left[1+(\alpha_1-\alpha_3)\Masgs^2\right]\frac{\rho K^{3/2}}{\Delta} 
  + \alpha_2\Masgs\tau_{ij}^{\ast}S_{ij}^{\ast} - \frac{16}{3}\alpha_4\Masgs^2 \rho K d\,,
\end{equation}
where
\begin{equation}
  \Masgs = \frac{\sqrt{2K}}{c_{\rm s}}
\end{equation}
is the turbulent Mach number associated with the SGS velocity fluctuation $\sqrt{2K}$.

A particularly simple SGS model can be formulated by neglecting all gravitational and
transport terms associated with subgrid-scale effects in equation~(\ref{eq:pde_energy_sgs}). 
If furthermore a balance between production and dissipation is assumed, then
\[
  \Sigma \simeq \rho(\epsilon + \lambda)
\]
implies
\begin{equation}
  \label{eq:sar_smag}
  C_{\nu}\left(1-\alpha_2\Masgs\right)\Delta K^{1/2}|S^{\ast}|^2
  - \frac{2}{3}\left(1-8\alpha_4\Masgs^2\right) K d
  \simeq C_{\epsilon}\left[1+(\alpha_1-\alpha_3)\Masgs^2\right]\frac{K^{3/2}}{\Delta}\,.
\end{equation}
Here, the eddy-viscosity closure~(\ref{eq:tau_eddy}) is substituted for $\tau_{ij}^{\ast}$. 
If the above algebraic equation is solved for $K$, the PDEs~(\ref{eq:dens_flt}), (\ref{eq:momt_flt_high_Re}), 
and~(\ref{eq:pde_energy_res}) form a closed system. 
The effect of the compressibility corrections is a reduction of the production due to
anisotropic shear, $\nu_{\rm sgs}|S|^{\ast}$, by the factor $(1-\alpha_2\Masgs)$ and an enhancement 
of the solenoidal dissipation rate $C_{\epsilon}K^{3/2}/\Delta$ by a factor that
increaes with the square of $\Masgs$ ($\alpha_1$ tends to be greater than $\alpha_3$).
An extension to a non-equilibrium model based on the dynamical equation~(\ref{eq:pde_energy_sgs}) for $K$
was exploited in \cite{MaierPhD,MaierIap09} for cosmological LES of the gas in galaxy clusters.
However, this model applies only if $\Masgs$ is small compared to unity, which is the case for
turbulence in the intracluster medium. On the other hand, the correction factors are close to unity for small 
$\Masgs$ and, given the many approximations involved, it is not clear whether
they have any significant effect. Apart from that, the model definitely breaks down
in the vicinity of accretion shocks and in the cooler regions of the intergalactic medium,
where $\Masgs$ can become large compared to unity.

In the limit $\Masgs\rightarrow 0$ and $d\rightarrow 0$, the classical Smagorinsky model for incompressible turbulence 
\cite{Smago63} follows from equation~(\ref{eq:sar_smag}). In this case, 
\begin{equation}
  \label{eq:energy_smag}
  K \simeq \frac{C_\nu}{C_\epsilon}\Delta^2|S|^2\qquad\mbox{and}\qquad
  \nu_{\rm sgs} = (C_{\rm S}\Delta)^2|S|\,,
\end{equation}
where $C_{\rm S} = (C_\nu^3/C_\epsilon)^{1/4}$. The corresponding equilibrium dissipation rate is
\begin{equation}
  \label{eq:diss_smag}
  \epsilon \simeq (C_{\rm S}\Delta)^2|S|^3\,.
\end{equation}
This expression has an important implication. One could calculate $\epsilon$ from ILES data
analogous to the viscous dissipation rate following from the Navier-Stokes equations, i.~e.,
\[
  \epsilon \sim \mathrm{\nu}_{\mathrm{eff}}|S|^{2}\,.
\]
Here, $\nu_{\mathrm{eff}}=VL/\Reyn_{\rm eff}$ is assumed to be the constant numerical viscosity, which
is given by the  effective Reynolds number $\Reyn_{\rm eff}$ of the simulation \cite{PanPad09}.
However, the above estimation of the dissipation rate is clearly at odds with equation~(\ref{eq:diss_smag}).
This can be understood as follows.
If $\Reyn_{\rm eff}\sim\mathrm{Re}$, where $\mathrm{Re}$ is the physical Reynolds number 
defined by equation~(\ref{eq:Re}), then $\nu|S|^{2}$ is the physical dissipation rate
in a direct numerical simulation of a fluid with microscopic viscosity $\nu$. 
In an LES with $\Reyn_{\rm eff}\ll\mathrm{Re}$, on the other hand, the Smagorinsky model implies a 
turbulent viscosity of the order $\Delta^2|S|$ for steady-state turbulence. The turbulent viscosity
is not a constant. In this case,
the dissipation rate is approximately given by equation~(\ref{eq:diss_smag}). This is a consequence of 
equation~(\ref{eq:sgs_diss}), which implies that $\rho\epsilon$ cannot be expressed as the contraction of the filtered
viscous stress tensor with the filtered rate-of-strain tensor. The dissipation rate  
is instead given by the filtered contraction of the two tensors. As shown in \cite{SchmFeder11}, the argument remains
valid even if the dissipation rate is calculated for LES of supersonic turbulence with the advanced
SGS model presented in the following section.
 
\subsection{The compressible subgrid-scale turbulence energy model}
\label{sec:K_eq}

To determine K, one can either invoke the equilibrium condition, such as in the Smagorinsky
model, or numerically solve the PDE~(\ref{eq:pde_energy_sgs}). The latter is called the SGS turbulence energy model \cite{Sagaut,Schu75,MoinSqui91,Yoshi91,Germano92,SchmNie06b,SchmFeder11}.
If gravitational terms are negligible, the turbulence energy equation can be explicitly written as
\begin{equation}
	\label{eq:pde_energy_close}
	\begin{split}
	\frac{\partial}{\partial t}\rho K + \vecnab\cdot\left(\rho\vecu K\right) = &\,
		C_1\rho\Delta(2K)^{1/2}|S^\ast|^2 - 4C_2\rho K\frac{u_{i,k}u_{j,k}S_{ij}^{\,\ast}}{|\vecnab\otimes\vecu|^2}
		-\frac{2}{3}\rho K d\\
		&- C_{\epsilon}\frac{K^{3/2}}{\Delta} + \vecnab\cdot\left[C_\kappa\rho\Delta K^{1/2}\vecnab K\right]\,
	\end{split}
\end{equation}
where the closure~(\ref{eq:flux_mixed}) for the production rate $\Sigma$, the dissipation rate $\epsilon$ defined by 
equation~(\ref{eq:diss_close}), and the gradient-diffusion closure for $\Fdiff + \Fpress$ were substituted
into equation~(\ref{eq:pde_energy_sgs}). 
The gradient-diffusion hypothesis, which is also known as Kolmogorov-Prandtl relation, 
is based on the assumption that the turbulent transport of $K$ is a diffusion 
process satisfying Fick's law (see, for example, \cite{Pope,SchmNie06b}):
\begin{equation}
	\label{eq:grad_diff}
	\Fdiff + \Fpress = \rho\kappa_{\rm sgs}\vecnab K\,,
\end{equation}
with a turbulent diffusivity
\begin{equation}
	\kappa_{\rm sgs} = C_{\kappa}\Delta \sqrt{K} = \frac{C_{\kappa}}{C_{\nu}}\nu_{\rm sgs}\,.
\end{equation}
The Prandtl number of turbulent transport, $C_{\kappa}/C_{\nu}$, is often assumed to be of the order unity. 
For a calibration of $C_{\kappa}$, see Section~\ref{sec:closure}. The pressure-dilatation $\lambda$ is assumed to be 
negligible in equation~(\ref{eq:pde_energy_close}) because no satisfactory closure is known for the highly compressible regime 
\cite{WoodPort06,SchmFeder11}.
For weakly compressible turbulence, equation~(\ref{eq:sarkar}) could be used, but the applicability of this
closure and an appropriate calibration of the coefficients $\alpha_1,\ldots,\alpha_4$ requires further investigation.
As pointed out in Section~\ref{sec:smag_sarkar}, the contribution from $\lambda$ is too small to significantly influence K
in the weakly compressible regime. In this case, the equation with the eddy-viscosity closure, i. e., $C_1 = C_{\nu}/\sqrt{2}$ and $C_2 = 0$, can be regarded as a sufficient model for most applications. In particular, this variant of the SGS turbulence
energy model was used for simulations of thermonuclear combustion in white dwarfs (Section~\ref{sec:SN_Ia}) and 
cosmological simulations (Section~\ref{sec:clusters}).

Negligible pressure-dilatation is also a reasonable assumption at high Mach numbers because the kinetic energy is large
compared to the internal energy and non-linear interactions between turbulent velocity fluctuations should be the dominant mode
of energy transfer. This is manifest in the closure~(\ref{eq:flux_eddy}) for $\Sigma$, which is solely constructed from 
the gradient of the resolved velocity field, but does not depend on density or pressure gradients. There are both 
theoretical and numerical studies in support of this conjecture. In \cite{Aluie11,Aluie13}, it is argued that a range of 
length scales exists, in which the kinetic and internal energies decouple and the flux through the kinetic energy cascade 
becomes asymptotically constant, while $\rho\lambda$ is subdominant. The computation of the different contributions to the total energy flux at varying length scales from supersonic turbulence data in \cite{KritWag13} confirms this conclusion. 
Based on an analytical theory for the two-point correlations of compressible turbulence \cite{GalBan11}, it is shown that 
the main contribution to the energy flux is
\begin{equation}
	F_{\parallel}(r) = \langle \delta(\rho\vecu)\cdot\delta\vecu\,\delta u_{\parallel} \rangle
\end{equation}
where $\delta\vecu$ is the velocity difference between two points separated by a distance $r$, $\delta u_{\parallel}$
is the longitudinal component of the velocity difference in the direction of $\vect{r}$, and the brackets
denote the ensemble average. The closure~(\ref{eq:flux_mixed}) for the $\Sigma$ has a similar structure,
with factors of $\rho K^{1/2}$ and derivatives of $\vecu$ corresponding to fluctuations on the grid scale. 

However, as pointed out in Section 2.5.2 of \cite{Garnier}, a subtlety arises in the presence of shocks
because the Rankine-Hugeniot conditions for jumps across shock fronts should be filtered in place of the PDEs. 
This entails SGS terms that are different from the terms in the filtered PDEs. However,
it is questionable whether any attempt to model these terms would be useful. 
The assumption that the unmodified jump conditions apply to the numerical solution 
amounts to a fallback from LES with an explicit SGS model to ILES.
Since shock-capturing schemes, such as PPM, fall back to stronger diffusion in the vicinity of 
shocks, this is probably the most reasonable thing one can do. Nevertheless, 
the closure~(\ref{eq:flux_mixed}) captures the non-linear interscale
transfer of energy due to supersonic turbulent velocity fluctuations. The SGS model outlined above
accounts for the statistical effect of shocks as well as vortices interacting with each other 
across the grid scale \cite{SchmFeder08,SchmFeder09}, 
while any SGS terms in the jump conditions would mainly correct 
geometric differences between smoothed shock fronts 
and the corresponding unfiltered fronts with substructure on smaller length scales 
(just like the turbulent flame fronts discussed in Section~\ref{sec:SN_Ia}).

\epubtkImage{}{
  \begin{figure}[htbp]
    \centerline{\includegraphics[width=0.67\textwidth]{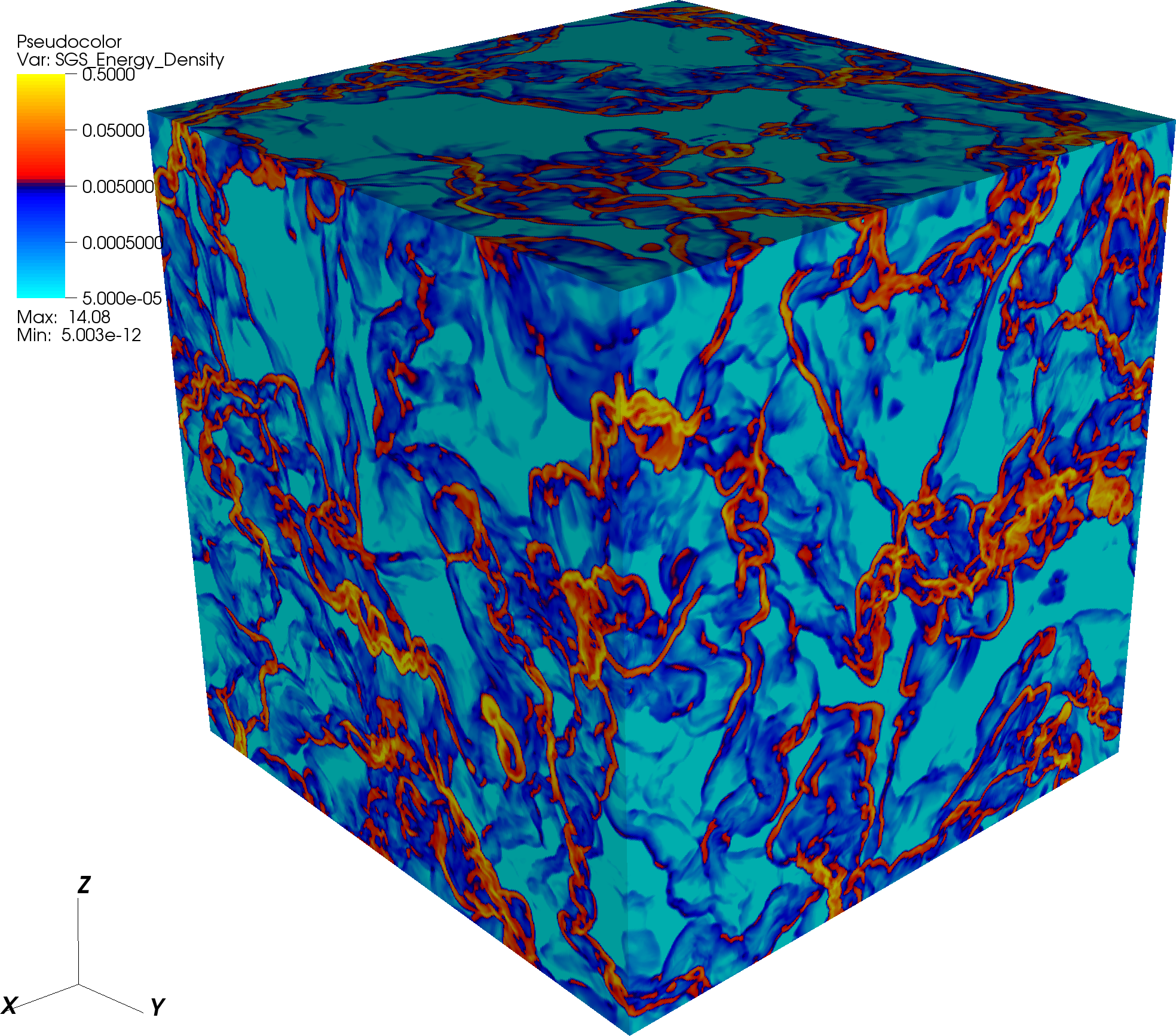}}
    \caption{Visualization of the SGS turbulence energy density $\rho K$ in a $512^{3}$ LES with solenoidal forcing
	\cite{SchmFeder11}.}
    \label{fig:ksgs_soln}
\end{figure}}

Indeed, equation~(\ref{eq:pde_energy_close}) for $\rho K$ was demonstrated to work very well in the highly compressible regime \cite{SchmFeder11}. As an example, Figure~\ref{fig:ksgs_soln} shows a visualization of $\rho K$ from an LES of isotropic supersonic turbulence, where solenoidal stochastic forcing maintains a root mean square Mach number of about $5$ in the statistically stationary regime.\epubtkFootnote{In this simulation, a quasi-isothermal equation of state is
	applied with an adiabatic exponent $\gamma=1.001$.} The numerical resolution is $512^3$. In the reddish regions of the plot, $K_{\mathrm{sgs}}$ is higher than the spatial average, while it is lower in the bluish regions. The structure of the numerically resolved turbulent flow is illustrated by the so-called denstrophy,
\[
	\Omega_{1/2}=\frac{1}{2}\left|\vecnab\times\left(\rho^{1/2}\vecu\right)\right|^2\,,
\] 
in Figure~\ref{fig:denstr_soln}. Since $\Omega_{1/2}$ combines density fluctuations and the rotation of the velocity, $\vecnab\times\vecu$, it indicates both small-scale compression and eddy-like motion \cite{KritNor07}. There is clearly a correlation between $\Omega_{1/2}$ and $\rho K$, which reflects the local interaction between resolved small-scale modes and subgrid-scale turbulence, as expressed by the production terms in equation~(\ref{eq:pde_energy_close}). This correlation is akin to the equilibrium  condition~(\ref{eq:energy_smag}) following from the Smagorinsky model for incompressible turbulence. Owing to the non-local effects in the PDE~(\ref{eq:pde_energy_close}), however, the SGS turbulence energy cannot be reliably estimated from local quantities such as $\Omega_{1/2}$ \cite{SchmFeder11}. In particular, turbulent diffusion smears out $\rho K$ in comparison to $\Omega_{1/2}$. 

\epubtkImage{}{
  \begin{figure}[htbp]
    \centerline{\includegraphics[width=0.67\textwidth]{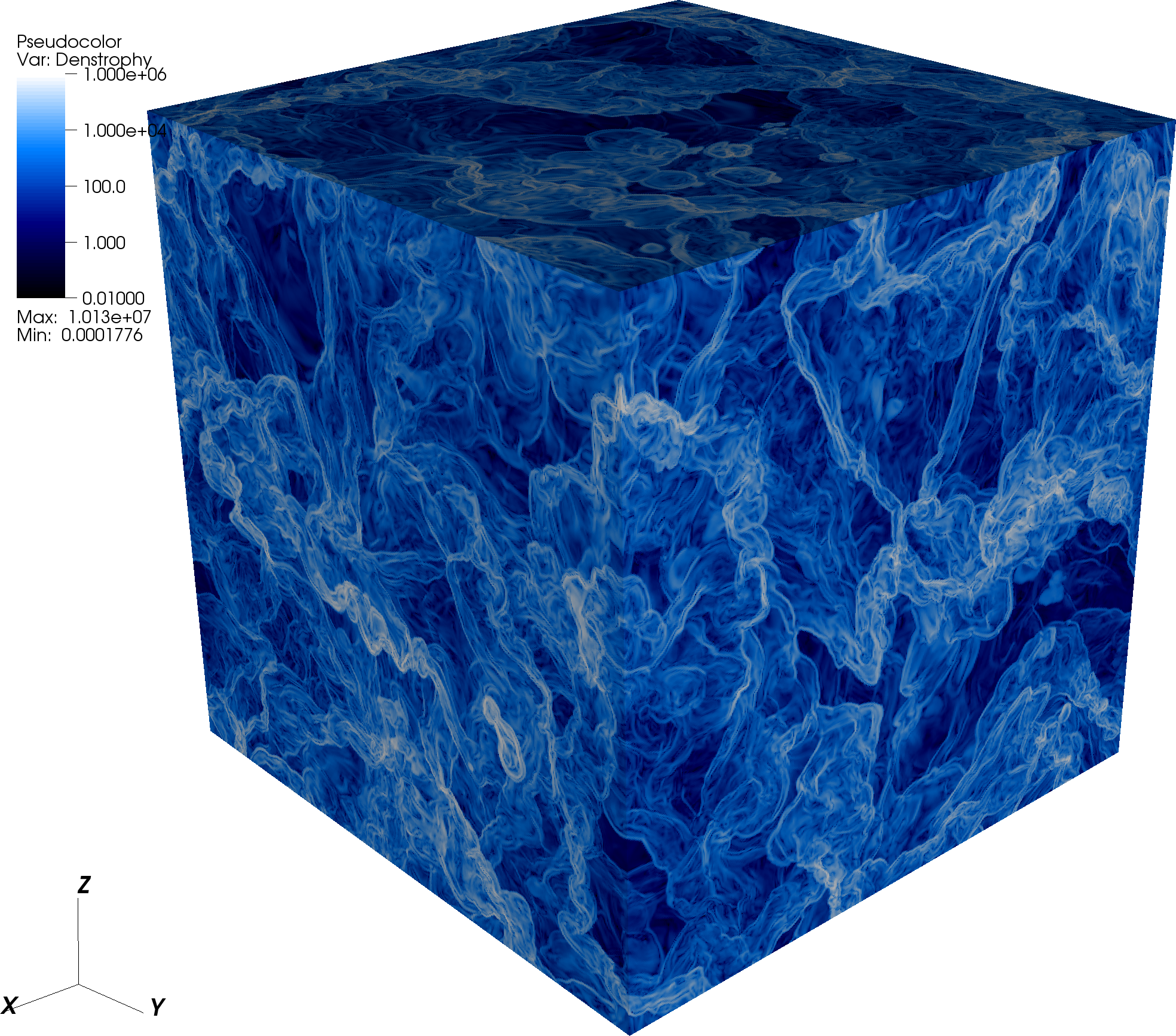}}
    \caption{Visualization of the the denstrophy $\Omega_{1/2}$ for the same LES as in Figure~\ref{fig:ksgs_soln}.}
    \label{fig:denstr_soln}
\end{figure}}

\epubtkImage{}{
  \begin{figure}[htbp]
    \centerline{\includegraphics[width=0.49\linewidth]{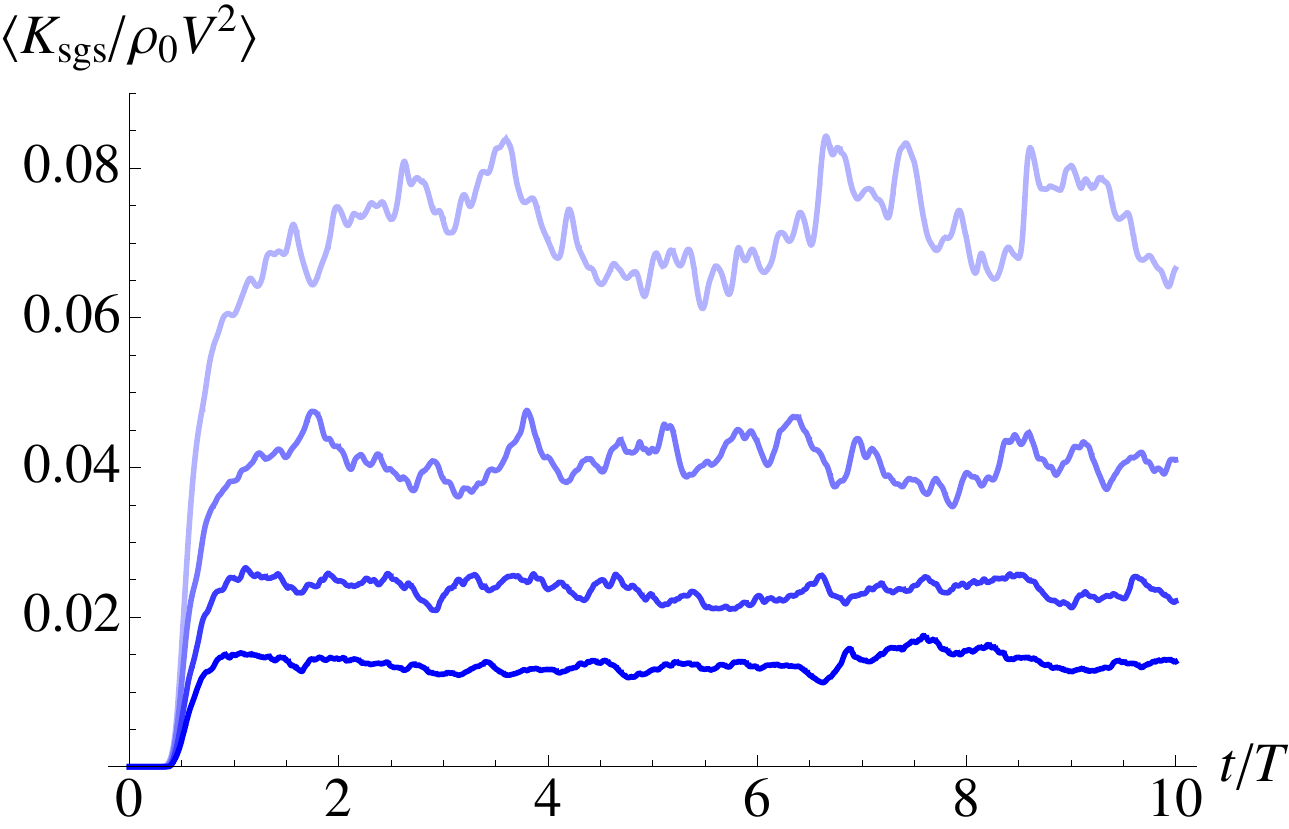} 
    	\includegraphics[width=0.49\linewidth]{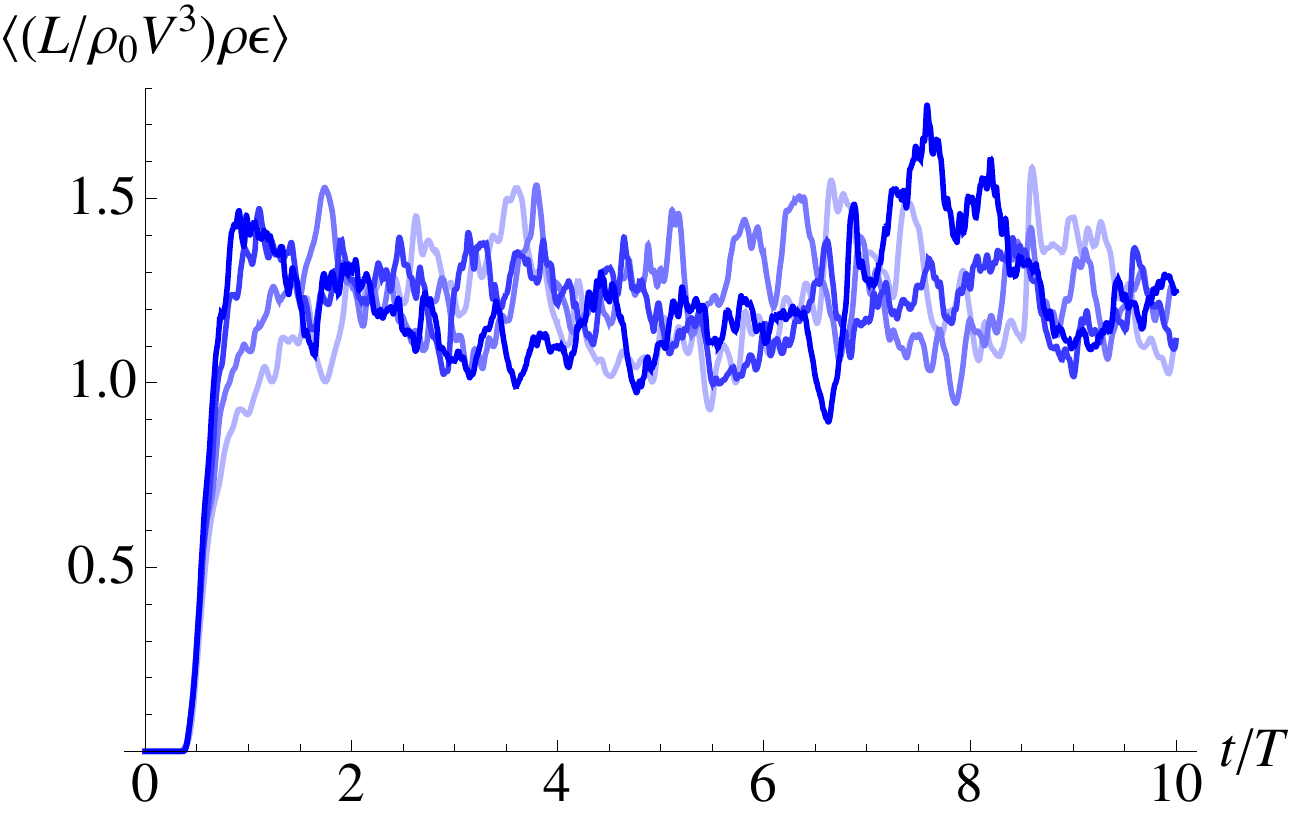}}
    \caption{Temporal evolution of the spatially averaged SGS turbulence energy (left) and the dissipation rate (right) 
        for forced supersonic turbulence \cite{SchmFeder11}. The grid scale $\Delta$ decreases from $L/32$ 
		(light colour) to $L/256$ (full colour).}
    \label{fig:statistics_soln}
\end{figure}}

\epubtkImage{}{
  \begin{figure}[htbp]
    \centerline{\includegraphics[width=0.5\linewidth]{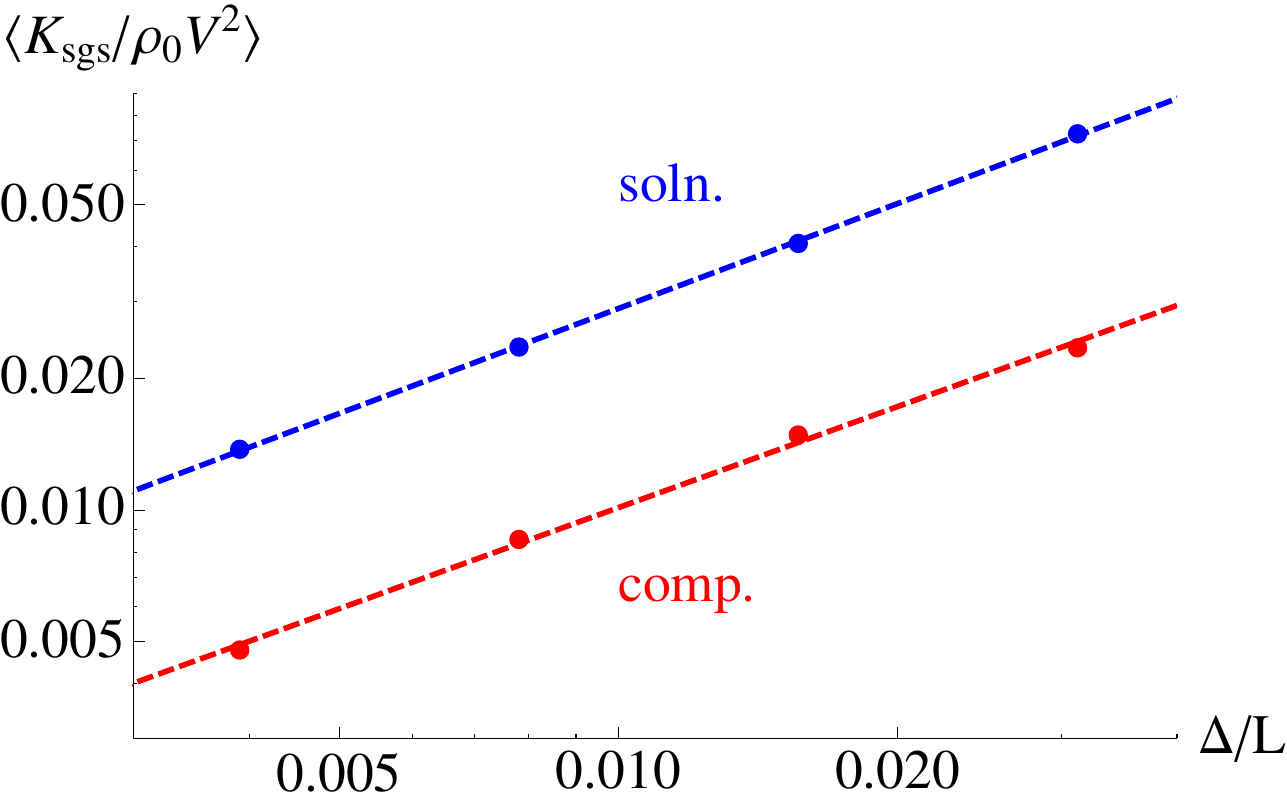}}
    \caption{Time-averaged mean values of the SGS turbulence energy in LES with different resolutions
    	(dots) and power-law fits (dashed lines) for solenoidal and compressive forcing \cite{SchmFeder11}.}
    \label{fig:K_power_law}
\end{figure}}

\epubtkImage{}{
  \begin{figure}[htbp]
    \centerline{\includegraphics[width=0.5\linewidth]{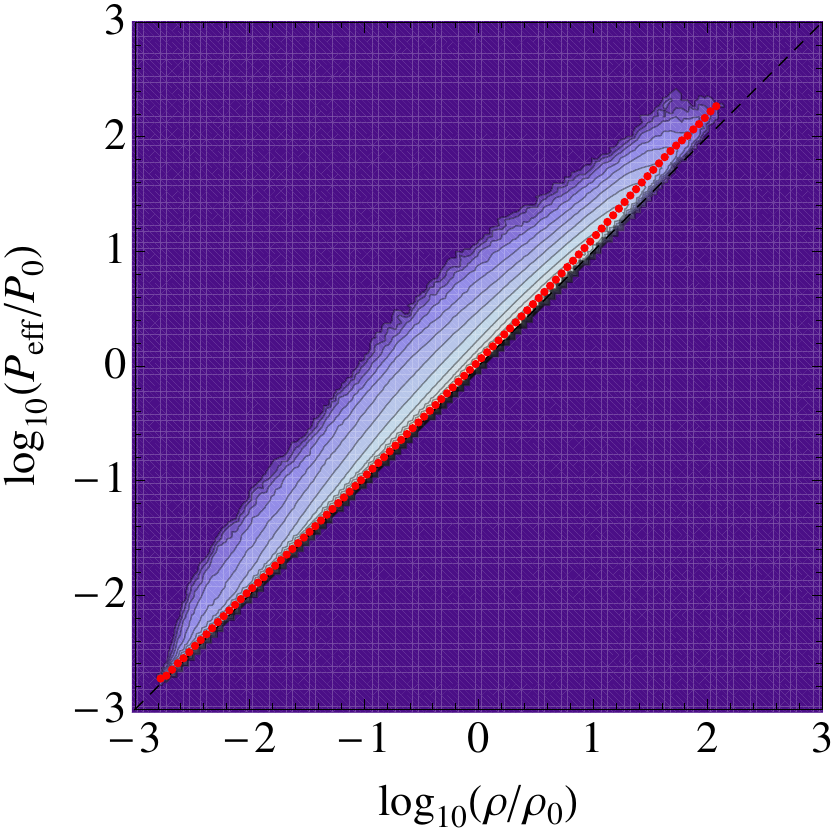}}
    \caption{Phase diagram of the effective pressure defined by equation~(\ref{eq:press_eff}) vs.\ the  mass density
    	\cite{SchmFeder11}. Both quantities are normalized by their mean values, $P_0$ and $\rho_0$.
		 The thin dashed line corresponds to the isothermal relation $P\propto\rho$, and the dotted line
		 indicates the mean effective pressure for given mass density.}
    \label{fig:press_eff}
\end{figure}}

A critical property is the scaling behavior of the SGS turbulence energy. For statistically stationary homogeneous turbulence,
the mean value of $\rho K$ should scale as a power of the grid resolution because the fraction of unresolved kinetic energy
changes as the the cutoff of the energy spectrum is shifted (see equation~\ref{eq:ksgs_kolmogorov}). This was verified in 
\cite{SchmFeder11} by running LES with different grid scales $\Delta$ and fixed forcing length $L$. The global spatial averages
$\langle\rho K\rangle$ in these simulations are plotted in Figure~\ref{fig:statistics_soln} (left panel) for $\Delta$ ranging 
from $L/256$ to $L/32$, where the case $\Delta=L/256$ corresponds to the $512^3$ simulation depicted in Figures \ref{fig:ksgs_soln} and~\ref{fig:denstr_soln}. Although there are substantial fluctuations, one can qualitatively see that $\langle\rho K\rangle$ decreases with $\Delta$. Time-averaging over the statistically stationary
regime yields mean values that are close to the power law 
\begin{equation}
	\label{eq:sgs_scaling}
	\langle \rho K\rangle \propto \Delta^{\alpha},
\end{equation}
with $\alpha\approx 0.799\pm0.009$ (see Figure~\ref{fig:K_power_law}). The scaling exponent is in between the Kolmogorov and
Burgers exponents and roughly comparable to the slope of the second-order structure functions with fractional mass-weighing 
reported in \cite{SchmFeder08}.

In the filtered momentum and energy equations~(\ref{eq:momt_flt}) and~(\ref{eq:pde_energy_res}), respectively, 
the trace of the SGS turbulence stress tensor acts as an additional turbulent pressure. The sum of the thermal
and turbulent pressures is sometimes called the effective pressure:
\begin{equation}
	\label{eq:press_eff}
	P_{\mathrm{eff}} = P + \frac{2}{3}\rho K\,.
\end{equation}
It is important to keep in mind that $P_{\mathrm{eff}}$ depends on the numerical resolution and
$P_{\mathrm{eff}}\rightarrow P$ in the limit $\Delta\rightarrow 0$ (DNS).
Figure~\ref{fig:press_eff} shows a phase plot of the effective pressure vs.\ the mass density for the highest-resolution case. One can see that the average of the effective pressure for a given mass density closely follows the isothermal relation $P\propto \rho$. Although the mean turbulent pressure $\frac{2}{3}\rho K$ is small compared to the thermal pressure for the resolution $\Delta=L/256$, the intermittency of turbulence can locally produce an effective pressure 
that exceeds the thermal pressure by one order of magnitude. Consequently, the turbulent pressure can become important for
compressible turbulence, particularly if there are other sources than the turbulent cascade. As an example,
turbulent feedback in galaxy simulations is discussed in Section~\ref{sec:galaxy}.
In addition to the turbulent pressure, the non-diagonal turbulent stresses 
$\tau_{ij}^{\ast}$ act on the resolved flow. In the case of the eddy-viscosity closure, $\tau_{ij}^{\ast}$ causes
a diffusion effect on top of the numerical diffusion, which occurs regardless of the compressibility of the flow.\epubtkFootnote{The diffusion on numerically resolved length scales due to the turbulent stresses $\tau_{ij}^{\ast}$ must not
	be confused with the subgrid-scale diffusion of $K$, which is given by equation~(\ref{eq:grad_diff}).}
For strongly diffusive numerical schemes, this effect is marginal. For high-resolution schemes, on the other hand,
the explicit turbulent stresses in LES can become significant. Moreover, the non-linear term in equation~(\ref{eq:tau_mixed}) 
modifies the diffusion-like tensor in the case of supersonic turbulence. Of course, adding the turbulent stresses in LES
does not merely degrade a high-resolution scheme to a more diffusive scheme because the diffusion is linked to the non-linear turbulent interactions across the grid scale. For non-turbulent flow, the turbulent stresses should vanish
if the SGS model is consistent. 

\subsection{Two equation models and gravity}
\label{sc:two_equation_grav}

In the framework of the Renolds-averaged Navier-Stokes equations (RANS), which are equivalent to the filtered
Navier-Stokes equation in the limit of a filter scale comparable the integral scale of the flow, the 
$K$-$\epsilon$ turbulence model can be used to calculate both the turbulence energy $K$ and the 
dissipation rate $\epsilon$. Two inhomogeneous PDEs of the advection-diffusion type 
determine $K$ and $\epsilon$ \cite{Pope}. In contrast to the simple dimensional closure~(\ref{eq:diss_close})
with a single coefficient of order unity, the diffusion and source terms in the equation for $\epsilon$
come with several additional closure coefficients. This type of model is commonly used for industrial and 
environmental flows. 

A two equation model of similar structure is proposed for buoyancy-driven flows in \cite{DiTip06}. The model predicts the energy $K$ and characteristic size $L$ of the dominant eddies produced by Rayleigh-Taylor and Richtmeyer-Meshkov instabilities. The evolution of these variables is given by the following two equations (in notation adapted to this review):
\begin{align}
	\frac{\partial}{\partial t}\rho L + \vecnab\cdot\left(\rho\vecu L\right) = &\,
		 \rho\sqrt{2K} + C_L\rho L d + \vecnab\cdot\left(\rho\kappa_L\vecnab L\right),\\
	\frac{\partial}{\partial t}\rho K + \vecnab\cdot\left(\rho\vecu K\right) = &\,
		\Gamma + \tau_{ij}^\ast S_{ij} - \frac{2}{3}\rho K d
		- C_{\epsilon}\frac{K^{3/2}}{L} + \vecnab\cdot\left(\rho\kappa_K\vecnab K\right)\,.
\end{align}
The production rate due to the Rayleigh-Taylor instability is basically given by
$\Gamma \propto \rho\sqrt{2K}\,g_{\rm eff}$.
For each grid cell, the buoyant acceleration $g_{\rm eff}$ is assumed to 
depend on the density contrast across the cell faces, the length scale $L$, 
and the components of gravity along the three coordinate axes. 
Apart from differences in the determination of $g_{\rm eff}$, the above expression for $\Gamma$ 
is the same as in equation~(\ref{eq:gamma_RT}) for the enhanced production of turbulence at
the interface between low-density and high-density material, which is applied to the
propagation of turbulent flame fronts in thermonuclear supernovae (see Section~\ref{sec:SN_Ia}).

The $K$-$L$ model outlined above was adopted as an SGS model for simulations of turbulence driven by active galactic nuclei 
in galaxy clusters \cite{ScanBruegg08}. In this case, turbulence is thought to be stirred by hot bubbles rising due to their
buoyancy in the ICM. These bubbles originate from the AGNs in the cluster. 
For this reason, production through the turbulent cascade is set to zero,
i.~e., $\tau_{ij}^\ast = 0$.\epubtkFootnote{In \cite{GrayScan11}, the model is extended with an eddy-viscosity closure for
	$\tau_{ij}^\ast$. 
} 
In contrast to LES based on the consistent decomposition derived in Section~\ref{sc:decomp_navier},
it follows from the very concept of the $K$-$L$ model that $K$ cannot be interpreted
as the kinetic energy associated with velocity fluctuations below the grid scale.
Since $K$ is the kinetic energy associated with the dominant eddies driven by the RT instability on a length scale $L$, 
where $L$ is a dynamical variable that can become larger than the grid scale, $K$ generally encompasses some fraction 
of the numerically resolved turbulence energy on top of the SGS turbulence energy 
(if $L$ falls below the grid resolution, on the other hand, $K$ will represent 
only a fraction of the SGS turbulence energy). Consequently, the $K$-$L$ model works as a hybrid model in these simulations, which stand somewhere between RANS and LES. As a result, the resolution-dependent small-scale structure of the RT-unstable bubbles is smeared out and only the coherent structure on large scales is captured in \cite{ScanBruegg08}. LES, 
on the other hand, resolve small-scale structure always down to the grid scale.

For the general case of self-gravitating turbulent gas, no satisfactory closure for $\Gamma$ has been found yet. 
A conceptual difficulty is that the acceleration caused by gravity is genuinely anisotropic, while 
SGS models such as the turbulence energy model are based on local isotropy. The usual solution to this problem is
to resolve the flow down to length scales that are not strongly affected by gravity. 
In AMR simulations, this is achieved by imposing a Truelove-like resolution criterion such that
a sufficiently large ratio between the local Jeans length and the grid scale is maintained \cite{TrueKlein97,FederSur11}.
Since the density of gravitationally unstable gas would increase indefinitely, 
excess mass is usually dumped into sink particles at the highest refinement level \cite{KrumKee04,FederBan10,WangLi10}. 
Thereby, collapsing gas is decoupled form the numerically computed gas dynamics. In a certain sense, 
a sink particle is nothing but an SGS model for a self-gravitating overdense cloud that collapses down to
scales below the minimal grid scale. Despite being a crude model, sinks particles are a reasonable approximation to
collapsed clouds because they mainly interact through accretion (i.~e., mass accumulation) with the numerically
resolved gas dynamics. A more complex situation is encountered if the objects represented by the sink particles
produce feedback onto the gas. An example is stellar feedback in galaxy simulations, which can be treated with the approach discussed in Section~\ref{sec:galaxy}.

\newpage


\section{Determination of Closure Coefficients}
\label{sec:closure}

One of the basic assumptions of the Kolmogorov theory is that
turbulence is statistically self-similar in the inertial subrange
(see, for example, \cite{Frisch}). With regard to subgrid-scale
closures, the self-similarity of turbulence implies that
dimensionless coefficients such as $C_{\nu}$ in equation~(\ref{eq:visc_sgs})
should be independent of the chosen filter scale. This is not only a
necessary condition for the feasability of LES, but it also allows for 
the calibration of closure coefficients by explicitly filtering turbulence
data. Since closures do not exactly match SGS terms, an improved approximation
can be achieved by so-called dynamical procedures, which estimate coefficients
from properties of the numerically resolved flow under the assumption of local
self-similarity.

\subsection{Hierarchical filtering}
\label{sec:hierarch_flt}

As a formal framework, let us consider an infinite series of
isotropic and time-independent filter operators
$\langle\ \rangle_{n}$. Each filter is defined by a
kernel $G_n(r)$ with filter length $\Delta_n$ (see Section~\ref{sec:separation}). 
We shall assume that $\Delta_0\sim L$, where $L$ is the integral length scale of the flow,
and 
\begin{equation}
  \label{eq:self_siml}
  \forall n\in\mathbb{N}_{0}: 
  G_{n}(\vect{x}) = \lambda^{3}G_{n-1}(\lambda\vect{x}),\quad\mbox{where\ $\lambda > 1$,}
\end{equation}
i.~e., $\langle\ \rangle_{n}$ for $n=0,1,2,\ldots$ is a self-similar hierarchy of filters.
Typical examples are the box filter defined by equation~(\ref{eq:box_filter}) or
the Gaussian filter, which has the kernel \cite{Sagaut,Pope}
\begin{equation}
  \label{eq:gauss_flt}
  G_{n}(r) = \left(\frac{6}{\pi\Delta_{n}^{2}}\right)^{3/2}
    \exp\left(-\frac{6r^{2}}{\Delta_{n}^{2}}\right)
\end{equation}
for isotropic filter lengths $\Delta_{n}=\Delta_{0}/\lambda^n$. Since $G_{n}(r)\rightarrow\delta(r)$
in the limit $n\rightarrow\infty$, $\langle\ \rangle_{\infty}$ is the idenity operator.
Because filtering in physical space corresponds to a multiplication with the transfer function
of the filter in Fourier space, it follows that 
\begin{equation}
  \label{eq:flt_twice}
  \langle\langle q\rangle_{m}\rangle_{n}\simeq\langle q\rangle_{n}\quad
  \mbox{if $\Delta_{n}\gg\Delta_{m}$}.
\end{equation}
For Gaussian filters, the validity of this approximation becomes immediately clear by
calculating the product of the transfer functions:
\begin{equation}
  \widehat{G}_{m}(k)\widehat{G}_{n}(k) = 
  \exp\left[-\frac{k^{2}(\Delta_{m}^{2}+\Delta_{n}^{2})}{24}\right] \simeq 
  \exp\left[-\frac{k^{2}\Delta_{n}^{2}}{24}\right] = \widehat{G}_{n}(k)\,.
\end{equation}

We can now apply the scale separation of the Navier-Stokes equations introduced 
in Section~\ref{sec:separation} at different levels of the filter hiearchy. In particular, the
filtered density field at the $n$-th level is $\langle\rho\rangle_n$, and the 
Favre-filtered velocity is given by
\begin{equation}
  \fvecu^{[n]} = \frac{\langle\rho\vecu\rangle_n}{\langle\rho\rangle_n}\,.
\end{equation}
By filtering twice at levels $m$ and $n$, we obtain 
\begin{equation}
  \fvecu^{[m][n]}\langle\langle\rho\rangle_m\rangle_{n} = 
  \langle\langle\rho\rangle_m\vecu^{[m]}\rangle_{n} =
  \langle\langle\rho\vecu\rangle_{m}\rangle_{n}\,.
\end{equation}
If the $n$-th level is much coarser than the $m$-th level, 
the asymptotic relation~(\ref{eq:flt_twice}) for $\Delta_{n}\gg\Delta_{m}$ implies
\begin{equation}
  \label{eq:vel_flt_twice}
  \fvecu^{[m][n]}\langle\langle\rho\rangle_m\rangle_{n} =
  \langle\langle\rho\vecu\rangle_{m}\rangle_{n} \simeq
  \langle\rho\vecu\rangle_n =
  \fvecu^{[n]}\langle\rho\rangle_n\,.
\end{equation}

The turbulence stress tensor on the length scale $\Delta_n$ of the $n$-th filter is defined by
\begin{equation}
    \label{eq:turb_stress_n}
    \tens{\tau}^{[n]} =
    -\langle\rho\vecu\otimes\vecu\rangle_{n} + \langle\rho\rangle_n \fvecu^{[n]}\fvecu^{[n]}.
\end{equation}
The stress tensors for two filter levels $m$ and $n$, where $\Delta_m<\Delta_n$, are
related by the Germano identity  
(see Section 3.3.3 in \cite{Sagaut} and \cite{Germano92,SchmidtPhD}):
\begin{equation}
  \label{eq:germano}
  \tens{\tau}^{[m][n]} = 
  \langle\tens{\tau}^{[m]}\rangle_{n} + \tens{\tau}^{[m,n]}\,.
\end{equation}
The stress tensor associated with the double-filtered variables
is defined by
\begin{equation}
  \begin{split}
  \tens{\tau}^{[m][n]} =
  &-\langle\langle\rho \vecu\otimes \vecu\rangle_{m}\rangle_{n}
  + \langle\langle\rho\rangle_{m}\rangle_{n}\fvecu^{[m][n]}\fvecu^{[m][n]}= \\
  &-\langle\langle\rho \vecu\otimes \vecu\rangle_{m}\rangle_{n}
  + \frac{\langle\langle\rho \vecu\rangle_{m}\rangle_{n}\otimes
          \langle\langle\rho \vecu\rangle_{m}\rangle_{n}}
         {\langle\langle\rho\rangle_{m}\rangle_{n}}
  \end{split}
\end{equation}
and
\begin{equation}
  \label{eq:leonard_stress_general}	
  \tens{\tau}^{[m,n]} =
  -\langle\langle\rho\rangle_m\fvecu^{[m]}\otimes\fvecu^{[m]}\rangle_{n} + 
  \frac{\langle\langle\rho\rangle_m\fvecu^{[m]}\rangle_{n}\otimes
        \langle\langle\rho\rangle_m \fvecu^{[m]}\rangle_{n}}
       {\langle\langle\rho\rangle_m\rangle_{n}}
\end{equation}
is the Leonard stress tensor, which is associated with velocity fluctuations 
in the intermediate range of length scales $\Delta_m \le \ell \le \Delta_{n}$. 
The Germano identity also holds for two arbitrary filters
in the hierarchy. In the limit $\Delta_{n}\gg\Delta_{m}$, the
contribution from $\langle\tens{\tau}^{[m]}\rangle_{n}$ becomes
negligible and
\begin{equation}
  \label{eq:turb_stress_asympt}
  \tens{\tau}^{[m][n]} \simeq \tens{\tau}^{[m,n]} \simeq \tens{\tau}^{[n]}\,,
\end{equation}
where the second relation follows from equation~(\ref{eq:flt_twice}).
As a consequence, the turbulent stresses associated with the scale $\Delta_{n}$
are not sensitive to the flow structure on much smaller scales. In particular,
it follows that $K^{[n]}\simeq K^{[m,n]}$ if $\Delta_m\ll\Delta_n$.

In \cite{SchmidtPhD,SchmNie06b,SchmFeder11}, Gaussian filters (see equation~\ref{eq:gauss_flt}) are applied to 
data form ILES of forced compressible turbulence for the verification of closures. 
The following line of reasoning is of central importance for estimating closure coefficients from 
finite-resolution data. To begin with, let us assume that $\rho$
and $\vecu$ are the physical density and velocity fields. Let us further assume that
the implicit filter of the ILES correspond to the filter level $m=I$, i.~e.,
$\langle\rho\rangle_I$ and $\fvecu^{[I]}$ represent the numerically computed density and velocity fields.
Now, if the numerical data are coarse-grained by an explicit filter $\langle\ \rangle_n$
in the inertial subrange,
the turbulence stress tensor $\tens{\tau}^{[I,n]}$ defined by equation~(\ref{eq:leonard_stress_general})
can be calculated. But a closure for the turbulent stresses on the the length scale $\Delta_n$ applies to 
$\tens{\tau}^{[n]}$, which is defined by equation~(\ref{eq:turb_stress_n}). 
If $\Delta_n$ is reasonably large compared to $\Delta_{I}$, however, one can make use of the approximation
$\tens{\tau}^{[n]}\simeq\tens{\tau}^{[I,n]}$ (see relation \ref{eq:turb_stress_asympt} for $m=I$). 
Owing to equation~(\ref{eq:vel_flt_twice}), the distinction between, on the one hand, 
the physical densities and velocities (or DNS data) and, on the other hand, the ILES data becomes immaterial. 
It is thus possible to calculate coarse-grained eddy-viscosity coefficients as:
\[
  C_\nu \simeq
  \frac{\tau_{ij}^{[n]\ast}S_{ij}^{[n]}}
       {\langle\rho\rangle_n\Delta_{n}\left(K^{[n]}\right)^{1/2}\left|S^{[n]\ast}\right|^2}\,.
\]
Of course, since the eddy-viscosity closure is not exact, the value of $C_\nu$ varies. 
But the mean value turns out to be roughly $0.05$ for different
filter lengths and simulation parameters \cite{SchmidtPhD}, in good agreement with other estimates in the literature.
The same method was used to determine the coefficient $C_{\kappa}$ for the gradient-diffusion closure~(\ref{eq:grad_diff}).
The result $C_{\kappa}\approx 0.4$ implies a Prandtl number $C_{\kappa}/C_{\nu}$
around $10$, contrary to the common assumption that the kinetic Prandtl number is of the order unity
\cite{Sagaut}.

\subsection{Dynamical procedures}
\label{sec:dyn_proc}

Subgrid-scale models in their standard form apply to statistically
stationary and isotropic turbulence. But turbulent flows in nature
often deviate from this idealization:
In terrestrial applications, flow inhomogeneities are inevitably caused by
boundary conditions (``walls''). In astrophysics, one of the major 
energy sources is gravity. It causes matter to clump (galaxies and clusters)
or to move under the action of central gravitational fields (stars),
which produces inherently inhomogeneous and anisotropic flows.
For example, turbulent convection in stars introduces a vertical anisotropy
of the flow. Turbulence driven by violent energy release (supernovae)
can also be highly inhomogeneous.

One of the solutions to this problem is to \emph{localize} closures, i.~e.,
to calculate local closure coefficients. This requires local estimators
that take properties of the flow in some small region as input.
Obviously, this works only if the size of this region is not significantly
affected by the flow inhomogeneity on larger scales. In other words,
the flow must be asymptotically homogeneous and isotropic at least
on length scales of the order of the grid scale. 
In this case, a so-called test filter $\langle\ \rangle_{\rm T}$ can be applied in LES, 
with a filter length $\Delta_{\rm T}$ that is a small multiple of the grid scale $\Delta$.
Test filters are usually implemented as discrete filters over several grid cells
(see Section 2.3.2 in \cite{Garnier}). A multi-dimensional test filter can be composed as a 
succession of one-dimensional filters.\epubtkFootnote{ 
	For filters with large stencils, test filtering can 
	nevertheless become too inefficient because of the access to remote blocks of
	memory. 
} The test filter length $\Delta_{\rm T}$ can be adjusted
by varying the weights of the cells. An optimal ratio $\gamma_{\rm T}=\Delta_{\rm T}/\Delta$
is given by the closest match between the filter transfer functions of the discrete and analytical
box filters with filter length $\Delta_{\rm T}$ \cite{VasilLund98}. 
For instance, a test filter with $\gamma_{\rm T}=2.771$ is optimal if a five-point stencil is used 
in each spatial dimension \cite{SchmidtPhD}.

By identifying $\Delta_m$ with $\Delta$ and $\Delta_{n}$ with $\Delta_{\rm T}$,
the Germano identity~(\ref{eq:germano}) allows us to express the
turbulence stress tensor on the length scale of the test filter as the sum of
the test-filtered SGS turbulence stress tensor and the Leonard
tensor for the intermediate velocity fluctuations (see also Section 4.3 in
\cite{Sagaut}):
\begin{equation}
  \label{eq:germano_test}
  \tens{T} = 
  \langle\tens{\tau}\rangle_{\rm T} + \tens{L}\,,
\end{equation}
Here, the Leonard tensor $\tens{L}$ associated with the
test filter is defined by
\begin{equation}
	\label{eq:leonard_test}
  \tens{L} =
  -\langle\rho\vecu\otimes\vecu\rangle_{\rm T} + 
  \frac{\langle\rho\vecu\rangle_{\rm T}\otimes\langle\rho\vecu\rangle_{\rm T}}
       {\langle\rho\rangle_{\rm T}}\,,
\end{equation}
where we use the simplified notation $\rho$ and $\vecu$ for the density
and velocity on the grid scale, as in Section~\ref{sec:turb_stress}.
Because of the scale-invariance of turbulence, \cite{GerPio91} proposed that
the eddy-viscosity closure~(\ref{eq:tau_eddy}) holds for both $\tens{T}$
and $\tens{\tau}$. In the case of the Smagorinsky model 
(see equation~\ref{eq:energy_smag} for $\nu_{\rm sgs}$), the corresponding tensors are:
\begin{align}
  \label{eq:prod_germ_lilly}
  \tau_{ij}^{\ast} &=
    2\rho (C_{\mathrm{S}}\Delta)^{2}|S|S_{ij} =: C_{\mathrm{S}}^2\beta_{ij}\,, \\
  T_{ij}^{\ast} &\simeq
    2\rho_{\rm T}(C_{\mathrm{S}}\Delta_{\rm T})^{2}|S_{\rm T}|(S_{\rm T})_{ij} =:
    C_{\mathrm{S}}^2\alpha_{ij}\,.
\end{align}
The rate-of-strain tensor $(S_{\rm T})_{ij}$ at the test filter level is
given by the symmetrized derivative of the test-filtered numerically resolved velocity field, 
$\partial_i\langle u_j\rangle_{\rm T}$, analogous to equation~(\ref{eq:strain_flt}). 
The variable $C_{\mathrm{S}}(\vecx,t)$ needs to be determined. This can be achieved by substituting
the above expressions for $\tau_{ij}^{\ast}$ and $T_{ij}^{\ast}$ into the trace-free part of the
Germano identity~(\ref{eq:germano_test}), 
which implies
\begin{equation}
  \label{eq:leonard_GL}
  L_{ij}^{\ast} \simeq C_{\rm S}^2\alpha_{ij}-\langle C_{\rm S}^2\beta_{ij}\rangle_{\rm T}\,.
\end{equation}
Under the assumption that $C_{\rm S}$ varies only little over the smoothing length of the test filter,
one can set
$\langle C_{\rm S}^2\beta_{ij}\rangle_{\rm T}\simeq C_{\rm S}^2\langle\beta_{ij}\rangle_{\rm T}$.
Since $L_{ij}$ can be evaluated from equation~(\ref{eq:leonard_test}),
minimalization of the residual error between $L_{ij}^{\ast}$ and the expression on the right-hand side
of equation~(\ref{eq:leonard_GL}) yields
\begin{equation}
  C_{\mathrm{S}}^2 = \frac{m_{ij}L_{ij}^{\ast}}{m_{ij}m_{ij}},
\end{equation}
where $m_{ij}=\alpha_{ij}-\langle\beta_{ij}\rangle_{\rm T}$. This is the 
Germano-Lilly dynamical procedure, which was applied, for example, in LES of turbulent channel flows
\cite{Pio93}.
In principle, this procedure could also be applied to the non-equilibrium model with
the turbulent viscosity defined by equation~(\ref{eq:visc_sgs}). In this case, the
turbulence energy associated with $\tens{T}$ is given by the contracted Germano
identity, 
\[
	-\frac{1}{2}T_{ii}=\langle\rho K\rangle_{\rm T}+\rho_{\mathrm{T}}K_{\mathrm{T}}\,,
\]
where $\rho K=-\tau_{ii}/2$ and $\rho_{\mathrm{T}}K_{\mathrm{T}} = -L_{ii}/2$.

However, the dynamical procedure as outlined above has several caveats.
In particular, the assumption of negligible variation of $C_{\mathrm{S}}$ over the the 
test filter length is found to be violated significantly. Moreover,
$C_{\mathrm{S}}$ diverges if $m_{ij}$ vanishes. Consequently, several
attempts were made to improve the dynamical procedure \cite{LiuMen94,PioLiu95,GhoLund95}.
A particularly simple modification was found by analyzing experimental measurements 
of turbulent velocity fluctuations in consecutive wave number bands $[k_{n-1},k_{n}]$, corresponding 
to a hierarchy of filters. By explicitly evaluating the turbulent stresses $\tens{\tau}^{[n]}$
associated with the wave numbers $k_n=\pi/\Delta_n$, the correlations with localized
closures were verified. Although some correlation between the turbulent
stresses at different filter levels was found, the correlation of $\tens{\tau}^{[n][n-1]}$
with the Leonard stresses $\tens{L}^{[n,n-1]}$ turned out to be significantly better.
This observation can be understood as a consequence of the locality of the energy
transfer \cite{Kraich76,Sagaut}, i.~e., the energy transfer across a certain wave number $k$ is mainly
caused by interactions in the narrow spectral band $[\frac{1}{2}k,2k]$. 
With regard to test filtering in LES, this implies that the eddy-viscosity closure should be applied
to $\tens{L}$ in place of $\tens{T}$. The localized coefficient $C_{\nu}(\vecx,t)$
of the turbulent viscosity is then given by \cite{KimMen99,SchmidtPhD,SchmNie06b,RoepSchm09}
\begin{equation}
  \label{eq:C_nu_test}
  C_{\nu} \simeq
  \frac{L_{ij}^{\ast}(S_{\rm T})_{ij}}
       {\rho_{\rm T}\Delta_{\rm T}K_{\mathrm{T}}^{1/2}|S_{\rm T}^{\ast}|^2}\,.
\end{equation}
where $K_{\mathrm{T}}=-L_{ii}/(2\rho_{\mathrm{T}})$ is the resolved kinetic energy on length scales 
$\Delta\le\ell\le\Delta_{T}$. Substitution of the above expression for $C_{\nu}$ into
equation~(\ref{eq:flux_eddy}) for the localized rate of production yields
\begin{equation}
  \label{eq:prod_locl}
  \Sigma=\tau_{ij}S_{ij} \simeq
  \frac{\rho\Delta}{\rho_{\rm T}\Delta_{\rm T}}  
  \left(\frac{K}{K_{\rm T}}\right)^{1/2}
  \left(\frac{|S^\ast|}{|S_{\rm T}^{\ast}|}\right)^2
  L_{ij}^{\ast}(S_{\rm T})_{ij} - \frac{2}{3}\rho K d\,,
\end{equation}
The above formula was used for simulations of turbulent thermonuclear combustion in white dwarfs
(see Section~\ref{sec:SN_Ia}). A generalization of the dynamical procedure to localize both $C_1$ and $C_2$ in the
closure~(\ref{eq:tau_mixed}) would be straight-forward, but has not been applied so far.
In this case, a linear system in the coefficients $C_1$ and $C_2$ has to be solved to minimalize the residual.

\epubtkImage{}{
  \begin{figure}[htbp]
    \centerline{\includegraphics[width=0.5\linewidth]{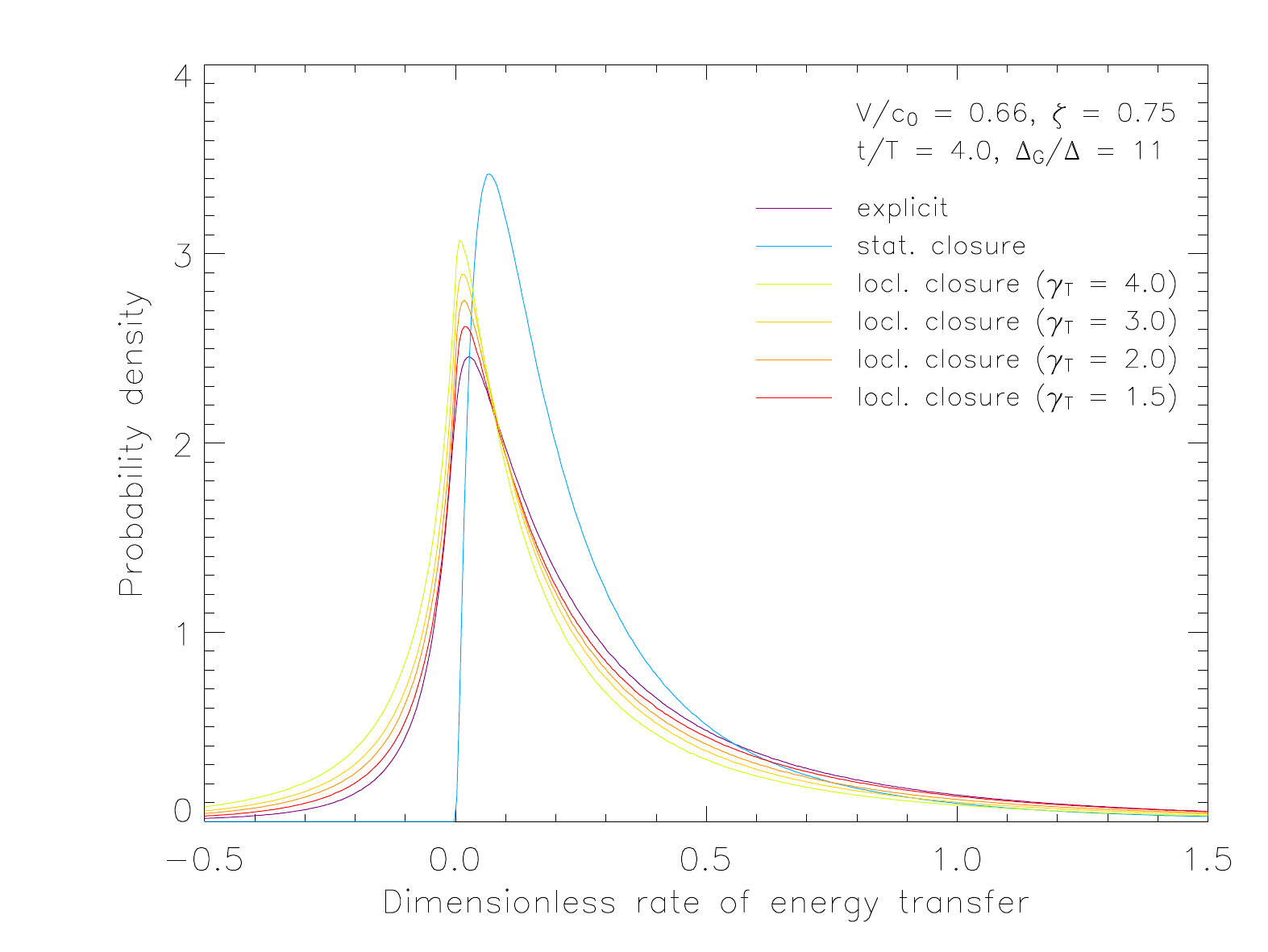} 
    	        \includegraphics[width=0.5\linewidth]{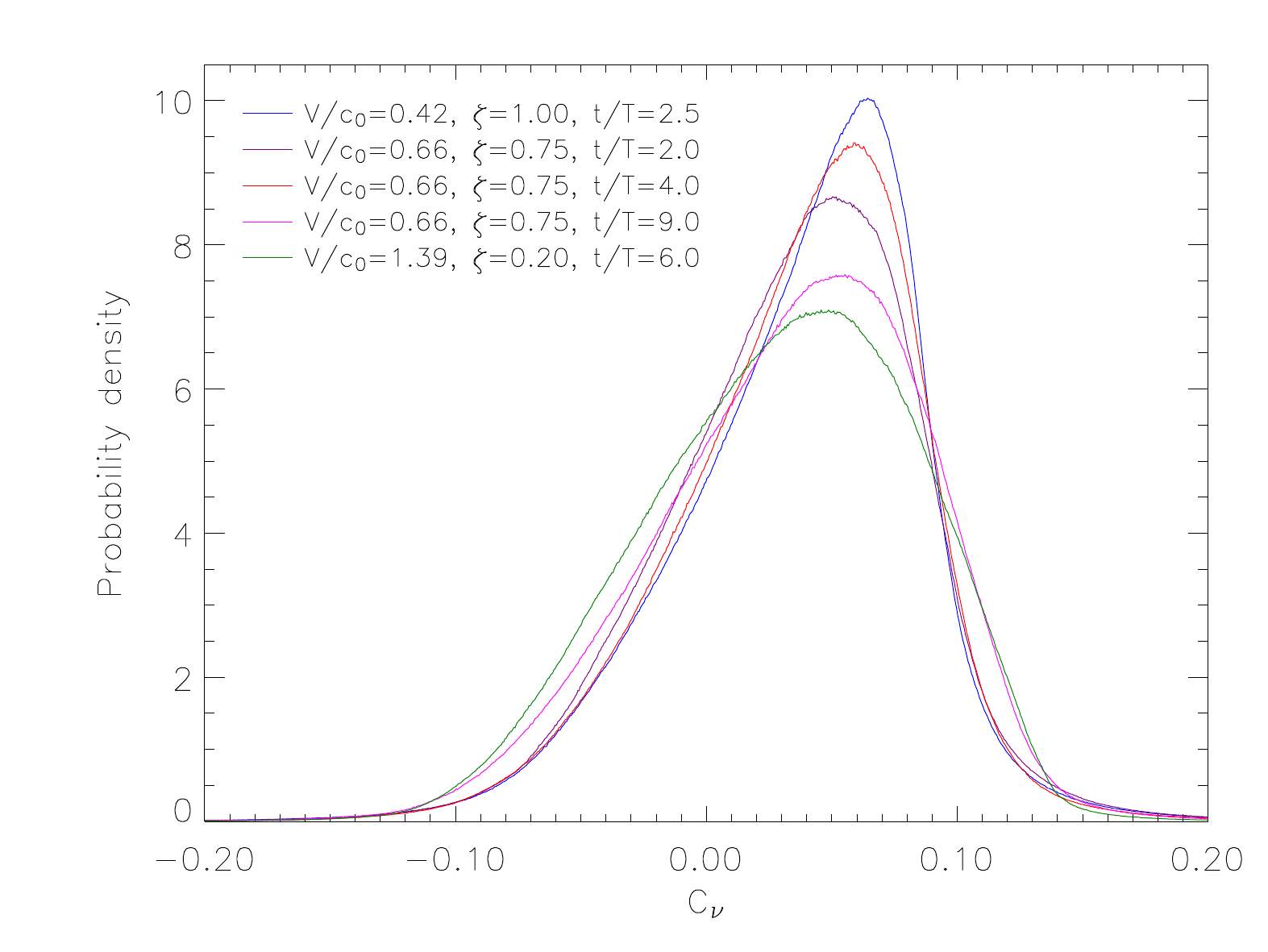}}
    \caption{Left: comparison of the probability density functions of the coarse-grained turbulence energy flux due to
    	anisotropic shear with different closures (left). Right plot: probability density functions of the
		localized closure coefficient $C_{\nu}$ obtained by test-filtering different coarse-grained ILES. 
		The ratio of the test filter length to the coarse-graining length is $\gamma_{\rm T}$. The static
		closure refers to the case with a constant coefficient. The parameters of the random forcing are
		the characteristic Mach number $V/c_0$ and the weight $\zeta$ of the 
		Helmholtz decomposition into solenoidal and compressive modes \cite{SchmidtPhD,SchmHille06,SchmNie06b}. }
    \label{fig:pdfs_eddy_visc}
\end{figure}}

For ILES of subsonic and transonic turbulence produced by stochastic forcing in periodic boxes \cite{SchmHille06}, 
the enhanced fidelity of the localized closure~(\ref{eq:prod_locl}) can be verified by coarse-graining the
data with hierarchical Gaussian filters $\langle\ \rangle_n$ as explained in Section~(\ref{sec:hierarch_flt}) \cite{SchmNie06b}. 
Let us first consider the turbulence energy flux produced by anisotropic shear on the length scale 
$\Delta_n$, assuming a turbulent viscosity with a constant coefficient $\langle C_\nu\rangle$, which is obtained
by averaging equation~(\ref{eq:C_nu_test}) over the domain of the flow.
If the closure with this coefficient were exact, we would have
\begin{equation}
	\label{eq:flux_eddy_coarse}
	\Sigma^{[n](\rm eddy)} + \frac{2}{3}\langle\rho\rangle_n K^{[n]} d^{[n]} = 
	\langle C_\nu\rangle\langle\rho\rangle_n\Delta_n\sqrt{K^{[n]}}\,|S^{[n]\ast}|^2\,.
\end{equation}
Here, the divergence term is added to $\Sigma^{[n](\rm eddy)}$ to express the flux associated with the
trace-free rate-of-strain tensor $S^{[n]\ast}$. However, coarse-grained data show that 
equation~(\ref{eq:flux_eddy_coarse}) is not very well satisfied.
The probability density functions of the expression on the right-hand side and the
explicitly calculated energy flux on the left-hand side are compared in the left plot in Figure~\ref{fig:pdfs_eddy_visc}. 
One can  see that the latter is negative in about $20\;\%$ of the domain (purple line), 
corresponding to backscattering from smaller to larger scales. This is excluded by the eddy-viscosity closure 
with a fixed coefficient (light blue line). In this case,
negative values of the total energy flux $\Sigma^{[n](\rm eddy)}$ are solely due to the divergence term. 
The bias toward positive fluxes can be avoided by localizing the eddy-viscosity closure \cite{SchmNie06b}. 
For a test filter $\langle\ \rangle_{n-1}$ with filter length $\gamma_{\rm T}=\Delta_{n-1}/\Delta_n>1$, 
the energy flux is given by the following analogue of equation~(\ref{eq:prod_locl}):
\begin{equation}
	\label{eq:flux_eddy_coarse_locl}
  \Sigma^{[n](\rm locl)} + \frac{2}{3}\langle\rho\rangle_n K^{[n]} d^{[n]} =
  \frac{\langle\rho\rangle_n\Delta_n}{\langle\rho\rangle_{n-1}\Delta^{[n-1]}}  
  \left(\frac{K^{[n]}}{K^{[n-1]}}\right)^{1/2}
  \left(\frac{|S^{[n]\ast}|}{|S^{[n-1]\ast}|}\right)^2
  L_{ij}^{[n-1]\ast}S_{ij}^{[n-1]}\,,
\end{equation}
The probability density functions that are plotted for different test filtering ratios $\gamma_{\rm T}$
in Figure~\ref{fig:pdfs_eddy_visc} (left plot) indicate a substantially improved match between the localized closure and
the explicitly calculated energy flux. Indeed, distributions of the localized closure coefficients show that $C_{\nu}$ has a negative branch (see right plot in Figure~\ref{fig:pdfs_eddy_visc}). This result suggests an improvement due to the 
dynamical procedure even in the case of homogeneous turbulence. On the other hand, the mean values of $C_{\nu}$ appear to be fairly robust for different forcing parameters.

\subsection{Global least squares method}
\label{sec:least_squares}

Closures can also be tested by analyzing correlations. This allows for the calibration of
the closure coefficients by least squares minimalization of the integrated residual \cite{SchmFeder11}.
For example, let us consider a generic closure with a single coefficient $C_1$ at the $n$-th filter
level:
\begin{equation}
	C_1 f^{[n](\rm cls)} =
	\Sigma^{[n](\rm cls)} + \frac{2}{3}\langle\rho\rangle_n K^{[n]} d^{[n]}\,,
\end{equation}
The global residual of the explicitly computed turbulence energy flux $\Sigma^{[n]}=\tau_{ij}^{[n]}S_{ij}^{[n]}$,
where $\tau_{ij}^{[n]}$ is defined by equation~(\ref{eq:turb_stress_n}), can be quantified 
by the squared error integrated over the whole domain $\mathcal{V}$ of the turbulent flow:
\begin{equation}
	\mathrm{err}^{2}(C_1) =
	\int_{\mathcal{V}} \left|\Sigma{[n]} + \frac{2}{3}\langle\rho\rangle_n K^{[n]} d^{[n]} -
		C_1 f^{[n](\rm cls)}\right|^{2}\dd^{3}x\,.
\end{equation}
The minimum of $\mathrm{err}^{2}(C_1)$ is obtained by setting the derivative with respect to $C_1$ equal to zero:
\begin{equation}
	\label{eq:coeff_lse}
	C_1=\frac{\int_{\mathcal{V}} f^{[n](\rm cls)}
	\left[\Sigma^{[n](\rm cls)} + \frac{2}{3}\langle\rho\rangle_n K^{[n]} d^{[n]}\right]\,\dd^{3}x}
		{\int_{\mathcal{V}} |f^{[n](\rm cls)}|^2\dd^{3}x}\,.
\end{equation}
In contrast to the dynamical procedure, the resulting closure coefficients are constants.

The method of least squares described above is applied in \cite{SchmFeder11} to various ILES of supersonic
isothermal turbulence produced by stochastic forcing \cite{SchmFeder09,FederRom10}. To coarse-grain the data, 
a Gaussian filter with a smoothing length $\Delta_4=L/16=32\Delta_{I}$ is used, where $\Delta_{I}=\Delta_9\equiv\Delta$
is the grid resolution of the ILEs.\epubtkFootnote{The simulation from \cite{SchmFeder09} was performed on a 
	$768^3$ grid. In this case, the filter length is $\Delta_4=L/16=24\Delta$.} 
Since the filter length is large compared to the grid resolution in this case,
it is advantageous to apply the filter operation in Fourier space.
For the eddy-viscosity closure, we have
\[
	f^{[n](\rm cls)} = \Delta_{n}\langle\rho\rangle_n\sqrt{2K^{[n]}}|S^{[n]\ast}|^2\,.
\]
The closure coefficients following from equation~(\ref{eq:coeff_lse}) are, for instance,
$C_1\approx 0.102$ for the $1024^3$ ILES with purely solenoidal (divergence-free) forcing, and 
$C_1\approx 0.092$ in the case of compressive (rotation-free) forcing. The corresponding value
of $C_{\nu}=\sqrt{2}C_1$ is about $0.14$. When comparing this value 
to the results in \cite{SchmidtPhD,SchmNie06b} (see Section~\ref{sec:hierarch_flt}), one has
to bear in mind that not only a different method is applied to determine the coefficients, but also that
the turbulence properties differ substantially.

The correlation diagram for $\Sigma^{[4](\rm cls)}$, with the least-squares coefficient $C_1$, 
versus $\Sigma^{[4]}$ in the case of solenoidal forcing is shown in Figure~\ref{fig:corrl_sol1024} (left plot). 
The overall correlation is actually quite good. A quantitative measure is the correlation coefficient
\begin{equation}
\begin{split}
   \mathrm{corr} &[\Sigma^{[n]},\Sigma^{[n](\rm cls)}] =\\
   &\frac{\int_{\mathcal{V}} \left(\Sigma^{[n]}-\langle\Sigma^{[n]}\rangle\right)
   \left(\Sigma^{[n](\rm cls)}-\langle\Sigma^{[n](\rm cls)}\rangle\right)\,\dd^{3}x}
   		{\mathrm{std}(\Sigma^{[n]})\,\mathrm{std}(\Sigma^{[n](\rm cls)})},
   \end{split}
\end{equation}
where $\mathrm{std}$ denotes the standard deviation and the angle brackets indicate an average over the whole domain.
The correlation coefficients of the eddy-viscosity closure are found to be $0.95$ and $0.93$ for solenoidal and
compressive forcing, respectively \cite{SchmFeder11}. However, it becomes apparent that the closure breaks
down for negative fluxes. This corresponds to the bias of the probability density function for the static closure
with an averaged coefficient in Figure~\ref{fig:pdfs_eddy_visc} (left plot).

\epubtkImage{}{
  \begin{figure}[htbp]
    \centerline{\includegraphics[width=0.5\linewidth]{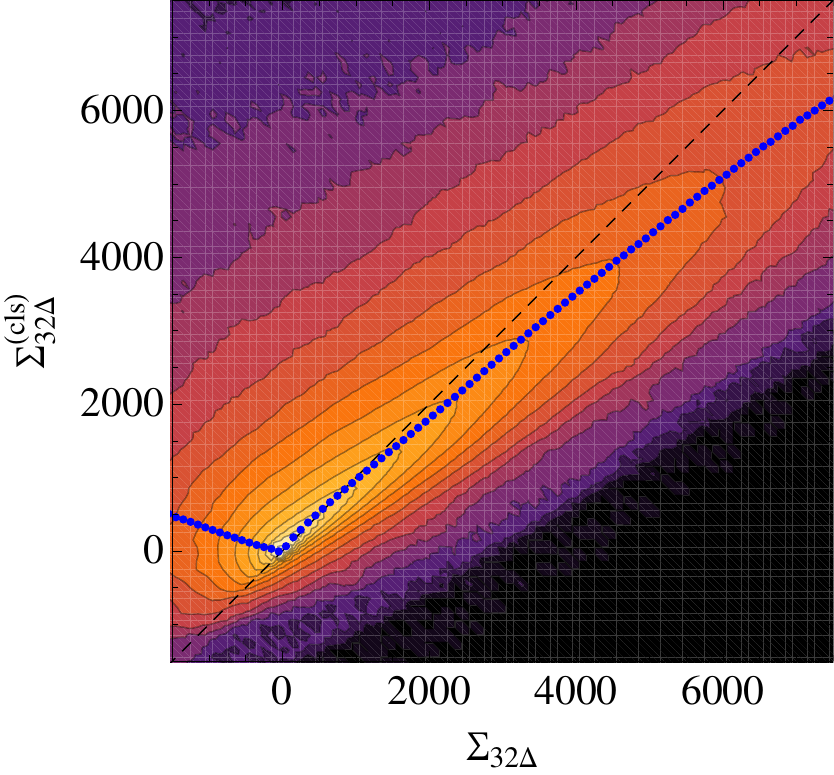} 
    	        \includegraphics[width=0.5\linewidth]{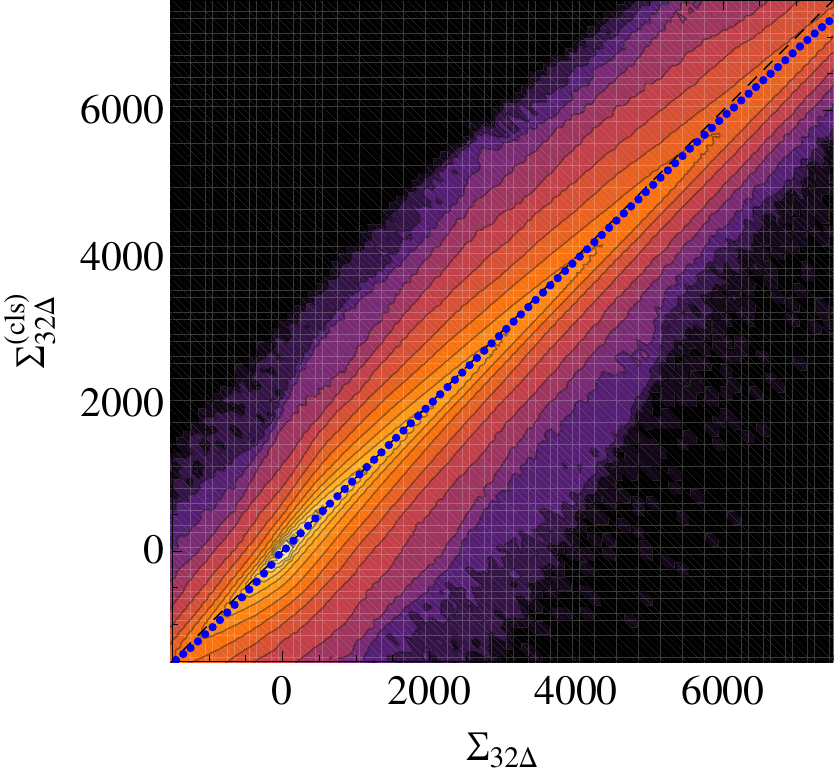}}
    \caption{Correlations of the coarse-grained turbulence energy flux with the
    	eddy-viscosity (left) and non-linear (right) closures for supersonic isothermal turbulence 
		produced by solenoidal forcing \cite{SchmFeder11}. The applied filter length is $32\Delta$, where
		$\Delta$ is the grid resolution. The average prediction of the closure for each bin is
		indicated by the blue dots.}
    \label{fig:corrl_sol1024}
\end{figure}}

Rather than applying the dynamical procedure, it is shown in \cite{SchmFeder11} that the
determinant closure~(\ref{eq:flux_det}) results in a largely improved approximation of negative turbulence
energy flux. This is a consequence of the varying sign of the determinant, $\det\tens{S}^\ast$, while
$\Delta K^{1/2}|S^\ast|^2$ is positive. However, the scatter of the determinant closure is high, particularly
for large positive flux. This is remedied by the non-linear closure~(\ref{eq:tau_nonlin}) for the turbulence stress tensor,
which produces an excellent correlation between $\Sigma^{[n](\rm cls)}$ and $\Sigma^{[n]}$,
as demonstrated by the right plot in Figure~\ref{fig:corrl_sol1024}. Since the formulae are analogous to 
the eddy-viscosity closure, we refer to \cite{SchmFeder11} for details. The correlation coefficients
are $0.991$ for both solenoidal and compressive forcing.

However, as explained in Section~\ref{sec:turb_stress}, the purely non-linear closure is not suitable for an SGS model.
This is why the least squares method was applied to the generalized closure with two coefficient, $C_1$ and $C_2$. For
\begin{equation}
	C_{1}f^{[n](\rm cls)} +  C_{2}g^{([n]\rm cls)} =
	\Sigma^{[n](\rm cls)} + \frac{2}{3}\langle\rho\rangle_n K^{[n]} d^{[n]},
\end{equation}
the closure coefficients are given by the linear system of equations
\begin{align}
	\label{eq:coeff_lse_2}
	\begin{split}
	\left(\int_{\mathcal{V}} |f^{[n](\rm cls)}|^2\dd^{3}x\right)\,C_1 + 
	&\left(\int_{\mathcal{V}} f^{[n](\rm cls)}g^{[n](\rm cls)}\dd^{3}x\right)\,C_2
		=\\
		&\int_{\mathcal{V}} f^{[n](\rm cls)}
		\left(\Sigma^{[n](\rm cls)} + \frac{2}{3}\langle\rho\rangle_n K^{[n]} d^{[n]}\right)\,\dd^{3}x\,, 
	\end{split}\\
	\begin{split}
	\left(\int_{\mathcal{V}}f^{[n](\rm cls)}g^{[n](\rm cls)} \dd^{3}x\right)\,C_1 + 
	&\left(\int_{\mathcal{V}} |g^{[n](\rm cls)}|^{2}\dd^{3}x\right)\,C_{2}
	=\\
	&\int_{\mathcal{V}} g^{[n](\rm cls)}
	\left(\Sigma^{[n](\rm cls)} + \frac{2}{3}\langle\rho\rangle_n K^{[n]} d^{[n]}\right)\,\dd^{3}x\,,
	\end{split}
\end{align}
where 
\begin{equation}
	f^{[n](\rm cls)} =
	\Delta_{n}\langle\rho\rangle_n\sqrt{2K^{[n]}}|S^{[n]\ast}|^2
	\qquad\mbox{and}\qquad
	g^{[n](\rm cls)}  =-4\langle\rho\rangle_n K^{[n]}
	\frac{u_{i,k}^{[n]}u_{j,k}^{[n]}S_{ij}^{[n]\ast}}
	{|\vecnab\otimes\vecu^{[n]}|^{2}}\,.
\end{equation}
The solution is $C_{1}\approx 0.02$ and $C_{2}\approx 0.7$, with a correlation coefficient $0.990$ \cite{SchmFeder11}.
These coefficients also yield good approximations to the turbulence energy flux for isothermal and adiabatic
turbulence simulations at lower Mach numbers \cite{SchmFeder09,SchmFeder11}. Moreover, the coefficients appear to
vary only little with the filter length, at least in the range from $\Delta_3=64\Delta$ to $\Delta_5=16\Delta$.
Choosing higher or lower filter lengths is not sensible because of the influence of the forcing ($\Delta_n$ must be small
compared to $\Delta_0=L$) and numerical dissipation ($\Delta_n\gg\Delta$).

\newpage


\section{Adaptive Methods}
\label{sec:AMR}

The most powerful technique for finite-volume codes to resolve localized
and anisotropic structures in a flow is adaptive mesh refinement (AMR) 
\cite{BergOli84,BergCol89}. Even with AMR, however, it is generally not possible
to fully resolve turbulence. This entails the problem that the numerically
resolved and unresolved turbulence energy fractions vary as
regions are refined or de-refined. In the following, it shown how
to address this problem in adaptively refined LES. In principle, 
global energy and momentum conservation can be achieved, while reducing the need 
for artificial changes in the internal energy, which is
the standard method to restore energy consistency between different
refinement levels in AMR simulations. Apart front that, localized and anisotropic
flow structures pose the problem that SGS models with constant coefficients
introduce systematic errors because they are usually calibrated for statistically
stationary and isotropic turbulence. Shear-improved SGS models
alleviate this problem by adjusting the non-linear energy transfer across the grid
scale to local flow conditions. This is possible by applying an adaptive temporal
filter, the so-called Kalman filter. 

\subsection{Energy- and momentum conservation in AMR simulations}
\label{sec:consrv}

In AMR simulations, data have to be transferred between different refinement levels by conservative interpolation
or averaging. For example, if a region is refined, data from coarser grids are interpolated to
finer grids. The same operation is used for filling ghost cells at the boundaries between a finer and a coarser level,
which is required to compute fluxes through the faces of adjacent finer and coarser cells. Moreover, block-structured
AMR codes usually average down the data from the highest-level grid to the all coarser levels. 
The mass density, momentum, and energy variables at two levels, say, $l$ and $l+1$, are in the simplest case
related by
\begin{align}
	\label{eq:mass_crs}	
	\rho_{\rm crs} :=&\,\overline{\rho} = \frac{1}{N}\sum_n\rho_{n}\,,\\
	\label{eq:vel_crs}	
	(\rho\vecU)_{\rm crs} :=&\,\overline{\rho\vecU} = \frac{1}{N}\sum_n(\rho\vecU)_{n}\,,\\
	\label{eq:energy_crs}	
	E_{\rm crs} :=&\,\overline{\rho E} =  \frac{1}{N}\sum_n (\rho E)_{n} =
		\frac{1}{N}\sum_n\left[(\rho e)_n + \frac{1}{2}\frac{(\rho U)_{n}^2}{\rho_{n}}\right]\,,
\end{align}
where $N$ grid cells at level $l+1$ are summed up to a single value in a coarse cell at level $l$. 
Obviously, these relations guarantee mass, momentum, and energy conservation.
For more sophisticated interpolation schemes, the fine-grid values have different weights $w_n$ in the
above sums. Without loss of generality, we assume $w_n=1$ in the following. Now, if we work out the
internal energy at the coarser level, we obtain
\[
	e_{\rm crs} 
	= E_{\rm crs} - \frac{1}{2}\frac{(\rho U)_{\rm crs}^2}{\rho_{\rm crs}}\\
	= \overline{\rho e} + \Delta(\overline{\rho K})\,.
\]
where the energy difference  
\begin{equation}
	\label{eq:delta_energy}
	\Delta(\overline{\rho K}) :=
	\frac{1}{N}\sum_{n}\frac{1}{2}\frac{(\rho U)_{n}^2}{\rho_{n}} -
	\frac{1}{2}\frac{(\rho U)_{\rm crs}^2}{\rho_{\rm crs}}
\end{equation}
is generally non-zero because of equation~(\ref{eq:vel_crs}). This
implies that
\[
	e_{\rm crs} \ne \overline{\rho e} = \frac{1}{N}\sum_n(\rho e)_{n}\,.
\]
To put it another way, the kinetic energy differences between refinement levels have to be compensated
by numerical cooling or heating in order to maintain energy conservation.

This can be alleviated by incrementing the SGS turbulence energy
by $\Delta(\overline{\rho K})$ when the cutoff scale is shifted from $\Delta_{l+1}$ to $\Delta_{l}=r\Delta_{l+1}$,
where $r>1$ is the refinement ratio: 
\begin{equation}
	\label{eq:rho_K_balance}
	(\rho K)_{\rm crs} = \overline{\rho K} + \Delta(\overline{\rho K}),\quad
	\mbox{where}\quad
	\overline{\rho K} = \frac{1}{N}\sum_{n}(\rho K)_n\,.
\end{equation}
We then have conservation of the total (resolved plus unresolved) kinetic energies,
\begin{equation}
	\label{eq:energy_balance_ftec}
	\frac{1}{2}\frac{(\rho U)_{\rm crs}^2}{\rho_{\rm crs}} + (\rho K)_{\rm crs} = 
	\frac{1}{N}\sum_n\left[\frac{1}{2}\frac{(\rho U)_{n}^2}{\rho_{n}} + (\rho K)_n\right]\,,
\end{equation}
and $e_{\rm crs}=\overline{\rho e}$.

As demonstrated in \cite{SchmAlm13}, equation~(\ref{eq:energy_balance_ftec}) works very well for fully developed 
homogeneous turbulence, but leads to erroneous projections of the SGS turbulence energy to coarser grids
if the flow structure is strongly inhomogeneous. A particular problem is posed by non-turbulent bulk flows such
as gas accretion into gravitational wells. In this case, applying the increment $\Delta(\overline{\rho K})$ 
defined by equation~(\ref{eq:delta_energy})
can greatly overestimate the energy difference associated with turbulent velocity fluctuations. A tentative
solution to this problem follows along similar lines as the re-scaling procedure used in \cite{MaierIap09}.  
By extrapolating the turbulent energy on the length scale $\Delta_{l+1}$ via a power law to
$\Delta_{l}$, we have 
\begin{equation}
	\label{eq:rhoK_pow}
	(\rho K)_{\rm crs} = \overline{\rho K}\left(\frac{\Delta_{l}}{\Delta_{l+1}}\right)^{2\eta}
	= \overline{\rho K}\,r^{2\eta}
\end{equation}
Substitution into equation~(\ref{eq:rho_K_balance}) yields
\[
	\Delta(\overline{\rho K}) = \left(r^{2\eta}-1\right)\overline{\rho K}.
\]
where $\eta=1/3$ in the case of Kolmogorov scaling. A shortcoming of this estimate is, of course, that
the turbulent velocity fluctuations follow power-law scaling only in a statistical sense. 
To avoid an overshoot of $\Delta(\overline{\rho K})$ if turbulence dominates the energy
difference between levels, it is necessary to set
\begin{equation}
	\label{eq:delta_rhoK_ctec_proj}
	\Delta(\overline{\rho K}) = 
	\min\left[\left(r^{2\eta}-1\right)\overline{\rho K},\,
	\frac{1}{N}\sum_{n}\frac{1}{2}\frac{(\rho U)_{n}^2}{\rho_{n}} -
	\frac{1}{2}\frac{(\rho U)_{\rm crs}^2}{\rho_{\rm crs}}\right]	
\end{equation}
Now, energy conservation can only be satisfied if $\Delta(\overline{\rho K})$
is complemented by a correction of the internal energy,
\begin{equation}
	\label{eq:delta_rhoe_ctec_proj}
	\Delta(\overline{\rho e}) = 
	\frac{1}{2}\frac{(\rho U)_{n}^2}{\rho_{n}} -
	\frac{1}{2}\frac{(\rho U)_{\rm crs}^2}{\rho_{\rm crs}}
	-\Delta(\overline{\rho K})\,.
\end{equation}
As a result, we have
\begin{align}
	\label{eq:energy_int_crse_ctec}
	(\rho K)_{\rm crs} &= \overline{\rho K} + \Delta(\overline{\rho K})\, ,\\
	(\rho e)_{\rm crs} &= \overline{\rho e} + \Delta(\overline{\rho e})\, ,\\
	(\rho E)_{\rm crs} &= \overline{\rho E} - \Delta(\overline{\rho K})\, ,
\end{align}
and the total energy is conserved because
\[
	(\rho E)_{\rm crs} + (\rho K)_{\rm crs} = \overline{\rho E} + \overline{\rho K}\,.
\]
For 
\[
	\left(r^{2\eta}-1\right)\overline{\rho K} \ll 
	\frac{1}{2}\frac{(\rho U)_{n}^2}{\rho_{n}} -
	\frac{1}{2}\frac{(\rho U)_{\rm crs}^2}{\rho_{\rm crs}}
\]
most of the resolved kinetic energy difference is compensated by internal energy, 
similar to the standard energy correction employed in AMR simulations without SGS model. 
Fully turbulent energy compensation, on the other hand, follows as limiting case if $\Delta(\overline{\rho K})$ 
is given by equation~(\ref{eq:delta_energy}), 

For grid refinement, the data from the parent grid are interpolated to 
fine-grid values $\rho_{n}^{\ast}$, $(\rho\vecU_{n})^{\ast}$, etc.\
such that
\begin{alignat*}{5}
	\frac{1}{N}\sum_n\rho_{n}^{\ast} &= \rho_{\rm crs}\,,&\qquad
	\frac{1}{N}\sum_n(\rho\vecU)_{n}^{\ast} &= (\rho\vecU)_{\rm crs}\,,&\qquad\\
	\frac{1}{N}\sum_n (\rho E)_{n}^{\ast} &= E_{\rm crs}\,,&\qquad
	\frac{1}{N}\sum_n (\rho e)_{n}^{\ast} &= e_{\rm crs}\,,&\qquad
	\frac{1}{N}\sum_{n}(\rho K)_{n}^{\ast} = (\rho K)_{\rm crs}\,.
\end{alignat*}
One can set $\rho_n=\rho_n^\ast$, $(\rho\vecU)_{n}=(\rho\vecU)_{n}^\ast$, and $(\rho E)_{n}=(\rho E)_{n}^{\ast}$, 
but then the kinetic energy difference resulting from the conservative interpolation of the momenta 
has to be compensated. By defining the energy corrections
\begin{align}
	\label{eq:delta_rhoK_ctec}
	\Delta(\overline{\rho K}) &= \left(1-r^{-2\eta}\right)(\rho K)_{\rm crs}\,,\\
	\label{eq:delta_rhoe_ctec}
	\Delta(\overline{\rho e}) &= 
	\frac{1}{N}\sum_{n}\frac{1}{2}\frac{[(\rho U)_{n}^{\ast}]^2}{\rho_{n}^{\ast}} -
	\frac{1}{2}\frac{(\rho U)_{\rm crs}^2}{\rho_{\rm crs}}
	-\Delta(\overline{\rho K})\,,
\end{align}
the interpolated values of the SGS turbulence and internal energies can be adjusted as follows:
\begin{align}
	\label{eq:energy_sgs_refined_ctec}
	(\rho K)_n &= (\rho K)_n^{\ast} - \frac{(\rho K)_n^{\ast}}{(\rho K)_{\rm crs}}\,\Delta(\overline{\rho K})
	=r^{-2\eta}(\rho K)_n^{\ast}\,,\\
	\label{eq:energy_int_refined_ctec}
	(\rho e)_n &= (\rho e)_n^{\ast} - \frac{(\rho e)_n^{\ast}}{(\rho e)_{\rm crs}}\,\Delta(\overline{\rho e})\,,
\end{align}
The conservation of total energy follows from
\begin{align*}
	\frac{1}{N}\sum_n\left[(\rho e)_n + \frac{1}{2}\frac{(\rho U)_{n}^2}{\rho_{n}} + (\rho K)_n\right]
	&=\frac{1}{N}\sum_n\left[(\rho e)_n^{\ast} + 
		\frac{1}{2}\frac{[(\rho U)_{n}^{\ast}]^2}{\rho_{n}^{\ast}} + (\rho K)_n^{\ast}\right]
		- \Delta(\overline{\rho e}) - \Delta(\overline{\rho K})\\ 
	&=e_{\rm crs} + \frac{1}{2}\frac{(\rho U)_{\rm crs}^2}{\rho_{\rm crs}} + (\rho K)_{\rm crs}\,,
\end{align*}
where the last equality follows by substituting equations~(\ref{eq:delta_rhoK_ctec}) and~(\ref{eq:delta_rhoe_ctec})
for $\Delta(\overline{\rho e})$ and $\Delta(\overline{\rho K})$, respectively. 
Equation~(\ref{eq:energy_sgs_refined_ctec}) is formally the same as the correction rule for the SGS turbulence energy in \cite{MaierIap09}. In contrast to the method outlined above, however, momentum conservation is not fulfilled by \cite{MaierIap09} because the full kinetic energy difference between the finer and coarser levels
is compensated by internal energy and then the velocities are re-scaled to compensate the power-law correction of the SGS turbulence energy (see also Section~\ref{sec:clusters}).

\subsection{Shear-improved model}
\label{sec:kalman}

For inhomogeneous and non-stationary flows, dynamical procedures can be applied to adjust SGS model coefficients to local
flow conditions (see Section~\ref{sec:dyn_proc}). A completely different idea was put forward by L\'{e}v\^{e}que to
calculate the turbulent stresses in LES of wall-bounded turbulence \cite{LevTosch07}. 
Rather than adjusting the eddy-viscosity coefficient $C_\nu$, the numerically resolved velocity field is decomposed
into a mean flow $\langle\vecu\rangle$ and turbulent fluctuations $\vect{u}'$. It is then argued that the turbulence energy
flux for the Smagorinsky model should linearly depend on the shear associated with the fluctuating component, i.~e.,
\[
	\epsilon\simeq\Sigma=(C_{\rm S}\Delta|S|)^2(|S|-|\langle S\rangle|).
\]
where $|\langle S\rangle|$ is the rate of strain of the mean flow. If the flow is laminar, $\langle\vecu\rangle\simeq \vect{u}$ and turbulence production vanishes because $|S|-|\langle S\rangle|\simeq 0$. For developed isotropic turbulence, on the other hand, $u'\gg u$. In this case, the standard Smagorinsky model with $\Sigma\propto|S|^3$ applies.
For intermediate cases, the model corrects the energy flux $\Sigma$ by taking into
account interactions of the grid-scale fluctuations with the mean shear. This is why it is called the shear-improved model.

As proposed in \cite{SchmAlm13}, this idea can be carried over to the SGS turbulence energy model (see Section~\ref{sec:K_eq})
by defining the shear-improved eddy-viscosity closure as
\begin{equation}
	\label{eq:tau_si}
	\tau_{ij} = 2\rho\left(\nu_{\rm sgs}S_{ij}^{\,\prime\ast}-\frac{1}{3}K\delta_{ij}\right)\,,
\end{equation}
where $S_{ij}^{\,\prime\ast}$ is the trace-free part of the rate-of-strain tensor associated with
the fluctuating component of the flow:
\begin{equation}
	S_{ij}^{\,\prime}=\frac{1}{2}\left(\frac{\partial u_i^\prime}{\partial x_j}+\frac{\partial u_j^\prime}{\partial x_i}\right)
	=S_{ij}-\langle S\rangle_{ij}\,.
\end{equation}
The turbulence energy flux is then given by
\begin{equation}
	\label{eq:flux_si}
	\Sigma = \tau_{ij}S_{ij} =
	 C_{\nu}\rho\Delta K^{1/2}\left(|S^{\ast}|^2-2\langle S\rangle_{ij}^{\ast}S_{ij}\right)-\frac{2}{3}\rho K d\,.
\end{equation}
Apart from the divergence term, the expression on the right-hand side is motivated by the generalized K\'{a}rm\'{a}n-Howarth equation in \cite{LevTosch07}.

A difficulty of implementing the shear-improved model is the computation of the mean flow $\langle\vecu\rangle$, which is an ensemble average
over the flow velocity $\vect{u}$. A practical solution is to find an approximation to the mean flow by smoothing $\vect{u}$
with a temporal low-pass filter. For statistically stationary turbulence, it is possible to use an exponentially weighted
recursive average. In component notation, an estimate 
for the mean velocity at time $t_{n+1}$ is calculated as weighted sum of the
estimate at the previous time step and the local velocity $u_i^{(n+1)}$ for each grid cell
(cell indices are omitted):
\begin{equation}
	\label{eq:filter_recursive}
	[u_i]^{(n+1)} = \left(1-\alpha_i^{(n+1)}\right)[u_i]^{(n)} + \alpha_i^{(n+1)}u_i^{(n+1)},	
\end{equation}
The weighing coefficients are defined by 
\[
	\alpha_i^{(n+1)} = \frac{2\pi(t_{n+1}-t_n)}{\sqrt{3}\,T_{\rm c}}\,.
\]
where $T_{\rm c}$ is the constant integral time scale of the flow \cite{CahuBou10}.
Changes occurring on time scales smaller than the smoothing scale $T_{\rm c}$ are
suppressed in $[u_i]$. Compared to dynamical procedures, this algorithm is very easy to implement and computationally much
cheaper.  
In \cite{LevTosch07}, it is demonstrated that the shear-improved Smagorinsky model with exponential smoothing
performs well in LES of plane-channel flow and reproduces data from direct numerical simulations.

However, the simple exponential smoothing algorithm  produces an estimate $[u_i]$ that lags behind the ensemble 
average $\langle u_i\rangle$ if the mean flow evolves. To address this problem, the so-called Kalman filtering technique 
is introduced in \cite{CahuBou10}. The Kalman filter adapts itself to an unsteady mean flow by dynamically 
adjusting the weights of the recursive relation~(\ref{eq:filter_recursive}), depending on the variances
of the mean flow evolution, $[u_i]^{(n)}-[u_i]^{(n-1)}$, and the detected deviation from the mean, $u_i^{(n)}-[u_i]^{(n)}$.
This is achieved by setting $\alpha_i^{(n+1)}$ equal to the so-called Kalman gain $K_i^{(n+1)}$, which is defined by the
ratio between the error variance of the smoothed component and the total error variance, 
including the fluctuating component. Since the error variances have to be evaluated
at time $t^{(n+1)}$, a predictor-corrector scheme is used:
\begin{enumerate}
\item Given the error variance $P_i^{(n)}$ at time $t^{(n)}$, the prediction for $t^{(n+1)}$ is
	\begin{equation}
		P_i^{(n+1)*} = P_i^{(n)} + \sigma^{2\,(n)}_{\delta[u_i]}\,,
	\end{equation}
	where
	\[
		\sigma_{\delta[u_i]}^{(n)} = \frac{2\pi\Delta t^{(n)}}{\sqrt{3}\,T_{\rm c}}\,u_{\rm c}\,.
	\]
	Here it is assumed that the typical correction of the mean flow, $\delta[u_i]^{(n)}=[u_i]^{(n)}-[u_i]^{(n-1)}$,
	is of the order $2\pi\Delta t^{(n)}u_{\rm c}/(\sqrt{3}\,T_{\rm c})$, where $\Delta t^{(n)}=t_n-t_{n-1}$
\item The Kalman gain is then given by
	\begin{equation}
		\label{eq:kalman_gain}
		\alpha_i^{(n+1)} = K_i^{(n+1)} = \frac{P_i^{(n+1)*}}{P_i^{(n+1)*} + \sigma^{2\,(n)}_{\delta u_i}}\,,
	\end{equation}	
	where
	\[
		\label{eq:sigma_fluc}
		\sigma^{2\,(n)}_{\delta u_i}= 
		\max\left(\left|\delta u_i^{(n)}\right|,0.1u_{\rm c}\right)u_{\rm c}
	\]	
	is the contribution of the fluctuating component $\delta u_i^{(n)}\equiv u_i^{\prime\,(n)}=u_i^{(n)}-[u_i]^{(n)}$ 
	to the error variance.
	The lower bound on $\sigma^{2\,(n)}_{\delta u_i}$ is necessary to obtain non-vanishing
	fluctuations from an initially smooth flow with $[u_i]=u_i$. 
\item The corrected error variance for the next step is given by
	\begin{equation}
		P_i^{(n+1)} = \left(1 - K_i^{(n+1)}\right)P_i^{(n+1)*}\,.
	\end{equation}
\end{enumerate}
In a statistically stationary state, the velocity fluctuations should be of the order 
$\sigma^{(n)}_{\delta u_i}\simeq u_{\rm c}$. In this case the Kalman filter corresponds to simple exponential smoothing:
\[
	\alpha_i^{(n+1)} \simeq \frac{\sigma_{\delta[u_i]}^{(n)}}{\sigma^{(n)}_{\delta u_i}} \simeq 
	\frac{2\pi\Delta t^{(n)}}{\sqrt{3}\,T_{\rm c}} \ll 1\,.
\]
The two filter parameters, $T_{\rm c}$ and $u_{\rm c}$, have to be chosen such that
$u_{\rm c}$ is roughly the integral velocity of turbulence if the flow enters a steady state and
$T_{\rm c}$ is the characteristic time scale over which the flow evolves. 
In \cite{CahuBou11},
LES of turbulence produced by the flow past a cylinder were shown to agree well with experimental data
if Kalman filtering is applied with $T_{\rm c}$ and $u_{\rm c}$ being set to the inverse of the expected
vortex-shedding frequency and upstream velocity, respectively. 

\epubtkImage{}{
  \begin{figure}[htbp]
    \centerline{\includegraphics[width=\textwidth]{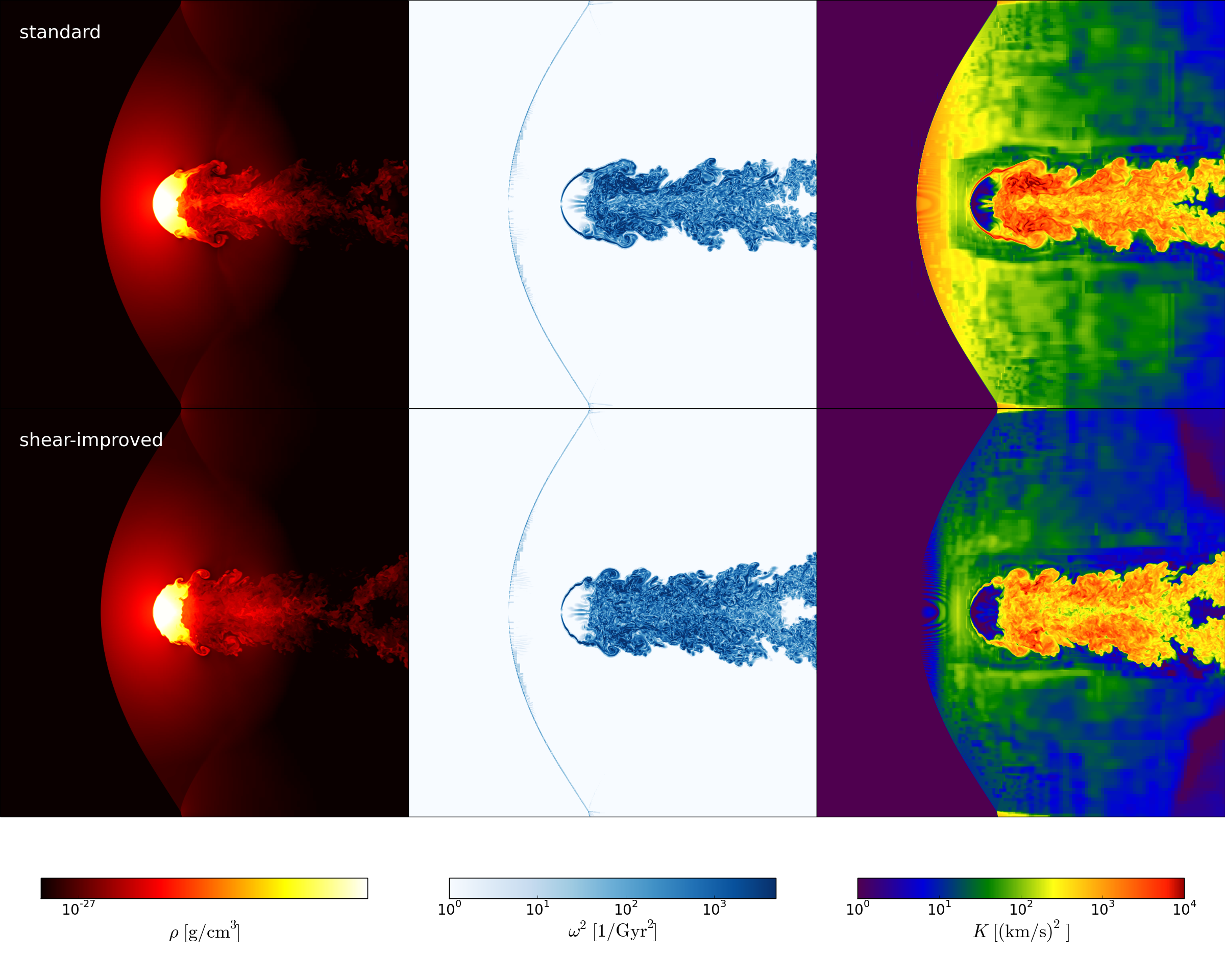}}
    \caption{Slices of the mass density (left), squared vorticity (middle), and specific SGS turbulence energy (right)
    	for AMR simulations of a gravitationally bound gas cloud in a wind after 3 Gyr of evolution \cite{SchmAlm13}.
		The box size is $4\;\mathrm{Mpc}$.}
    \label{fig:msc_slices}
\end{figure}}

A similar computational problem in astrophysics is an isothermal spherical cloud, which is bound by a static gravitational
potential, in a homogeneous wind. The initial density profile of the cloud is determined by hydrostatic equilibrium.
As mass is stripped from the cloud by the wind, a turbulent wake forms in the downstream direction. This problem
was originally investigated as a simple model for the infall of small subcluster into
the ICM of a much more massive cluster, computed in the frame of reference attached to the center of mass of the
subcluster \cite{IapiAdam08}. LES with the shear-improved SGS model are presented in \cite{SchmAlm13}, 
where the Kalman filter parameters are given by the velocity of the wind and the turn-over time of the largest eddies. 
Figure~\ref{fig:msc_slices} compares the gas density and flow structure for two runs, one with the standard SGS model and the
other one with the shear-improved model. In both cases, a $128^3$ root grid and three levels of refinement are used.
The AMR control variables are the squared vorticity and the compression rate \cite{IapiAdam08,SchmFeder09}.
The latter is defined by the substantial time derivative of the velocity divergence $d$ and tracks down 
the bow shock in front of the cloud. The top panels in Figure~\ref{fig:msc_slices} show that
the strain associated with the bow shock produces substantial SGS turbulence energy if the standard SGS model
is applied. One can also discern the down-scaling of the SGS turbulence energy that is transported with the wind,
as it enters the refined regions around the cloud. This is based on the algorithm explained in the previous section.
Kelvin-Helmholtz instabilities between the low-density wind and the high-density gas in the cloud cause vortex shedding,
which in turn produces turbulence. This becomes manifest in large values of $\omega$
and $K$ in the turbulent wake that extends from the cloud towards the right. In contrast to the turbulent wake,
however, the shear experienced by the gas when it is passing the bow shock is not associated with turbulence. 
As a result, the steep increase of $K$ is largely a spurious effect induced by the 
eddy-viscosity closure~(\ref{eq:tau_eddy}). Indeed, the SGS turbulence energy in the shocked wind is substantially reduced 
if the shear-improved closure~(\ref{eq:tau_si}) is applied (bottom right panel in 
Figure~\ref{fig:msc_slices}). In comparison to the
standard SGS model, production is also suppressed at the front side of the cloud and, hence, vortex
shedding is the dominant source of turbulence production. This has a clearly visible impact on the 
structure of the turbulent wake.

In general, an appropriate prior choice of the filter parameters might not be obvious. Nevertheless, they can be calibrated \emph{a posteriori} by performing low-resolution test runs. This is shown for cosmological simulations in \cite{SchmAlm13}. 
In this case, the Kalman filter also has the merit of separating the turbulent flow in clusters from
the gravity-driven bulk flows (see Section~\ref{sec:cosmo_turb_vel}).

\newpage


\section{Astrophysical Applications}
\label{sec:astro}

\subsection{Thermonuclear combustion in white dwarfs}
\label{sec:SN_Ia}

Among the various possible scenarios for supernovae of type Ia are 
thermonuclear explosions of gas accreting white dwarfs in 
close binary systems \cite{HilleNie00,RoepFink11,HilleKrom13}. 
If the mass of a white dwarf approaches the Chandrasekhar limit, 
explosive carbon and oxygen burning is ignited \cite{NonaAsp12}.
Owing to the degeneracy of white dwarf matter, the thermonuclear 
reaction zones propagate as thin flame fronts, whose thickness
$\delta_{\rm f}$ and propagation speed $s_{\rm f}$ are determined by 
the very high thermal conductivity of the fuel. This mode of burning is called 
deflagration. Since the burned material has lower density than the fuel, 
it rises because of its buoyancy. Consequently, the energy released by 
thermonuclear deflagration drives convection. Since eddies
exert strong shear on rising bubbles of burning material, 
they are deformed into mushroom-like shapes and Kelvin-Helmholtz instabilities 
at the surface are rapidly producing turbulence \cite{MalNona13}.
Eventually, this results in a very complex flame front with a fractal structure that
cannot be resolved in numerical simulations over the full range of length scales.

This problem was addressed by performing LES with
an effective propagation speed of the turbulent flame front (\cite{ReinHille02,SchmNie06a,
RoepHille07}, to mention just a few examples). In these simulations, the flame front 
is tracked by means of the level set method \cite{OshSeth88,ReinHille99}, which is able to follow
complex topological changes by determining the interface between fuel and burned material
as the spatial surface for which a signed distance function is zero.
The evolution of the distance function from given initial conditions is
given by the advection velocity relative to the fluid plus the flame propagation
speed. In a fully resolved simulation, the flame propagation speed
would be given by the microscopic flame speed $s_{\rm f}$ (also called laminar
flame speed). If the flame front is underresolved, however, the wrinkling and folding 
of the front by turbulent eddies below the grid scale leads to an enhanced rate
of energy release. In this case, the relevant time scale is not the
microscopic diffusion time scale but the turn-over time of numerically
unresolved eddies. Simple dimensional reasoning implies that
$s_{\rm f}$ has to be replaced by the turbulent flame speed
$s_{\rm t}=\sqrt{2K}$ \cite{NieHille95,NieKer97,Peters99}, where $K$ is the SGS turbulence
energy given by equation~(\ref{eq:pde_energy_close}).
In \cite{SchmNie06c}, the formula
\begin{equation}
  \label{eq:flame_speed_turb}
  s_{\mathrm{t}}= s_{\mathrm{f}} 
  \sqrt{1 + C_{\mathrm{t}}\frac{2K}{s_{\mathrm{f}}^2}}.
\end{equation}
is proposed for a smooth transition between laminar and turbulent flame propagation.
A further complication comes from the pronounced anisotropy of turbulence
at the flame front \cite{CirSchm08,MalNona13}. While the burned material inside the flame 
is highly turbulent, there is little or no turbulence in the fuel just outside the flame. In a way,
this is similar to walls in terrestrial flows. For this reason, it is important to apply
the dynamic produced explained in Section~\ref{sec:dyn_proc} to locally calculate the 
eddy-viscosity coefficient, which determines the rate of production of $K$.

The minimal length scale for which the flame front is affected by 
turbulence is given by the Gibson scale $\ell_{\mathrm{G}}$.
If $v^{\prime}(\ell)$ is the mean turbulent velocity fluctuation
on the length scale $\ell$, then $\ell_{\mathrm{G}}$ is implicitly
given by the condition 
\begin{equation}
  v^{\prime}(\ell_{\mathrm{G}}) = s_{\mathrm{f}}\,,
\end{equation}
i.~e., the turbulent velocity fluctuation on the Gibson scale
equals the microscopic flame speed. For $\ell\ll\ell_{\mathrm{G}}$,
the turn-over time associated with $v^{\prime}(\ell_{\mathrm{G}})$
is much longer than the crossing time of the
flame front through an eddy of size $\ell$. On these scales, the flame front is
virtually unaffected by turbulence. If $\ell\gg\ell_{\mathrm{G}}$,
on the other hand, eddies will significantly distort the flame front 
on length scales between $\ell_{\mathrm{G}}$ and $\ell$.
In LES of a thermonuclear supernovae, the Gibson scale $\ell_{\mathrm{G}}$ is much
smaller than computationally feasible grid resolutions $\Delta$ during most
of the deflagration phase and, consequently, a turbulent flame speed model has
to be applied.\epubtkFootnote{To be more precise, the notion of a flame front propagating
	with the turbulent flame speed $s_{\rm t}$ applies to the so-called flamelet regime,
	for which $\ell_{\rm f}\ll\ell_{\rm G}$. 
	When $\ell_{\rm G}$ becomes comparable to $\ell_{\rm f}$,
	turbulence affects the internal structure of the flame and distributed
	burning sets in \cite{NieKer97,RoepHille05,Schm07}.}

As an illustration of the level set method with the turbulent flame speed~(\ref{eq:flame_speed_turb}), 
Figure~\ref{fig:SN_flame_front} shows snapshots of the flame front for a thermonuclear supernova simulation from
\cite{SchmNie06a}. The turbulent velocity fluctuations on subgrid scales,
$\sqrt{2K}$, are shown as color shades at the flame surface. The asymptotic
value of turbulent flame speed is $s_{\rm t}\simeq C_{\rm t}\sqrt{2K}$ if
$s_{\rm t}\gg s_{\rm f}$. In this simulation, a Poisson process is used
to randomly place small ignition spots in the high-density core of the white dwarf.
The statistics of this process is based on a simple model for temperature
fluctuations produced by convection in pre-ignition phase.
Although the number of ignition resulting from this model appears to be
by far too large in the light of recent numerical studies of the ignition process 
\cite{NonaAsp12}, the simulation nevertheless demonstrates in an exemplary manner how
the turbulent deflagration progresses. At early time (a), one can see a large number of small bubbles
generated by the stochastic ignition process 
at distances of the order $100\;{\rm km}$ from the center of the white dwarf. 
As the burning bubbles are rising from the centre, they begin to form the typical
Rayleigh-Taylor mushroom shapes (b). At this point, the effective flame
propagation speed is already dominated by turbulence. After about half a second (c), the 
turbulent flames mostly have merged into a single structure of about $1000\;{\rm km}$
diameter. Then the burning front rapidly expands to much larger radii and causes the white dwarf to explode
after roughly one second (d). 

\epubtkImage{}{
  \begin{figure}[htbp]
    \mbox{\subfigure[$t=0.2\,\mathrm{s}$]{\includegraphics[width=0.45\linewidth]{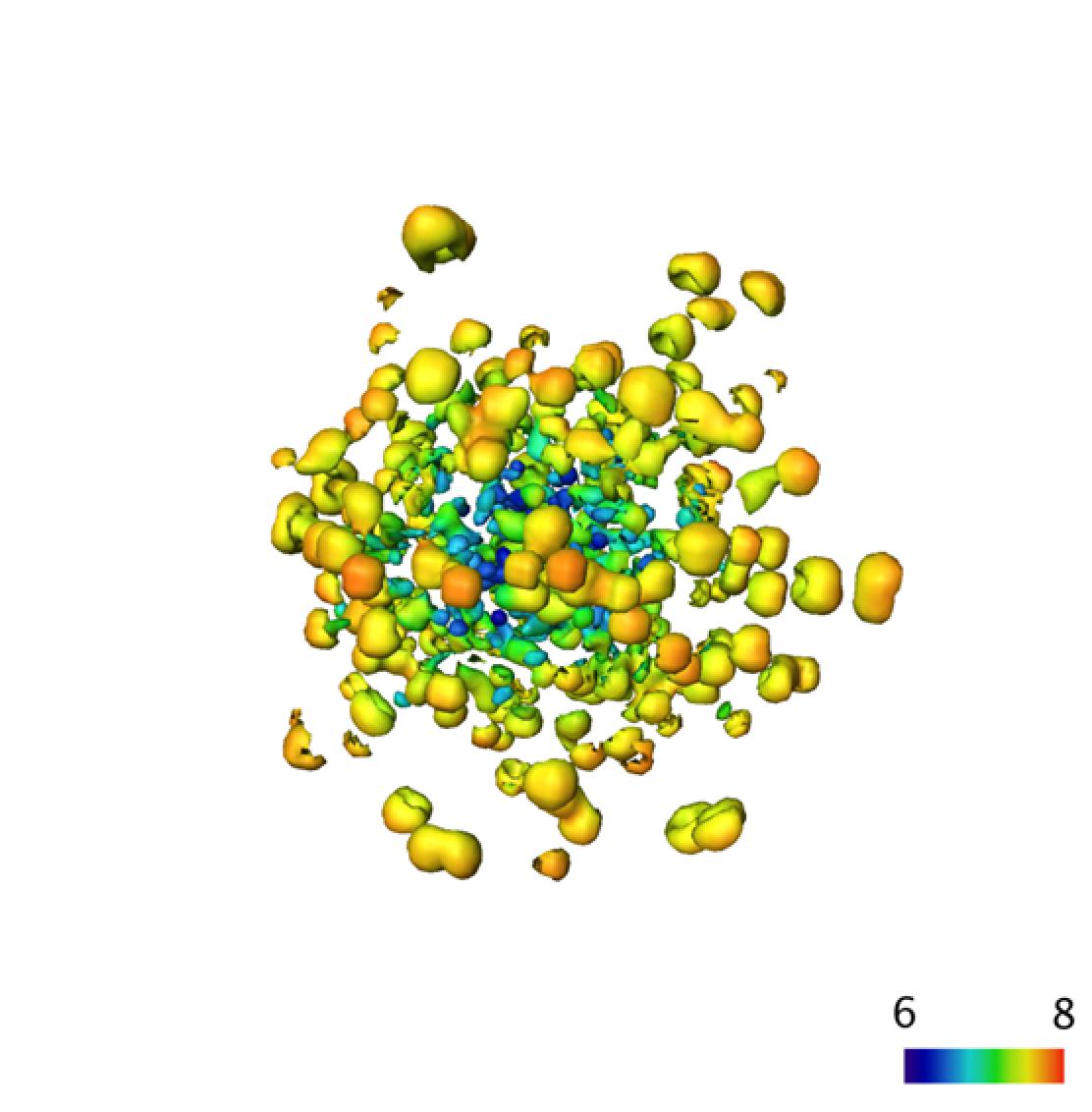}}
          \subfigure[$t=0.45\,\mathrm{s}$]{\includegraphics[width=0.45\linewidth]{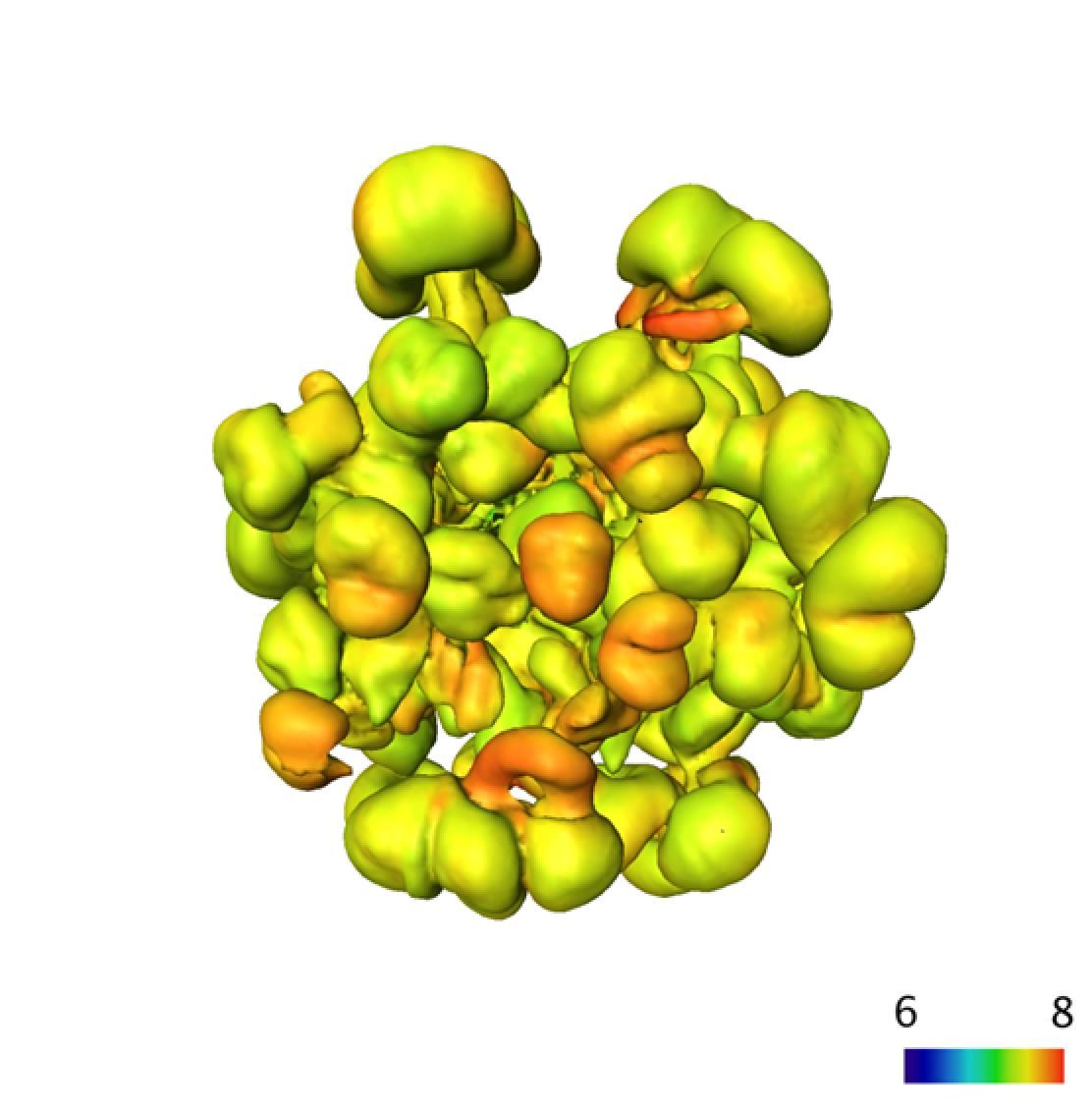}} }
    \mbox{\subfigure[$t=0.6\,\mathrm{s}$]{\includegraphics[width=0.45\linewidth]{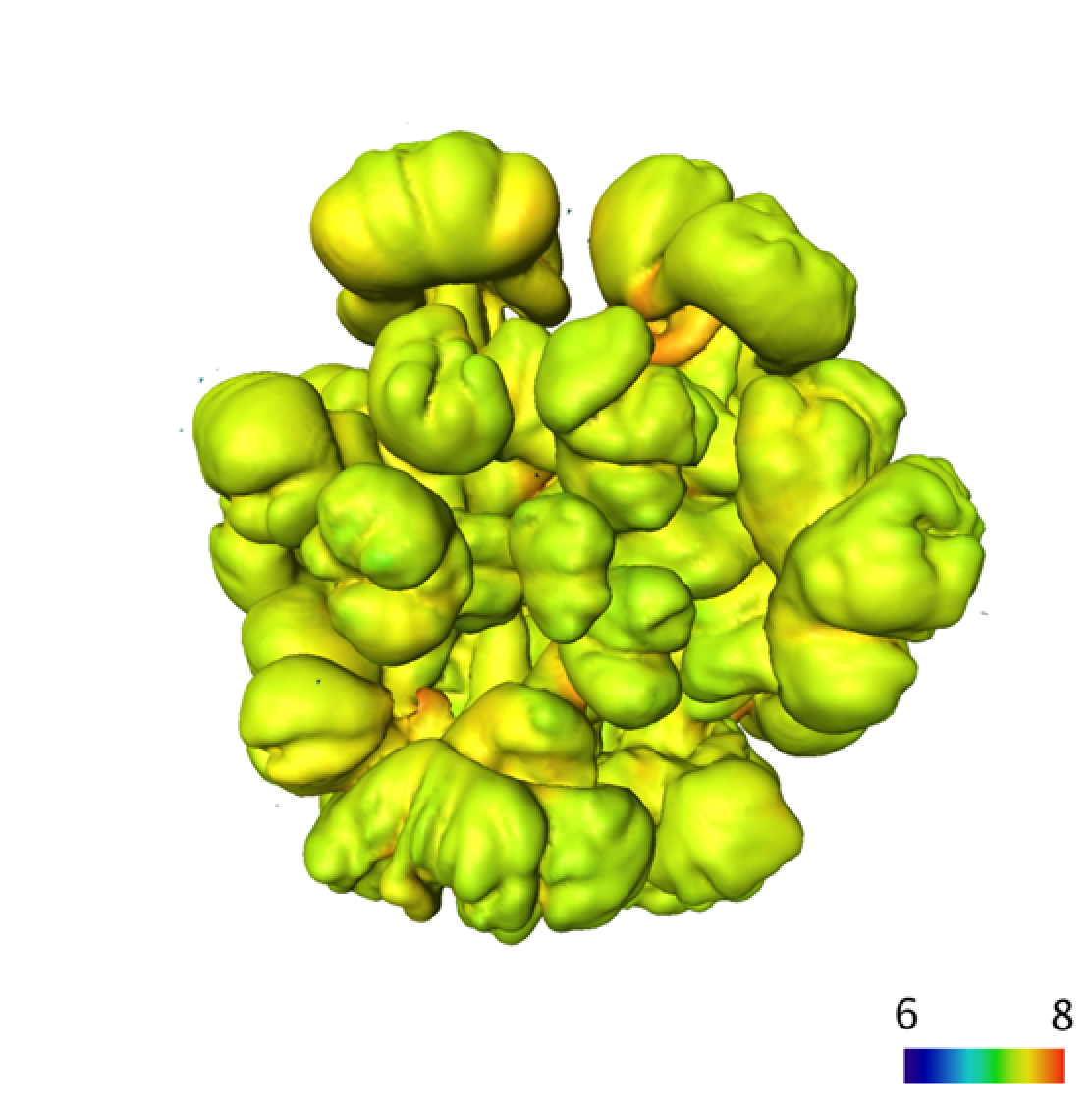}}
          \subfigure[$t=0.9\,\mathrm{s}$]{\includegraphics[width=0.45\linewidth]{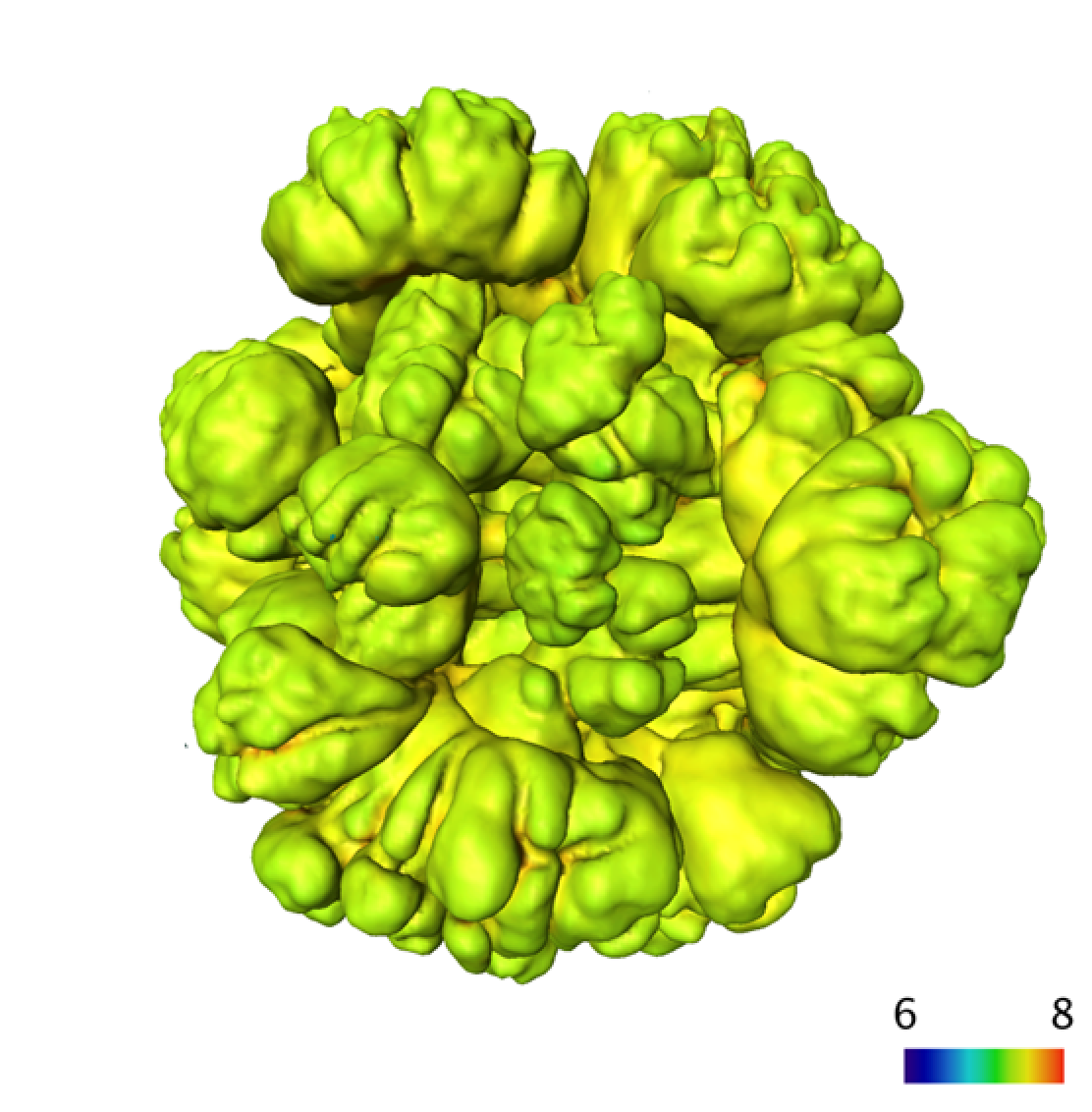} }}
    \caption{Evolution of the flame front in a thermonuclear supernova simulation. 
      The colour scale indicates the logarithm of the magnitude of turbulent velocity fluctuation predicted by
      the SGS model in units of cm/s (i.~e., 6 and 8 corresponds to $10$ and $1000\;\mathrm{km/s}$, respectively). A similar simulation
      is presented in \cite{SchmNie06a}. The spatial scale is adjusted to the bulk expansion of the white
      dwarf by means of a co-moving grid technique \cite{RoepHille05b}. }
    \label{fig:SN_flame_front}
\end{figure}}

Since the the asymptotic rising velocity of a perturbation of size $l$ due to its buoyancy
is given by the Sharp-Wheeler relation \cite{Sharp84}
\begin{equation}
	\label{eq:vel_sharp}
	v_{\mathrm{RT}}(\ell)=0.5\sqrt{\ell g_{\mathrm{eff}}},
\end{equation}
where $g_{\mathrm{eff}}$ is the effective gravity associated with the density contrast
at the interface between burned and unburned material,
$s_{\rm t}\simeq v_{\mathrm{RT}}(\Delta)$ was used as a simple turbulent flame speed
model (e.~g., \cite{GamKhok03}). In \cite{SchmNie06c}, buoyancy is included as an
ad-hoc source term in the SGS turbulence energy equation~(\ref{eq:pde_energy_sgs}):
\begin{equation}
	\label{eq:gamma_RT}
	\Gamma \propto \rho g_{\mathrm{eff}}\sqrt{2K}
\end{equation}
Here, $\Gamma$ is defined such that it vanishes everywhere except for the
close vicinity of the flame front. Moreover, $\Gamma$ is assumed to be non-zero only 
if the so-called fire polishing length $\lambda_{\mathrm{fp}}=4\pi s_{\rm f}^2/g_{\mathrm{eff}}$ 
is smaller than the grid resolution $\Delta$.
The fire polishing length is the smallest length scale
on which perturbations in the flame front are Rayleigh-Taylor unstable. 
If the buoyancy term dominates over the turbulent cascade,
$\sqrt{2K}\sim v_{\mathrm{RT}}(\Delta)$ is obtained as asymptotic solution. However,
different numerical studies indicate Kolmogorov scaling on small scales
\cite{ZingWoos05,RoepHille07,CirSchm08}, corresponding to $\Gamma\ll\Sigma$ in equation~(\ref{eq:pde_energy_sgs}).
Although the impact of the turbulent flame speed model on the energy release was demonstrated
in various simulations \cite{NieHille95,SchmNie06c}, it tends to become less significant if the burning is
resolved down to very small scales. This has become possible with the application of
of AMR and the power of contemporary computing facilities \cite{MaWoos13}.

\epubtkImage{}{
  \begin{figure}[htbp]
    \centerline{\includegraphics[width=\linewidth]{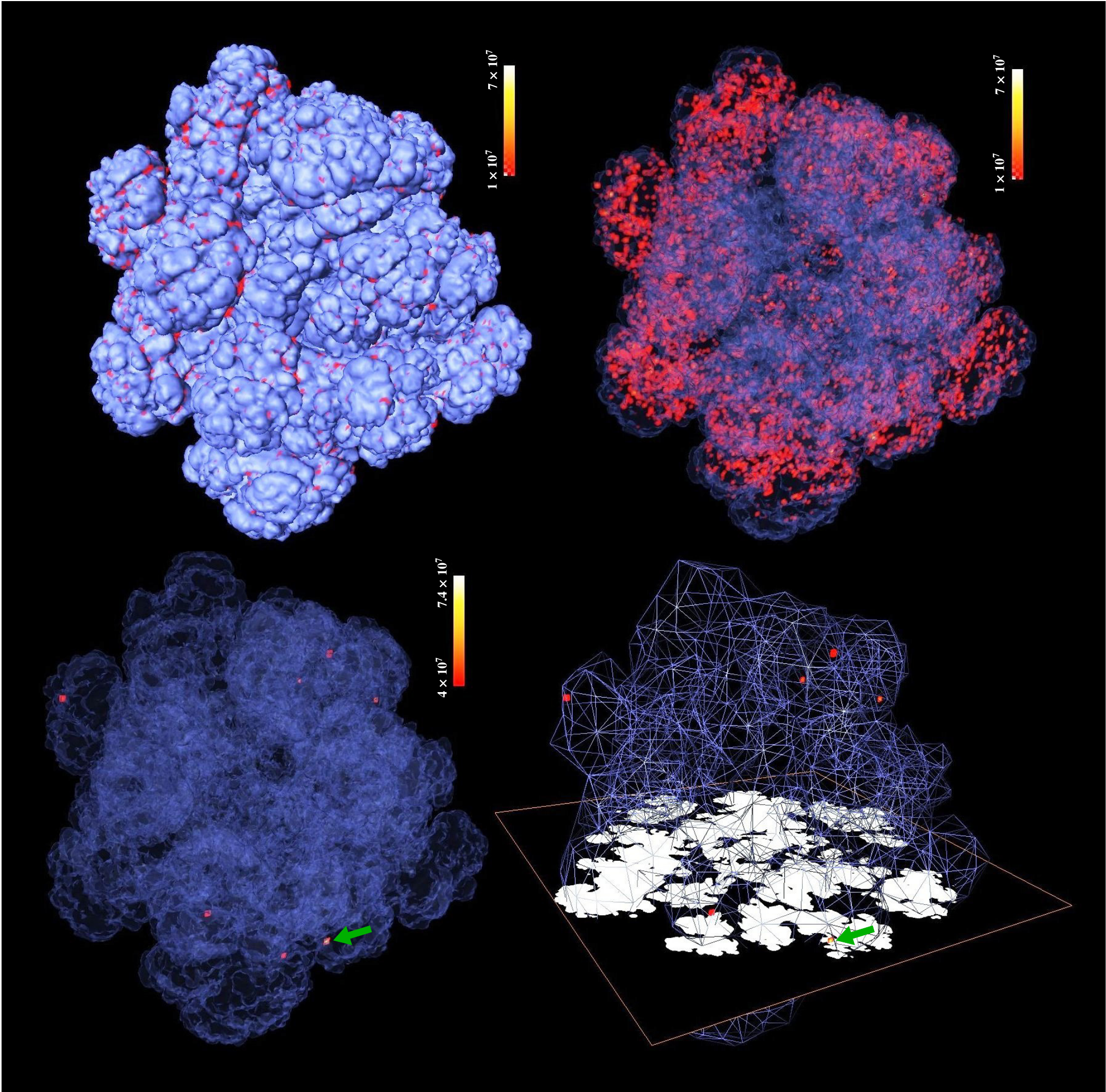} }
    \caption{Flame front in a high-resolution simulation of a thermonuclear supernova \cite{Roepke07}.
    	High turbulent velocity fluctuations occur in the reddish regions. Extremely strong fluctuations
		that could trigger a transition to a detonation are indicated by the green arrows. 
		By courtesy of Friedrich R\"{o}pke.}
    \label{fig:flame_turb}
\end{figure}}

However, since the predictions from pure deflagration models are not consistent with observational
properties of the majority of type Ia supernovae, a transition from the 
deflagration phase to a supersonic detonation is now considered as the most
likely explosion mechanism of Chandrasekhar-mass white dwarfs \cite{RoepFink11,HilleKrom13}.
A theoretical explanation of such a transition is a long standing problem. 
One possibility is that very strong local velocity fluctuations at the onset of distributed
burning might trigger the transition to a detonation \cite{NieKer97,KhokhOran97,LisHille00,Woosley07}.
The statistical distribution of turbulent velocity fluctuations following from the
SGS turbulence energy $K$ in a high-resolution simulation (see Figure~\ref{fig:flame_turb}) shows that such events might indeed occur \cite{Roepke07}. The very wide tail indicates the strong intermittency of turbulence at
the flame front. On the basis of this result, a criterion was recently proposed to
set off detonations in LES of thermonuclear supernovae \cite{CiaSeit13}. 

\subsection{Galaxy simulations} 
\label{sec:galaxy}

Numerical simulations of galaxies, particularly from cosmological initial conditions, cannot fully resolve processes
such as star formation and feedback from supernova (SN) explosions. As suggested in \cite{JoungLow09,ScanBruegg10,BraunSchm13}, 
a production term $\Sigma_{\star}$ due to feedback can be incorporated as an additional source term in the SGS 
turbulence energy equation~(\ref{eq:pde_energy_sgs}). The simplest model for this term is
\begin{equation}
	\label{eq:sigma_SN}
	\Sigma_{\star}= C_{\star}\frac{\rho e_{\rm SN}}{\tau_{\rm ff}}\,,
\end{equation}
where $e_{\rm SN}$ is the typical energy released by
a core-collapse supernova and $\tau_{\rm ff}\propto\rho^{-1/2}$ is the free-fall time scale. The parameter $C_{\star}$ controls
the effective time scale of the feedback. The above expression for $\Sigma_{\star}$ follows from the simple 
Kennicutt-Schmidt relation $\dot{\rho}_{\star}\propto\rho^{-3/2}$ for the star formation rate, 
where the factor $C_{\star}\rho/\tau_{\rm ff}$ is some fraction of $\dot{\rho}_{\star}$. Apart from stellar feedback, 
turbulence is produced by instabilities in the ISM, such as
gravitational, thermal, and hydrodynamical instabilities. Typically, energy is injected on numerically
resolved length scales by these instabilities
and transferred to smaller scales by the turbulent cascade.
If we simply assume that a turbulent velocity of magnitude $V$
is produced on the length scale $L$, the energy flux through the turbulent cascade is of the order $\Sigma\sim \rho V^3/L$
\cite{BraunSchm12}. 
In LES, $\Sigma$ is defined in terms of the grid scale and the shear of the numerically resolved flow via the
closure~(\ref{eq:flux_mixed}). As a very crude model, let us consider the local equilibrium between production and dissipation, i.~e., $\Sigma+\Sigma_{\star}\sim\rho\epsilon$. By substituting equations~(\ref{eq:flux_mixed}), (\ref{eq:sigma_SN}), 
and $\epsilon\sim\rho K^{3/2}/\Delta$, the equilibrium condition reads
\[
	C_{1}\Delta (2K)^{1/2}|S^\ast|^2
	- 4C_2 K\frac{u_{i,k}u_{j,k}S_{ij}^{\,\ast}}{|\vecnab\otimes\vecu|^2} - \frac{2}{3} K d
	+ C_{\star}\frac{e_{\rm SN}}{\tau_{\rm ff}} \sim \frac{K^{3/2}}{\Delta}\,.
\]
This relation implies the turbulent pressure 
\begin{equation}
	\label{eq:energy_eq_fb}
	P_{\rm eq,tot} \sim \rho\Delta^{2/3}
	\left(\frac{K}{\tau} + C_{\star}\frac{e_{\rm SN}}{\tau_{\rm ff}}\right)^{2/3}\,,
\end{equation}
where 
\begin{equation}
	\tau = \frac{\rho K}{\Sigma}\,,
\end{equation}
is the dynamical time scale associated with energy transfer through the cascade. 
Depending on the energy ratio $K/e_{\rm SN}$ and 
the ratio of the dynamical and feedback time scales, $\tau/\tau_{\rm ff}$, the production of SGS turbulence energy
is dominated by the turbulent cascade or by supernovae. 

The star formation rate, which in turn determines the feedback rate, can be parameterized by the
gas density, temperature, and turbulence intensity \cite{KrumMcKee05,PadNord09,HenneChab11,FederKless12,PadHaug12}.
The two main parameters appearing in these parameterizations are the turbulent Mach number $\mathcal{M}_{\star}$ and 
the virial parameter $\alpha_{\star}$ \cite{BertMcKee92}, which correspond to the ratios of the turbulence energy to the internal and gravitational energies:
\begin{equation}
  \mathcal{M}_{\star} = \sqrt{3}\,\frac{\sigma(\ell_{\star})}{c_{\rm s}}
  \quad\mbox{and}\quad
  \alpha_{\star} = \frac{15\sigma^2(\ell_{\star})}{\pi G\rho\ell_{\star}^2}\,.
\end{equation}
Here, $\sigma(\ell)$ is the turbulent velocity dispersion on the length scale $\ell$, 
$c_{\rm s}$ the speed of sound, and $G$ Newton's constant.
A major problem in galaxy simulations is the choice of a characteristic scale $\ell_{\star}$ of star formation.
In \cite{BraunSchm12} it is argued that $\ell_{\star}$ is given by the Jeans length for gravitational 
instability, i.~e.,
\begin{equation}
	\ell_{\star} = \lambda_{\rm J}:= c_{\rm s}\left(\frac{\pi}{\gamma G\rho}\right)^{1/2}\,,
\end{equation}
where $\gamma$ is the adiabatic exponent of the gas. Since the SGS model predicts the specific turbulence energy
$K=3\sigma^2(\Delta)/2$ on the grid scale $\Delta$, it is possible to estimate the turbulent velocity
dispersion in gravitationally unstable, star-forming clouds as 
\begin{equation}
    \label{eq:sigma_c}
	3\sigma_{\star}^2 = 2K\left(\frac{l_{\star}}{\Delta}\right)^{2\eta}\,,
\end{equation}
where the scaling exponent $\eta$ is in the range between $1/3$ and $1/2$, depending on the compressibility
of the gas and the intermittency of turbulence \cite{KritNor07,SchmFeder08,SchmFeder09,HenneFalg12}.
The star formation rate is then given by
\begin{equation}
	\label{eq:sfr}
	\dot{\rho}_{\star} = \mathrm{SFR}(\alpha_{\star},\mathcal{M}_{\star})\frac{\rho}{\tau_{\rm ff}}\,.
\end{equation}
The mass fraction $\mathrm{SFR}(\alpha_{\star},\mathcal{M}_{\star})$ that is converted into stellar mass
per free-fall time is model-dependent. Basically, this factor accounts for the gravo-turbulent fragmentation of 
star-forming clouds.\epubtkFootnote{The assumption $\ell_{\star} = \lambda_{\rm J}$ does not imply that 
	a cloud of size $\ell_{\star}$ collapses and is turned into stars as a whole. 
	The internal structure of star-forming clouds with strong density fluctuations due to supersonic turbulence 
	is usually not resolved in galaxy simulations. For this reason, the Jeans length for the
	gas density in a grid cell serves only as a reference length scale for the mean properties of the
	numerically unresolved star-forming clouds. If the clouds are partially resolved, however, this
	assumption needs revision.  
} For example, a power-law function based on data from numerical simulations is proposed in \cite{KrumMcKee05}. 
In \cite{PadNord09} it is assumed that 
$\mathrm{SFR}(\alpha_{\star},\mathcal{M}_{\star})$ is determined by the log-normal probability density function,
whose width depends on the turbulent Mach number $\mathcal{M}_{\star}$ \cite{PadNord07,FederRom10},
and a critical density that is controlled by the virial parameter $\alpha_{\star}$. 
The different parameterizations are compared and calibrated by numerical data in \cite{FederKless12}. 
A simple star formation law, for which $\mathrm{SFR}(\alpha_{\star})$ is solely determined by the virial parameter, 
is proposed in \cite{PadHaug12}.

A star formation rate proportional to the density of molecular hydrogen, $f_{\rm H_2}\rho$ instead of the total gas density 
$\rho$, was suggested on grounds of the observed tight correlation between the star formation rate and the column density of molecular hydrogen \cite{KrumMcKee09,GnedTass09}. Since molecular hydrogen forms only in the cold phase of the interstellar
medium, the star formation law~(\ref{eq:sfr}) is further modified by replacing $\rho$ and $c_{\rm s}$ with the mean mass
density and speed of sound, respectively, in the cold phase \cite{BraunSchm12}. Since the separation into cold and warm
phases cannot be fully resolved in galaxy simulations, a model such as \cite{SpringHern03} has to be employed to estimate
the fractional densities of the two phases. This entails additional turbulence energy production by
cooling instabilities on length scales below the grid resolution \cite{BraunSchm12,IwaInut13}. 
If the resulting internal driving due to feedback and cooling instabilities, 
$\Sigma_{\rm int}=\Sigma_{\star} + \Sigma_{\rm TI}$,
along with the closure~(\ref{eq:flux_mixed}) for the compressible turbulence energy flux $\Sigma$ are incorporated into equation~(\ref{eq:pde_energy_close}), a full model for the numerically unresolved turbulence energy budget in 
galaxy simulations is obtained \cite{BraunSchm13}. 

\epubtkImage{}{
  \begin{figure}[htbp]
    \centerline{\includegraphics[width=0.65\textwidth]{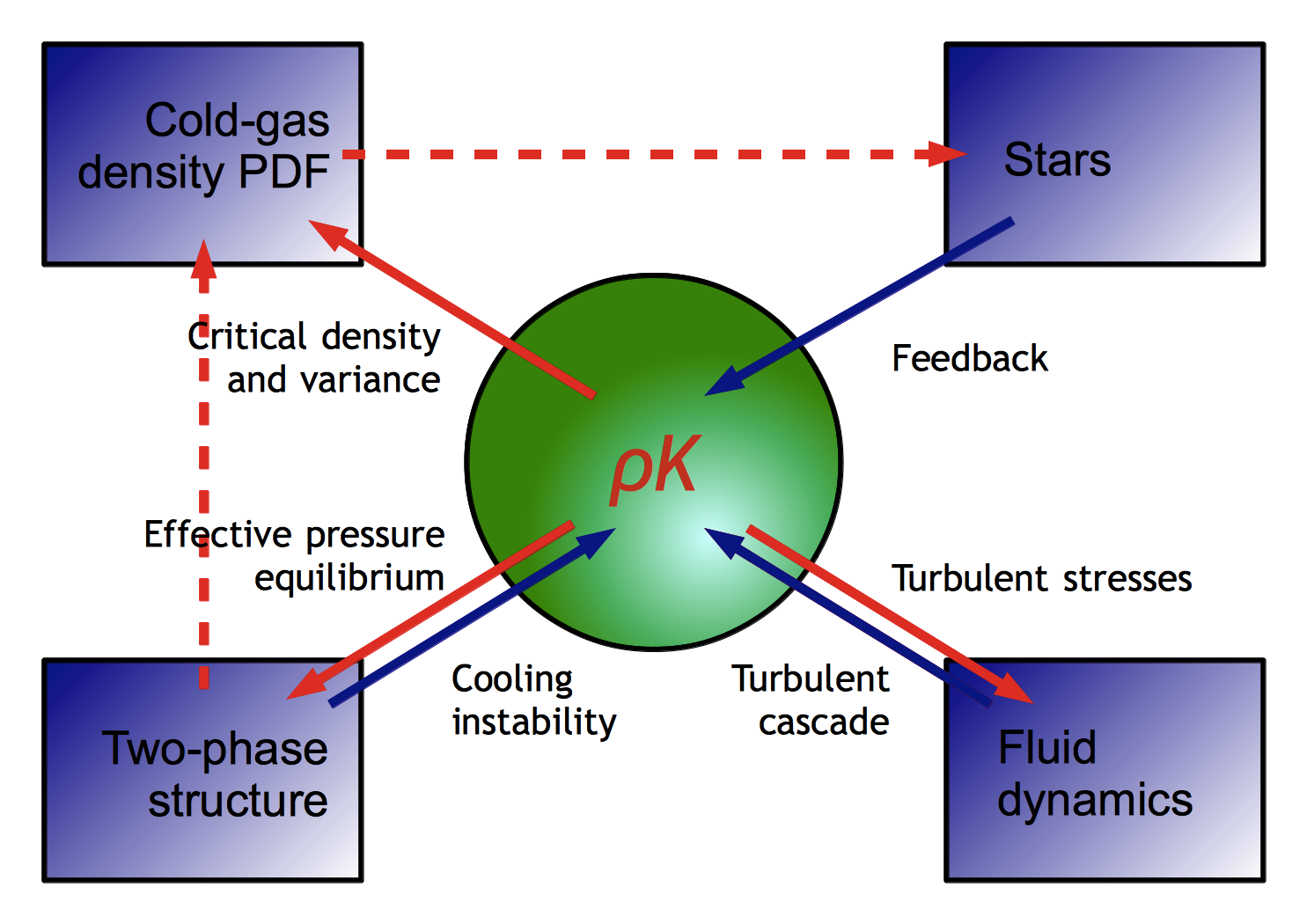}}
    \caption{The SGS turbulence energy $\rho K$ has several functions in the multiphase
   		 model for the turbulent ISM in disk galaxies \cite{BraunSchm13}. It determines the shape and width
		 of the mass density PDF of cold clumps and the effective pressure equilibrium between the cold
		 and warm phases. Via the density PDF, it influences the star formation rate, which acts back
		 on $\rho K$ through feedback. Other sources of $\rho K$ are the turbulent cascade and cooling instabilities.}
    \label{fig:multiphase_model}
\end{figure}}

In combination with the two-phase model for the gas and a particular flavor of the star formation and feedback models outlined above, this SGS turbulence energy model has recently been applied to adaptively refined LES of isolated disk galaxies. As schematically shown in Figure~\ref{fig:multiphase_model}, the computation of the SGS turbulence energy $\rho K$ 
plays a central role. In Figure~\ref{fig:galaxy_slices} (left), an equatorial slice of the gas density illustrates the fragmentation of the inner gaseous disk into dense star-forming clumps
(bluish regions) in one of these simulations. The dilute gas with densities below $1\;\mathrm{cm^{-3}}$ (reddish regions) is partially heated by supernovae. The fragmentation of the disk produces turbulence, as indicated by the slice
of $\rho K$ in Figure~\ref{fig:galaxy_slices} (right). Particularly large turbulent energies are found in regions with strong
feedback, which suggest that the non-thermal feedback from supernovae plays a significant role in the production
of SGS turbulence energy. This is indeed confirmed by the statistics of the production terms plotted in 
Figure~\ref{fig:galaxy_driving} (left). This plot shows profiles of
\[
	\frac{\Sigma_{\rm int}}{\Sigma_{\rm tot}}=\frac{\Sigma_{\star} + \Sigma_{\rm TI}}{\Sigma+\Sigma_{\star} + \Sigma_{\rm TI}}
	\qquad\mbox{and}\qquad
	\frac{\Sigma}{\Sigma_{\rm tot}}=\frac{\Sigma}{\Sigma+\Sigma_{\star} + \Sigma_{\rm TI}}\,,
\]
which are calculated by averaging $\Sigma$, $\Sigma_{\rm int}$, and $\Sigma_{\rm tot}$ over bins of SGS turbulence energy 
per unit mass. The maximum of $\Sigma/\Sigma_{\rm tot}$ around $K\sim 100\;\mathrm{(km/s)^2}$ implies that
the turbulent cascade maintains a ground level corresponding to a turbulent velocity dispersion around $10\;\mathrm{km/s}$, 
while the internal driving caused by supernovae excites much stronger turbulent velocity fluctuations. Remarkably,
the average turbulence energy is quite close to the equilibrium value implied by the balance between
production and dissipation. 
This can be seen in Figure~\ref{fig:galaxy_driving} (right), where the ratio of the equilibrium pressure 
to the dynamical turbulent pressure,
\begin{equation}
	\label{eq:press_eq}
	\frac{P_{\rm eq,tot}}{P_{\rm sgs}}:=\frac{1}{K}\left(\frac{\Delta\Sigma_{\rm tot}}{\rho}\right)^{2/3}\,.
\end{equation}
is plotted against $P_{\rm sgs}=\frac{2}{3}\rho K$. 
Also shown is the asymptotic equilibrium pressure 
\begin{equation}
	\label{eq:press_eq_int}
	\frac{P_{\rm eq,int}}{P_{\rm sgs}}:=\frac{1}{K}\left(\frac{\Delta\Sigma_{\rm int}}{\rho}\right)^{2/3}
\end{equation}
if $\Sigma_{\rm tot}\simeq\Sigma_{\rm int}\gg\Sigma$, i.~e., internal driving dominates. 
This is the case for high turbulence intensity. Since 
$P_{\rm eq,int}/P_{\rm sgs}$ is only slightly above unity, turbulence is close to equilibrium in this regime.
For lower SGS turbulence energies, equation~(\ref{eq:energy_eq_fb}) tends to overestimate the turbulent pressure,
which indicates that turbulence production exceeds dissipation. In this case, 
the turbulent cascade contributes significantly to the production. For comparison, also profiles of the 
thermal pressures of the cold and warm phases are plotted. On the one hand, $P_{\rm eq,tot}$
is small compared to the warm-phase pressure, except for the highly turbulent regions with strong feedback, 
where $P_{\rm w}\sim P_{\rm eq,tot}\sim P_{\rm sgs}$. 
On the other hand, $P_{\rm eq,tot} \gg P_{\rm c}$. The pressure equilibrium between the cold and warm phases,
which is one of the basic assumptions of the model, is therefore dominated by the turbulent pressure in the cold
clumps and the thermal pressure in the warm medium. 

\epubtkImage{}{
  \begin{figure}[htbp]
    \centerline{\includegraphics[width=0.49\linewidth]{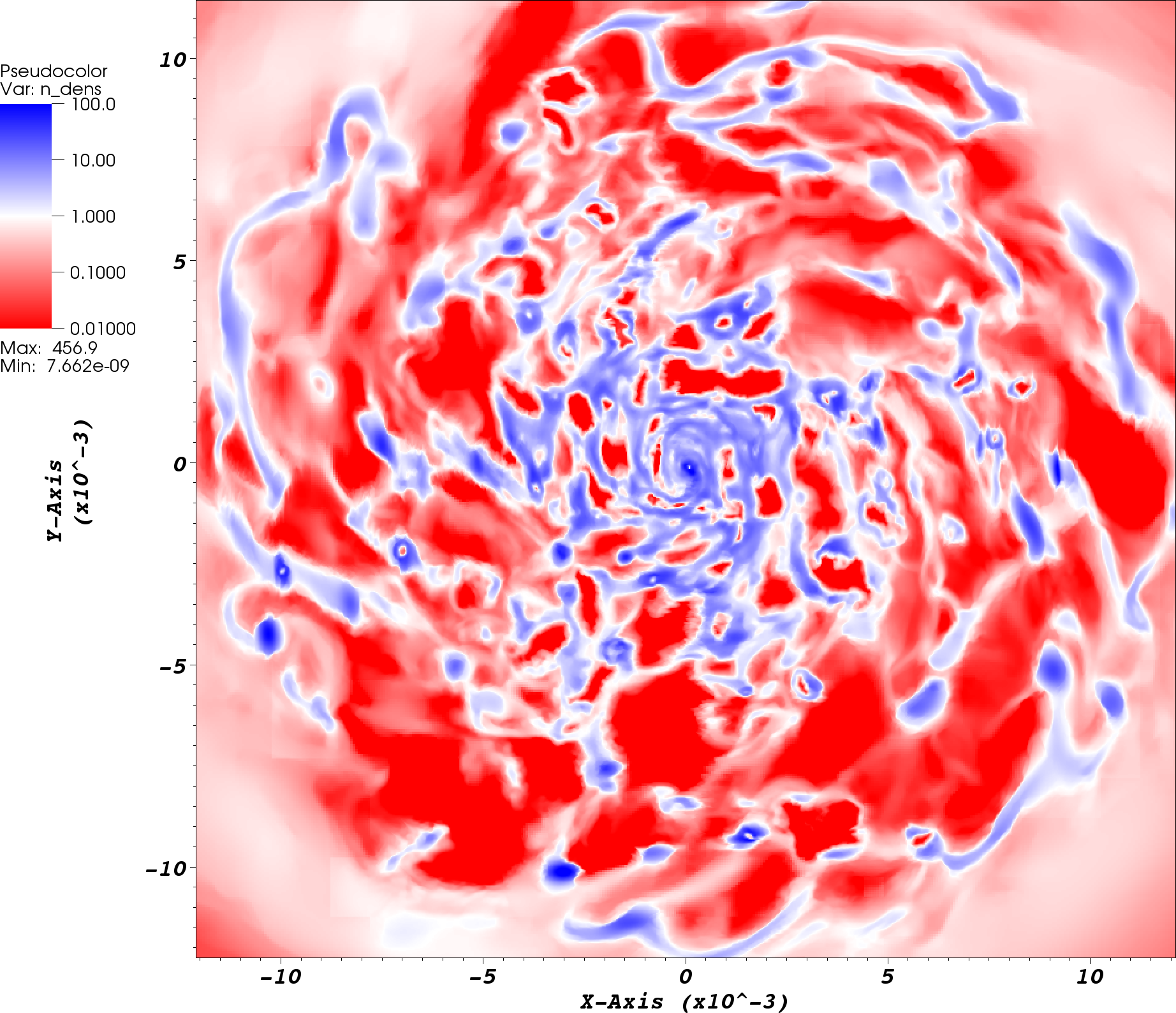}\quad
    	        \includegraphics[width=0.49\linewidth]{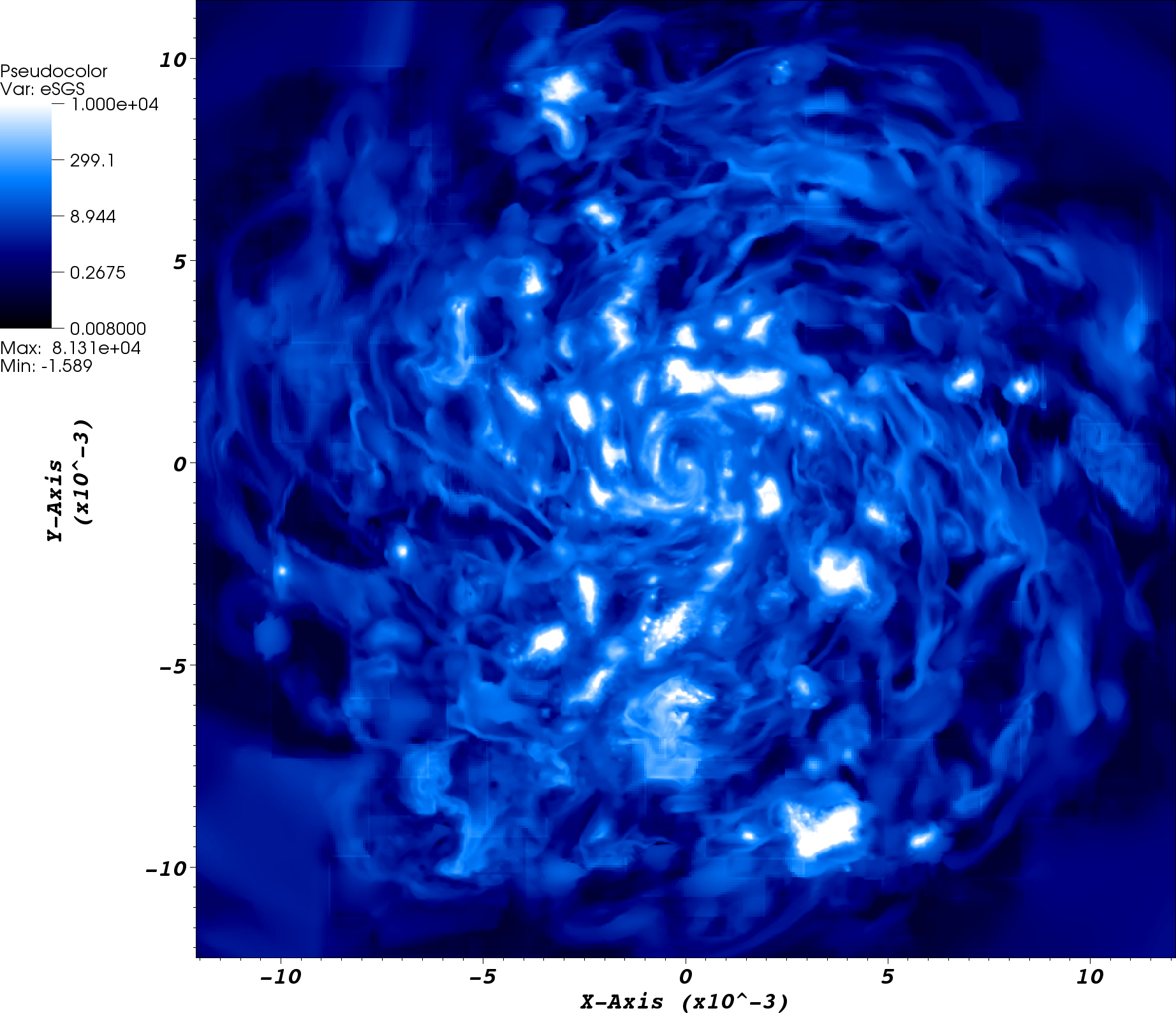}}
    \caption{Projections of the number density (left) and SGS turbulence energy (right) on the
    	equatorial plane in an isolated disk galaxy simulation \cite{BraunSchm13}.}
    \label{fig:galaxy_slices}
\end{figure}}

\epubtkImage{}{
  \begin{figure}[htbp]
    \centerline{\includegraphics[width=0.49\textwidth]{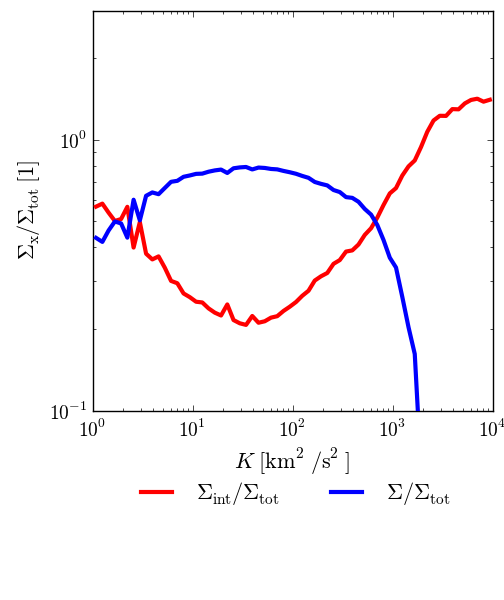}\quad
    	\includegraphics[width=0.49\textwidth]{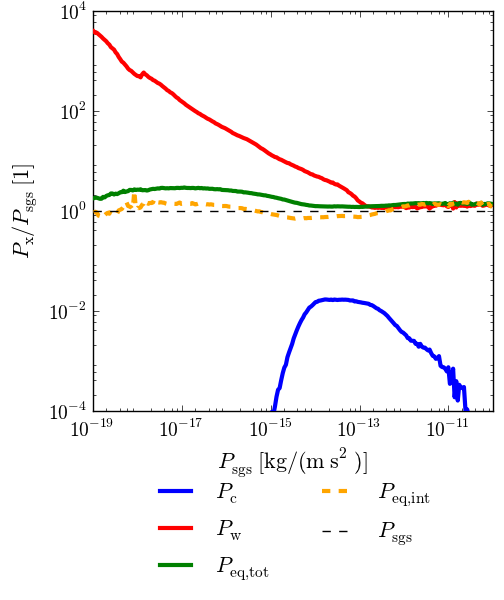}}
    \caption{Left: profiles of the rate of production through the turbulent cascade, $\Sigma$,
    	and internal driving due to numerically unresolved feedback from supernovae and thermal instabilities, 
		$\Sigma_{\rm int}$, relative to $\Sigma_{\rm tot}=\Sigma+\Sigma_{\rm int}$
		for the disk galaxy simulation shown in Figure~\ref{fig:galaxy_slices} \cite{BraunSchm13}.
		Right: profiles of the pressure ratios $P_{\rm X}/P_{\rm sgs}$, where $P_{\rm sgs}=\frac{2}{3}\rho K$ is
		the numerically unresolved turbulent pressure, for the thermal pressures $P_{\rm c}$ and $P_{\rm w}$
		of the cold and warm phases and the equilibrium pressures $P_{\rm eq,tot}$ and 
		$P_{\rm eq,int}$ defined by equations~(\ref{eq:press_eq}) and~(\ref{eq:press_eq_int}), respectively.
		By courtesy of Harald Braun.}
    \label{fig:galaxy_driving}
\end{figure}}

\subsection{Cosmological simulations} 
\label{sec:clusters}

Since cosmological scale structure formation produces a strongly clumped medium through gravitational contraction or
collapse, the numerical simulation of turbulence on cosmological scales is particularly challenging. Potential production mechanisms of turbulent flows in the baryonic gas are mergers between dark-matter halos and the accretion of gas into halos, 
but also feedback form active galactic nuclei and winds produced by strongly star-forming galaxies \cite{DoBy08,KravtBor12}.
A largely open question concerns whether the gaseous component of halos is in a state of developed turbulence. 
While there is no doubt about turbulence in the interior of galaxies, there is no direct observational evidence yet for
turbulence on larger scales, particularly in the intracluster medium (ICM). Theoretical and numerical studies
suggest that magnetic fields play a key role in the physical dissipation mechanism and, possibly, the onset 
of instabilities, but the associated length scales are highly uncertain \cite{FerrGov08,Brueggen13}.
Notwithstanding these uncertainties, turbulent flows resulting from cosmological structure formation 
are numerically investigated by computing the evolution of an ideal fluid subject to the gravitational potential of 
dark matter, which is modeled as a collisionless $N$-body system \cite{BorKravt11}. To follow gravitational collapse, 
AMR is an essential if Eulerian grid codes are used. As pointed out in Chapter~\ref{sec:AMR},
turbulence is generally underresolved in a clumpy medium, even at very high refinement levels. 

This problem was addressed for the first time by combining LES and AMR in \cite{MaierIap09}.\epubtkFootnote{In this article, the acronym FEARLESS (Fluid mEchanics with Adaptively Refined Large Eddy SimulationS) was introduced.} 
The main idea is to solve the filtered Euler equations~(\ref{eq:cosmo_mass_les}--\ref{eq:cosmo_energy_les}) and 
the SGS turbulence energy equation~(\ref{eq:cosmo_k_les}) in co-moving coordinates. For grid refinement and de-refinement,
\cite{MaierIap09} apply the power-law relation~(\ref{eq:rhoK_pow}) between the SGS turbulence energies at different
refinement levels. In contrast to the algorithms outlined in Section~\ref{sec:consrv}, however, their implementation
is not conservative by construction. The SGS turbulence energy decrement $-\Delta(\overline{\rho K})$ in the
case of refinement from a coarser level, for example, is simply compensated by adjusting the
interpolated momenta $(\rho \vecU)_{n}^{\ast}$ such that
\begin{equation}
	\label{eq:energy_balance_maier}
	\frac{1}{N}\sum_{n}\frac{1}{2}\frac{(\rho U)_{n}^2}{\rho_{n}} =
	\frac{1}{N}\sum_{n}\frac{1}{2}\frac{[(\rho U)_{n}^{\ast}]^2}{\rho_{n}} +
	\Delta(\overline{\rho K}),
\end{equation}
which follows by setting
\[
	(\rho K)_n =r^{-2\eta}(\rho K)_n^{\ast}
	\qquad\mbox{and}\qquad
	(\rho U)_{n} = (\rho U)_{n}^\ast
	\sqrt{1+\frac{2\rho_n(\rho K)_n^{\ast}}{[(\rho U)_{n}^{\ast}]^2}\left(1-r^{-2\eta}\right)}\;.
\]
While the first relation is identical to equation~(\ref{eq:energy_sgs_refined_ctec}),
the second relation, when substituted into the total energy balance,
implies 
\begin{equation}
	\Delta(\overline{\rho e}) = 
	\frac{1}{2}\frac{[(\rho U)_{n}^{\ast}]^2}{\rho_{n}} -
	\frac{1}{2}\frac{(\rho U)_{\rm crs}^2}{\rho_{\rm crs}}\,.
\end{equation}
This is just the standard method of compensating the difference of the kinetic energies between refinement
levels entirely by internal energy, which is applied in addition to the 
kinetic energy transfer~(\ref{eq:energy_balance_maier}).
Analogous relations are applied to average the data from a finer to a coarser level. 
The rationale of equation~(\ref{eq:energy_balance_maier}) is that the resolved turbulent
velocity fluctuations should increase from a coarser to a finer level.
However, as argued in Section~\ref{sec:consrv} and in \cite{SchmAlm13}, conservative interpolation of the momenta 
to a finer level entails already an increase of the resolved kinetic energy that encompasses and, in some cases, 
even overestimates the scale-dependence of the numerically resolved turbulence energy. Both energy and momentum conservation 
is guaranteed by compensating this increase with $\Delta(\overline{\rho K})$ and $\Delta(\overline{\rho e})$
defined by equations~(\ref{eq:delta_rhoK_ctec}) and~(\ref{eq:delta_rhoe_ctec}), respectively.

\epubtkImage{}{
  \begin{figure}[htbp]
    \centerline{\includegraphics[width=0.9\linewidth]{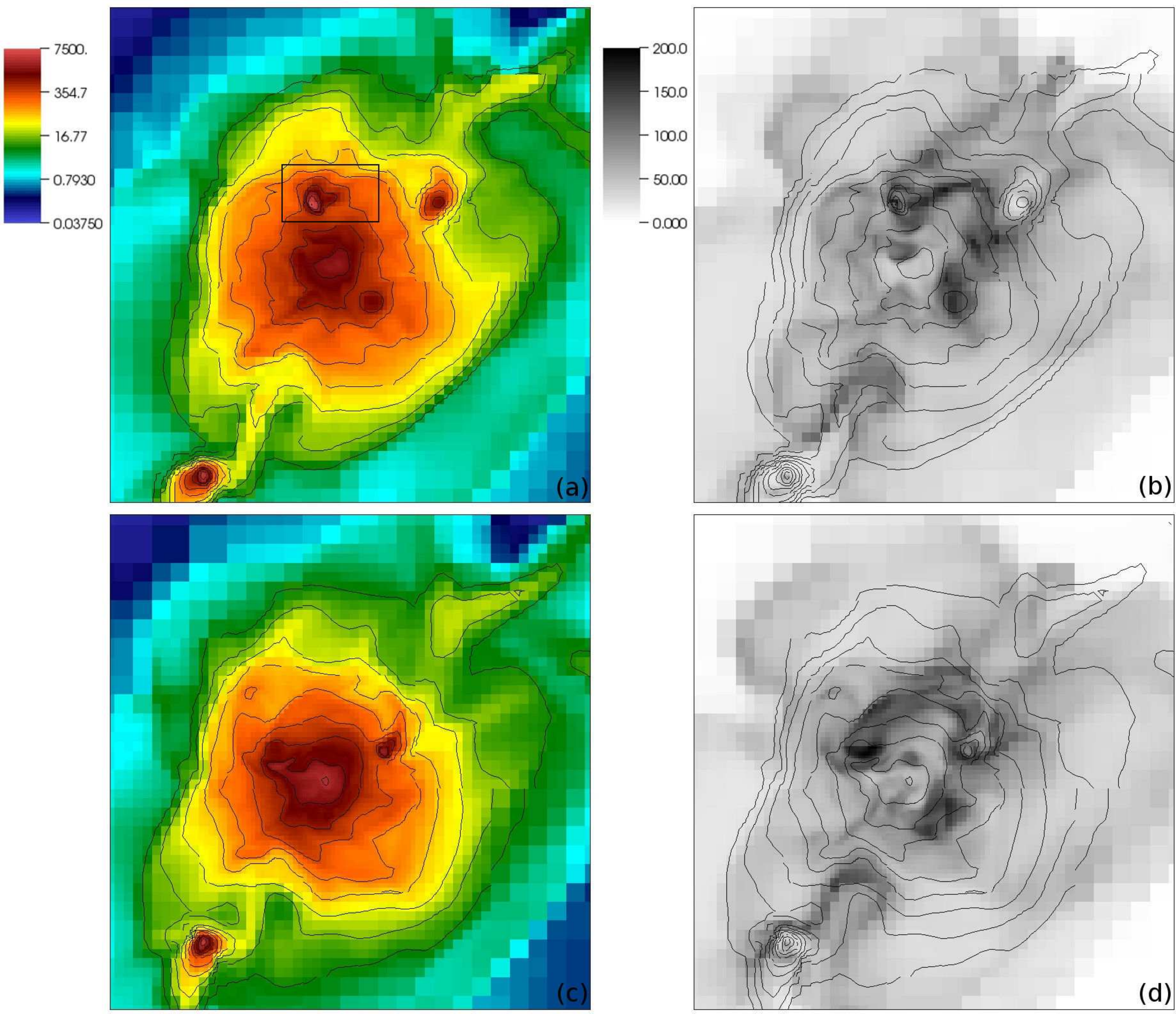}}
    \caption{Slices of baryonic overdensity (left) and the turbulent velocity at
    	the scale of the highest-resolution level in km/s (right) in
    	a cluster simulation at redshifts $z=0.05$ (top) and $z=0.0$ (bottom) \cite{MaierIap09}.
		The comoving size of the shown region is $6.4\;\mathrm{Mpc}\;h^{-1}$. By courtesy of Luigi Iapichino.}
    \label{fig:cluster_merger}
\end{figure}}

Despite its shortcomings, the method based on \cite{MaierIap09} has its merits as a first approximation.
For example, Figure~\ref{fig:cluster_merger} shows the density of the baryonic gas and the small-scale 
turbulence in an adiabatic cosmological simulation of
a cluster from \cite{MaierIap09}. As indicator of turbulence, the magnitude of the numerically unresolved 
turbulent velocity fluctuation $\sqrt{K}$ is scaled down from refinement level $l$ to the minimal cell size at
the maximal level $l_{\rm max}$ via the power law factor $r^{(l-l_{\rm max})\eta}$.
It is particularly interesting that the infall of a subhalo, which is marked by the small box in slice (a),
produces a pronounced turbulent wake, quite
similar to the idealized scenario discussed in Section~\ref{sec:kalman}. However, a shear-improved model
was not used in this simulation. Even at $z=0$, the turbulent velocity slice (d) in Figure~\ref{fig:cluster_merger}
shows a clear trace of the minor merger, which is not discernible in the density slice (c).

\subsubsection{Turbulence production and support against gravity}

Adiabatic simulations of cosmological structure formation with Enzo \cite{Enzo13} show 
that the SGS turbulence energy traces different production mechanisms of turbulence in the intracluster medium (ICM) and the warm-hot intergalactic medium (WHIM) \cite{IapSchm11}. While the former is found
at high densities and temperatures and is dominated by weakly compressible, subsonic turbulence, the WHIM is constituted
by gas of lower density, which is undergoing compression by shocks. The plot in Figure~\ref{fig:icm_whim} shows different
histories of the mean thermal and SGS turbulence energies in the ICM and WHIM. While the SGS turbulence energy
in the ICM peaks around redshift $z=1$, which can be interpreted as a consequence of mergers between galaxy clusters,
the energy in the WHIM gradually rises as a result of the continuous turbulence production by accretion shocks. These
are caused by the infall of low-density gas into the potential wells of clusters and filaments, which accelerates
the gas to supersonic speed. 

\epubtkImage{}{
  \begin{figure}[htbp]
    \centerline{\includegraphics[width=0.5\linewidth]{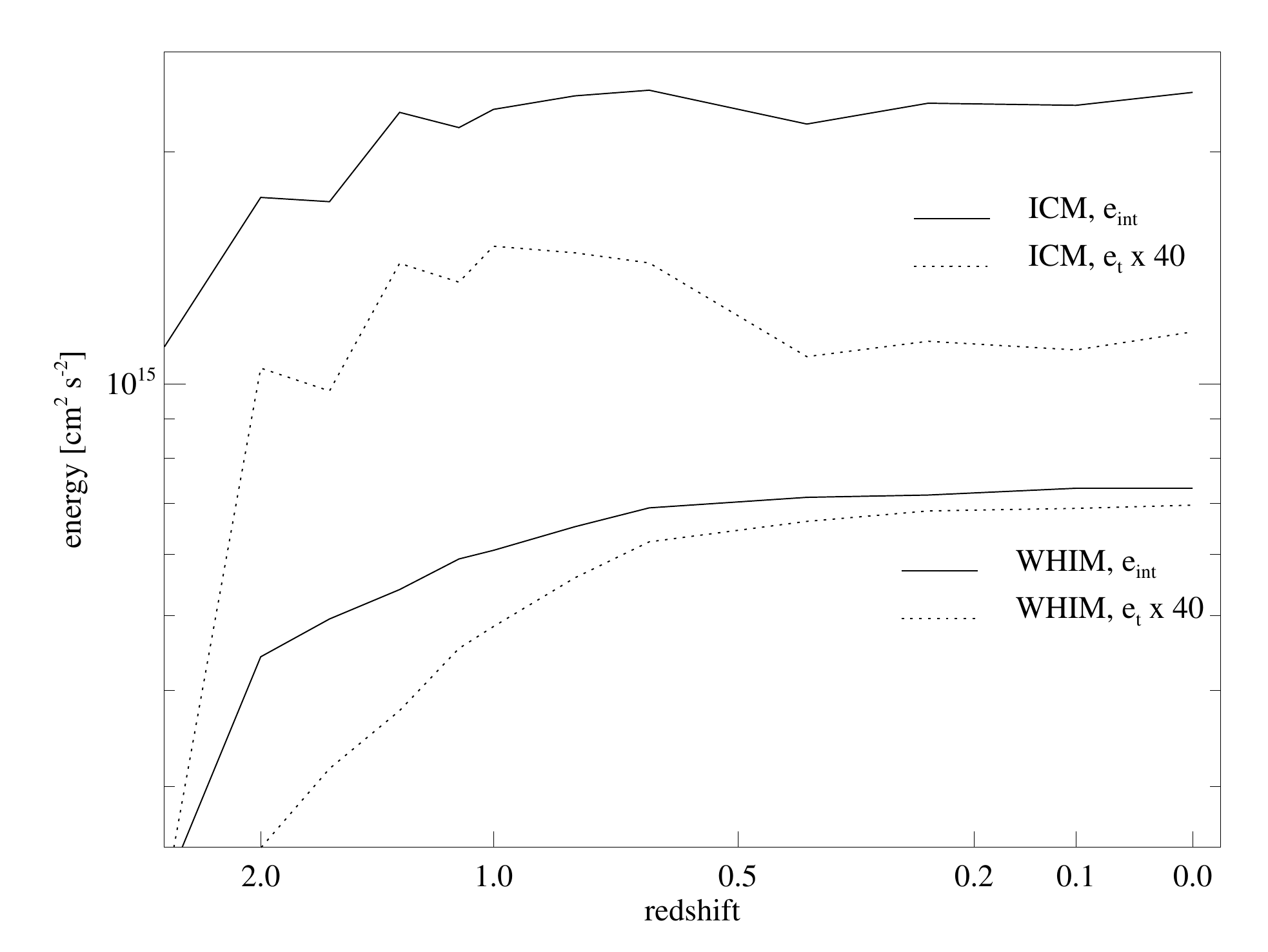}}
    \caption{Evolution of the mean internal (solid) and SGS turbulence (dashed) energies
    	in the ICM and the warm-hot intergalactic medium (WHIM) of a galaxy cluster \cite{IapSchm11}.
		By courtesy of Luigi Iapichino.}
    \label{fig:icm_whim}
\end{figure}}

In addition to other indicators, \cite{IapSchm11} analyses the relative importance of the turbulent and thermal
pressures for the support of the gas against gravity. As shown in \cite{SchmColl13},
the local support of self-gravitating gas due to the thermal pressure $P$ is given by
\begin{equation}
	\Lambda_{\rm therm} = 
	-\frac{1}{\rho}\nabla^2 P 
    + \frac{1}{\rho^{2}}\vecnab\rho\cdot\vecnab P\,.
\end{equation}
The above expression is derived by taking the divergence of the equation for the gas velocity, which
results in an equation for the rate of compression of a fluid parcel, $-\DD d/\DD t$, where $\DD/\DD t$ signifies the
substantial time derivative and $d$ the divergence of the flow. 
In the weakly compressible limit, the thermal support reduces to $\Lambda_{\rm therm}\simeq -(\nabla^2 P)/\rho$.
By considering equation~(\ref{eq:momt_flt_high_Re}), it immediately follows that the influence of turbulence below 
the grid scale can be expressed as
\begin{equation}
	\Lambda_{\rm sgs} = 
	-\frac{1}{\rho}\nabla^2 P_{\rm sgs} 
    + \frac{1}{\rho^{2}}\vecnab\rho\cdot\vecnab P_{\rm sgs}\,,
\end{equation}
where $P_{\rm sys}=\frac{2}{3}\rho K$ is the turbulent pressure on the grid scale. However, $\Lambda_{\rm sgs}$
does not account for the effect of the non-diagonal stresses $\tau_{ij}^\ast$. The contribution of numerically
resolved turbulence, on the other hand, is given by
\begin{equation}
	\Lambda_{\rm turb} = \frac{1}{2}\left(\omega^{2}-|S|^{2}\right)\,,
\end{equation}
where $\omega=\vecnab\times\vecu$ is the vorticity and $|S|$ the rate of strain (see equation~\ref{eq:strain}).
The first term on the right-hand side is associated with the pressure-like support caused by turbulent eddies,
the second term with the compression of the gas by shocks. Figure~\ref{fig:icm_whim} shows 
a mass-weighted histogram of the ratio
\begin{equation}
	r_{\rm tp}=\frac{\rho\left(\omega^{2}-|S|^{2}\right)-2\nabla^2 P_{\rm sgs}}{2\nabla^2 P}
\end{equation}
for different overdensities of the baryonic gas. The ratio $r_{\rm tp}$ is an approximation to the ratio of
the turbulent and thermal support functions, $(\Lambda_{\rm turb}+\Lambda_{\rm sgs})/\Lambda_{\rm therm}$, 
which was used for the sake of comparability with \cite{ZhuFeng10}. The distribution of $r_{\rm tp}$
plotted in Figure~\ref{fig:turb_to_therm} shows two peaks, one at intermediate densities and the other at higher densities,
which can be associated with the WHIM and the ICM. There is a trend of decreasing $r_{\rm tp}$ toward high
densities. This reflects the smaller Mach numbers of turbulence in the ICM, but a caveat is
that only positive contributions to the support are taken into account. Since it is demonstrated in
\cite{SchmColl13,LatSchl13b} that $\Lambda_{\rm turb}$ can be predominantly negative in the presence of shocks, 
the question of the turbulent support of the gas in clusters needs to be revisited. 

\epubtkImage{}{
  \begin{figure}[htbp]
    \centerline{\includegraphics[width=0.6\linewidth]{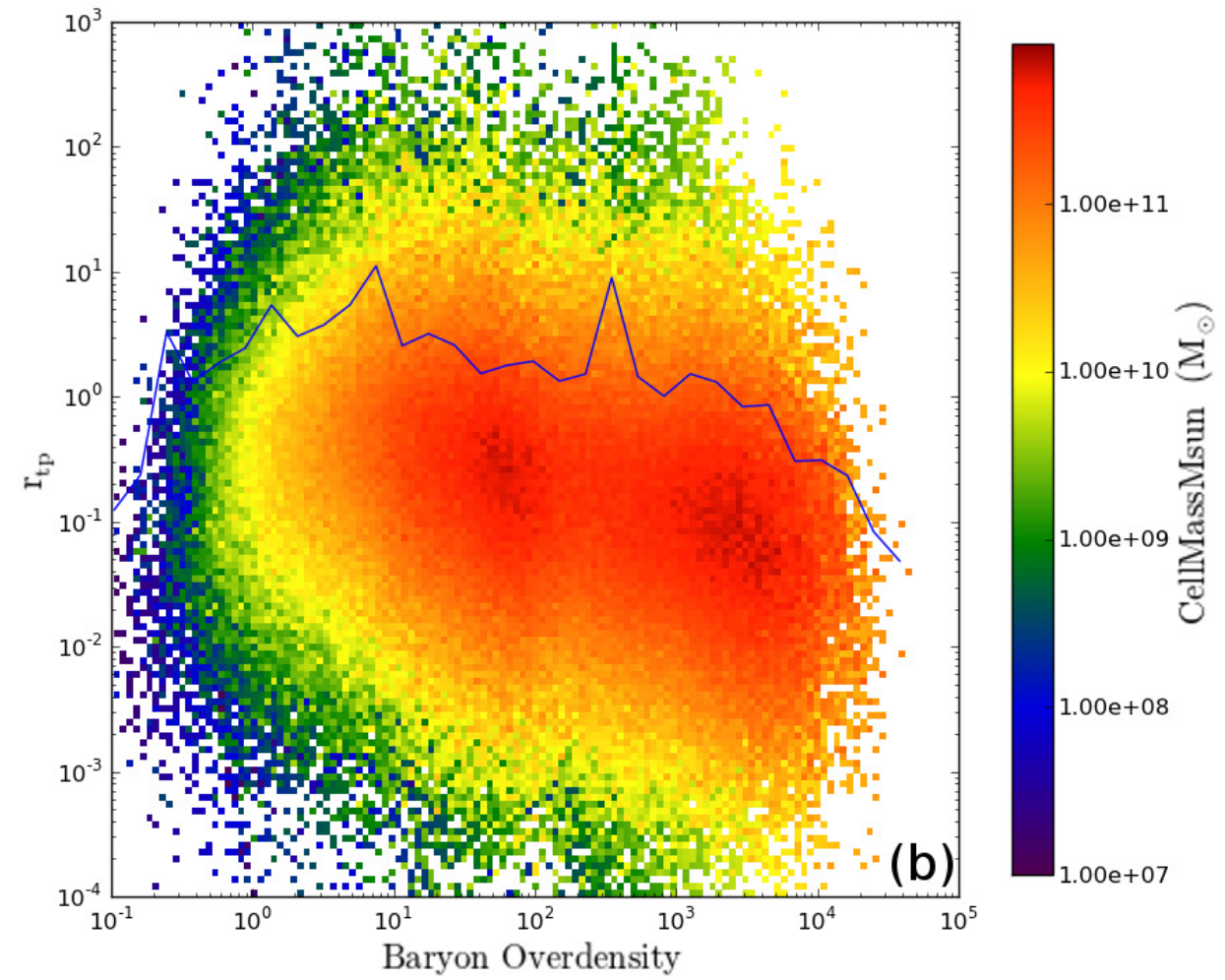}}
    \caption{Mass-weighted histogram of the turbulent-to-thermal support ratio vs. 
		baryon overdensity (the solid line indicates the mean value for given overdensity) 
		\cite{IapSchm11}. By courtesy of Luigi Iapichino.}
    \label{fig:turb_to_therm}
\end{figure}}

\subsubsection{Gravitational collapse of gas in primordial halos}

The Enzo implementation of the SGS model for cosmological fluid dynamics was also applied to the gravitational collapse 
of gas clouds in primordial atomic cooling halos. This scenario is potentially relevant for the formation of 
intermediate-mass black holes through direct collapse and subsequent accretion \cite{LatSchl13a,LatSchl13c,LatSchl13d}. 
The resulting black holes are possibly the seeds for supermassive black holes
at the centers of galaxies. Direct collapse into a supermassive star, which subsequently collapses into an 
intermediate-mass black hole, can occur if the molecular hydrogen formation is
suppressed through photodissociation by a Lyman-Werner radiation background. 
To follow the collapse of the primordial gas down to AU scales, deep-zoom in simulations
with many level of refinements have to be performed. 
In \cite{LatSchl13c}, both LES and ILES of several different halos of masses of the order $10^7\;M_{\odot}$
are presented. One of the most remarkable results of this study is that the additional turbulent viscosity
produced by the SGS model favors disk-like structures around collapsed objects in LES. 
This can be seen from the comparison of density
slices in the innermost regions around the density peaks. Figure~\ref{fig:halos_iles} shows that more or less compact
collapsed structures are produced in ILES. In contrast, more extended and disk-like structures are found in
the corresponding LES, as shown in Figure~\ref{fig:halos_les}. 
 
Apart form morphological differences, the SGS model also has a significant influence on the accretion of mass
by the protostars that are formed through the collapse of gas clouds in the halos. 
As mentioned in Section~\ref{sc:two_equation_grav}, it is possible to follow the evolution of gravitationally
bound dense objects, such as protostars, by inserting sink particles \cite{FederBan10,WangLi10}. 
This method is applied in \cite{LatSchl13d} to follow the accretion
history of protostars in atomic cooling halos. The resulting time evolution of the accretion rate is plotted
in Figure~\ref{fig:mass_accr} for simulations of three different halos, using both ILES and LES. As one can see, the
accretion rates reach peak values around $10\;M_{\odot}\mathrm{yr}^{-1}$ roughly within $10^4\;\mathrm{yr}$.
This value agrees with the theoretical expectation for Bondi-Hoyle accretion. A comparison of ILES and LES suggests
a systematically higher accretion rate for LES. This trend is confirmed by calculating the cumulative masses of the
sink particles (see right plot in Figure~\ref{fig:mass_accr}), which reach masses above $10^5\;M_{\odot}$. As a result, 
the SGS model favors the formation of higher black hole masses.

\epubtkImage{}{
  \begin{figure}[htbp]
    \centerline{\includegraphics[width=\textwidth]{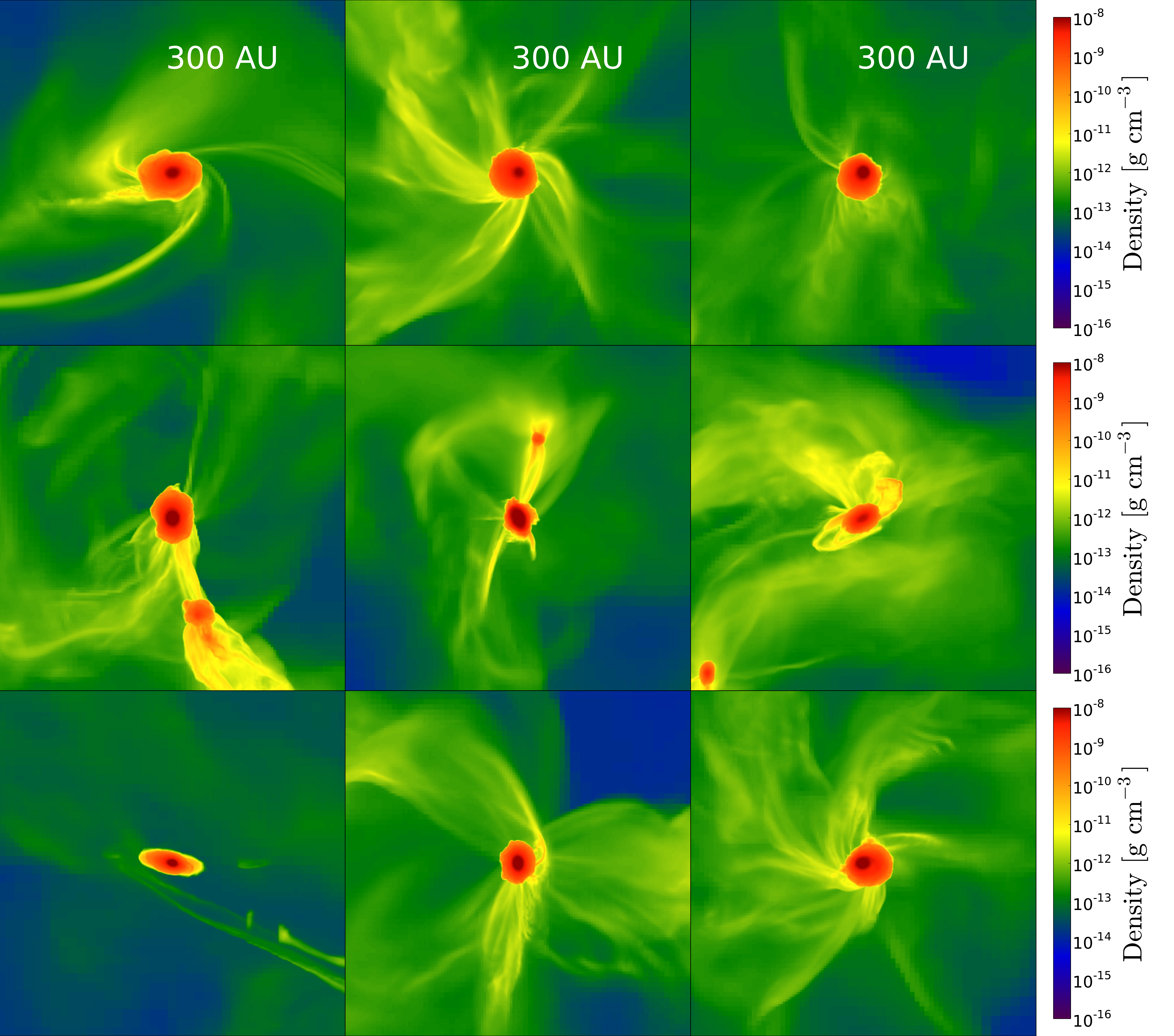}}
    \caption{Gas density in the central 300 AU of collapsing atomic cooling halos computed with ILES \cite{LatSchl13c}. By courtesy of Muhammad Latif.}
    \label{fig:halos_iles}
\end{figure}}
 
\epubtkImage{}{
  \begin{figure}[htbp]
    \centerline{\includegraphics[width=\textwidth]{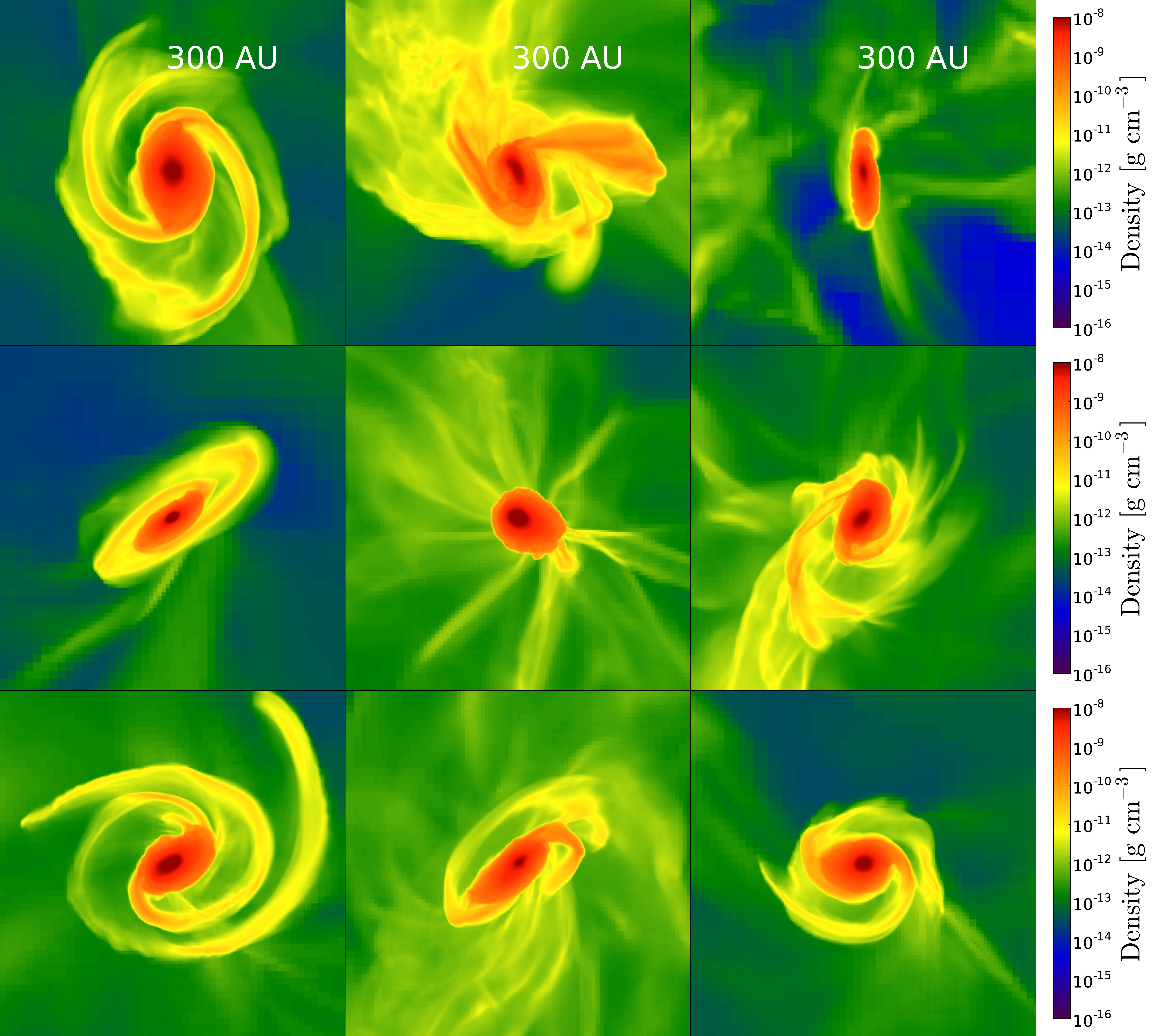}}
    \caption{The same halos as in Figure~\ref{fig:halos_iles} computed with an explicit SGS model \cite{LatSchl13c}. 
    	By courtesy of Muhammad Latif.}
    \label{fig:halos_les}
\end{figure}}

\epubtkImage{}{
  \begin{figure}[htbp]
  	\centerline{\includegraphics[width=0.45\textwidth]{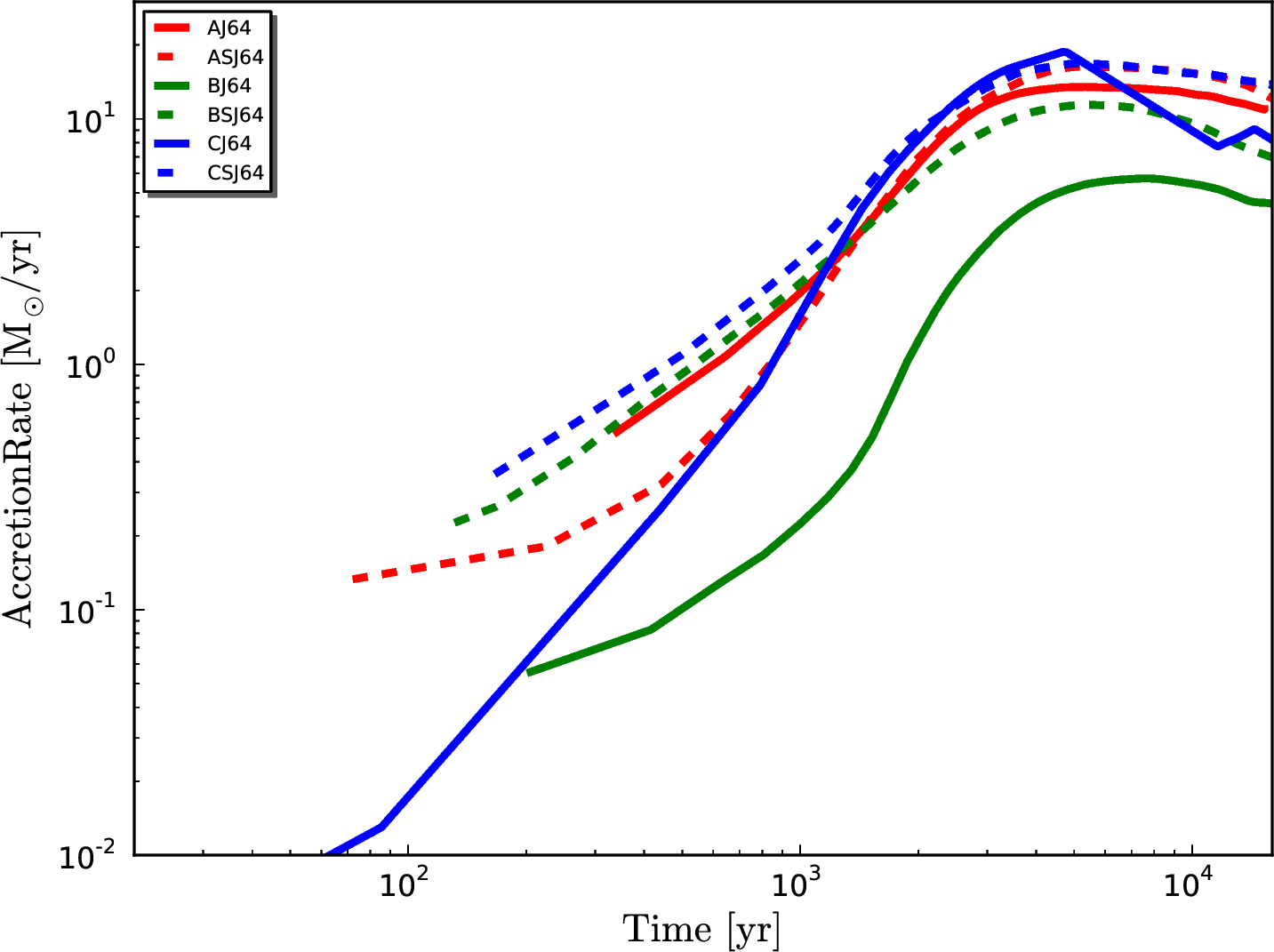}\quad
    	\includegraphics[width=0.5\textwidth]{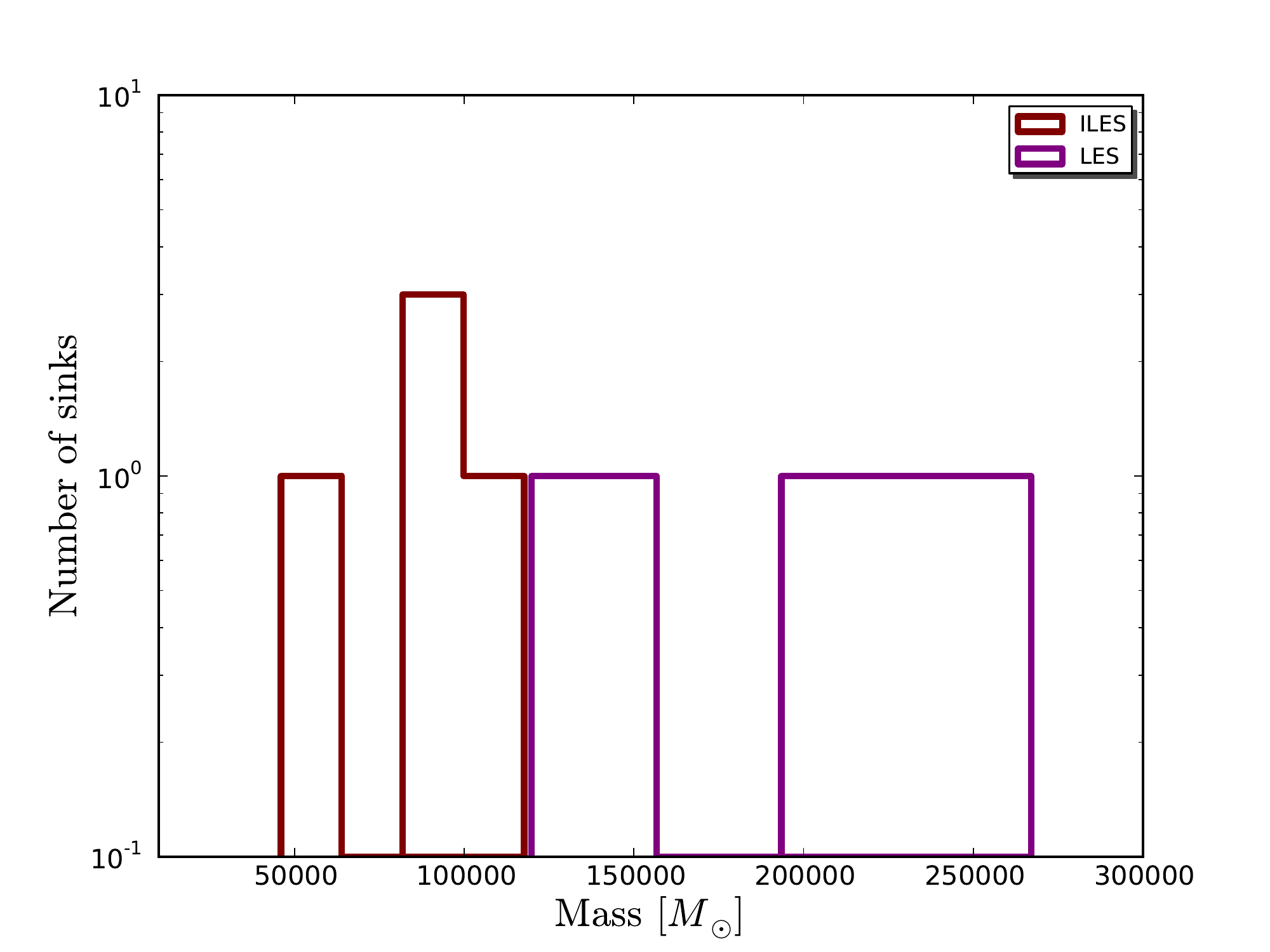}}
    \caption{Left plot: accretion rates of the most massive sink particles in simulations
    	of three different halos. Right plot: Comparison of the mass distributions of the sink 
		particles in LES and ILES \cite{LatSchl13d}. By courtesy of Muhammad Latif.}
    \label{fig:mass_accr}
\end{figure}}
\subsubsection{Turbulent velocity dispersion}
\label{sec:cosmo_turb_vel}

A key question that arises in connection with analyzing turbulence production by cosmological structure formation
concerns the turbulent velocity dispersion $\sigma_{\rm turb}$. 
In \cite{IapViel13}, $\sigma_{\rm turb}$ is defined as $\sigma_{\rm turb}=\sqrt{2K}$, 
which implies that $\sigma_{\rm turb}$ is identified with the
magnitude of the turbulent velocity fluctuations on the grid scale of the simulation. Although this variable
obviously depends on numerical resolution, it is nevertheless useful to infer the dependence on various factors
that influence the production of turbulence. For example, Figure~\ref{fig:doppler_broad} (left plot) shows the ratio of the
turbulent and thermal pressures, where $P_{\rm t}=P_{\rm sgs}=\frac{2}{3}\rho K$, and the
Doppler broadening parameter $b_{\rm t}=\sigma_{\rm turb}/\sqrt{3}=\sqrt{2K/3}$ \cite{EvoFerr11}
for a cosmological simulation with radiative background and cooling in a box of $10\;\mathrm{Mpc}\;h^{-1}$
comoving size. The analysis is carried out for data cubes at redshift $z=2.0$. 
For the intergalactic medium (IGM), which is usually defined by moderate baryonic
overdensities $\rho/\rho_0$ in the range from 1 to about 50, $b_{\rm t}$ is found to increase with density 
from roughly $1$ to $10\;\mathrm{km/s}$. The WHIM, on the other hand, has a flat
turbulent velocity dispersion, corresponding to $P_{\rm t}/P\sim 0.1$. The phase plot in 
Figure~\ref{fig:doppler_broad} shows that the ratio $b_{\rm t}/b$  varies
over several orders of magnitude for different densities and temperatures and reaches
peak values $\sim 1$. The WHIM is associated with gas that is heated by accretion shocks
to temperatures between $10^5$ and $10^7\;\mathrm{K}$. For this reason, the WHIM tends to be more
turbulent then the diffuse gas in the IGM. The phase diagram also shows that the
distinction between gas in the IGM ($1\le \rho/\rho_0\le 50$) and the WHIM ($10^5\;K\le T\le 10^7\;K$) is not
mutually exclusive and somewhat arbitrary. 

\epubtkImage{}{
  \begin{figure}[htbp]
  	\centerline{\includegraphics[width=0.53\textwidth]{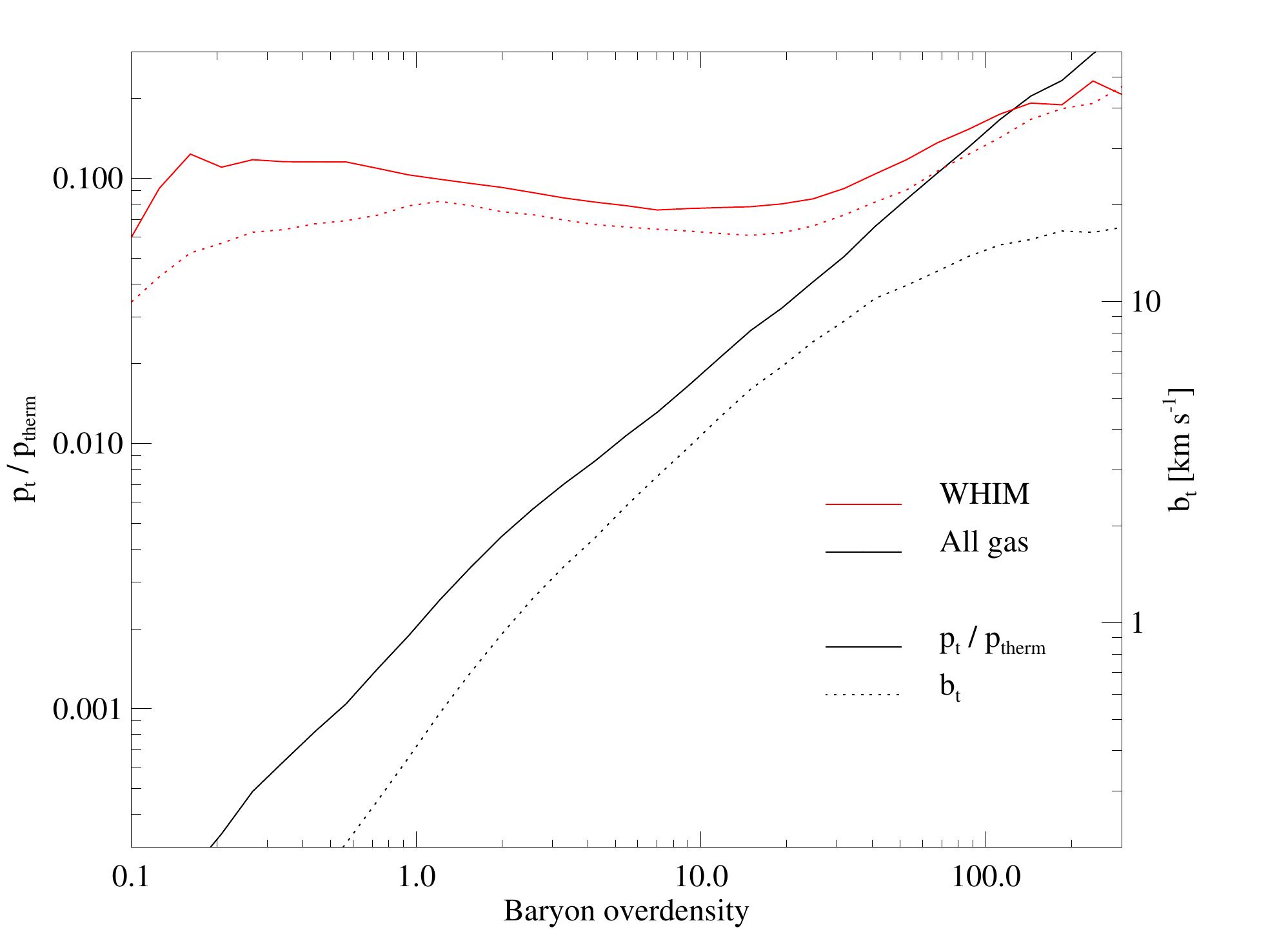}\quad
    	\includegraphics[width=0.45\textwidth]{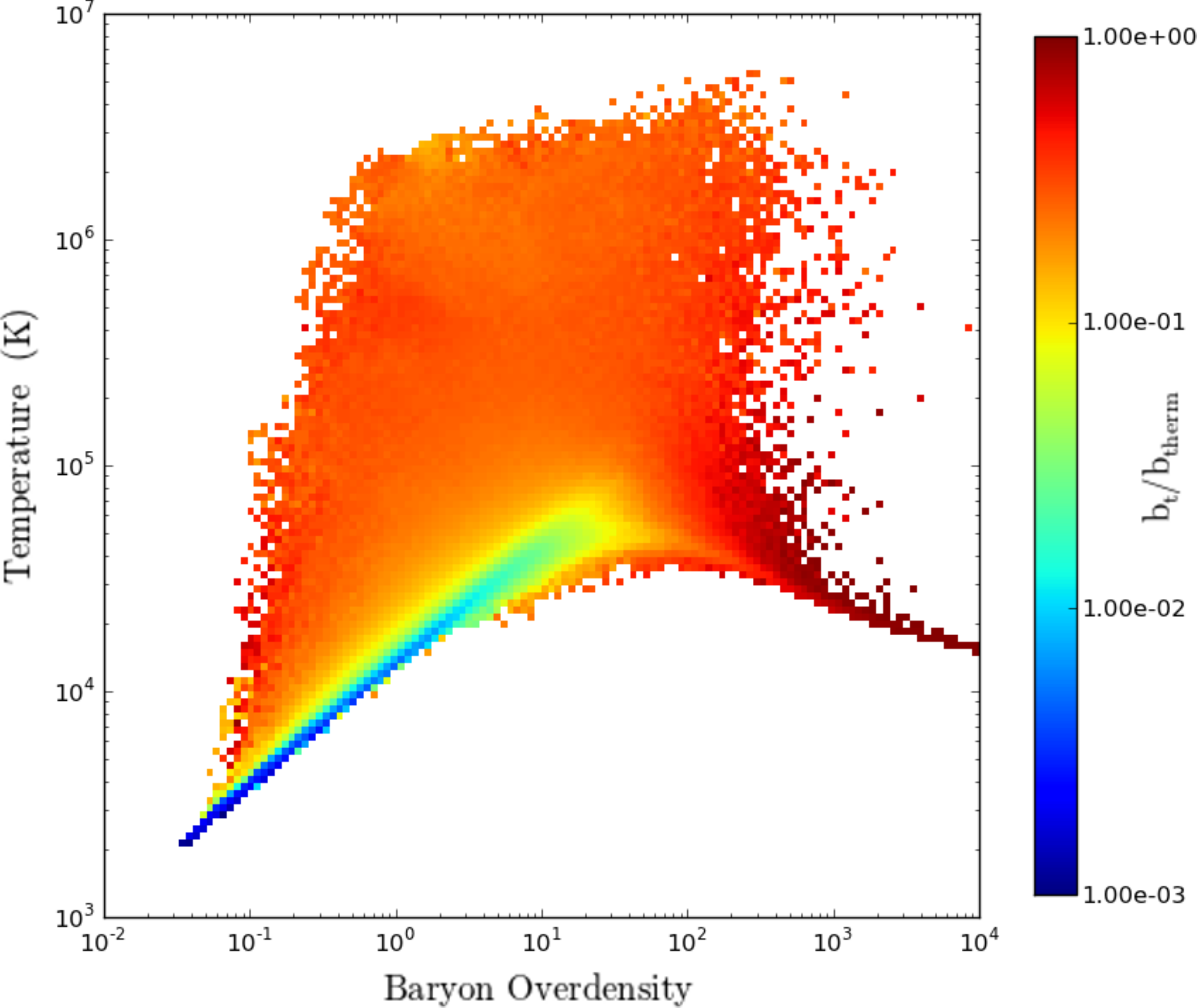}}
    \caption{Left plot: Average ratios of turbulent and thermal pressures vs.\ baryonic overdensity in 
    	a $10\;\mathrm{Mpc}\;h^{-1}$ box with heating and cooling at $z=2.0$. Right plot: 
		Ratio of turbulent to thermal Doppler broadening for bins of temperature 
    	and baryonic overdensity \cite{IapViel13}. By courtesy of Luigi Iapichino.}
    \label{fig:doppler_broad}
\end{figure}}

A resolution-dependent turbulent velocity dispersion is avoided by the
$K$-$L$ hybrid model for Rayleigh-Taylor-driven turbulence (see Section~\ref{sc:two_equation_grav}), however, 
at the cost of smearing out turbulent structures over most of the numerically resolved dynamical range. 
This model was used to simulate the production of turbulence by AGN feedback in galaxy clusters 
\cite{ScanBruegg08,BrueggScan09}. Another caveat of both the $K$-$L$ model and the SGS model based on \cite{MaierIap09} 
are the constant closure coefficients. Strictly speaking, constant-coefficient models are applicable only to statistically homogeneous and stationary turbulence. Turbulence produced by cosmological structure formation and AGNs, however,
is highly inhomogeneous and non-stationary.

\epubtkImage{}{
  \begin{figure}[htbp]
    \centerline{\includegraphics[width=\textwidth]{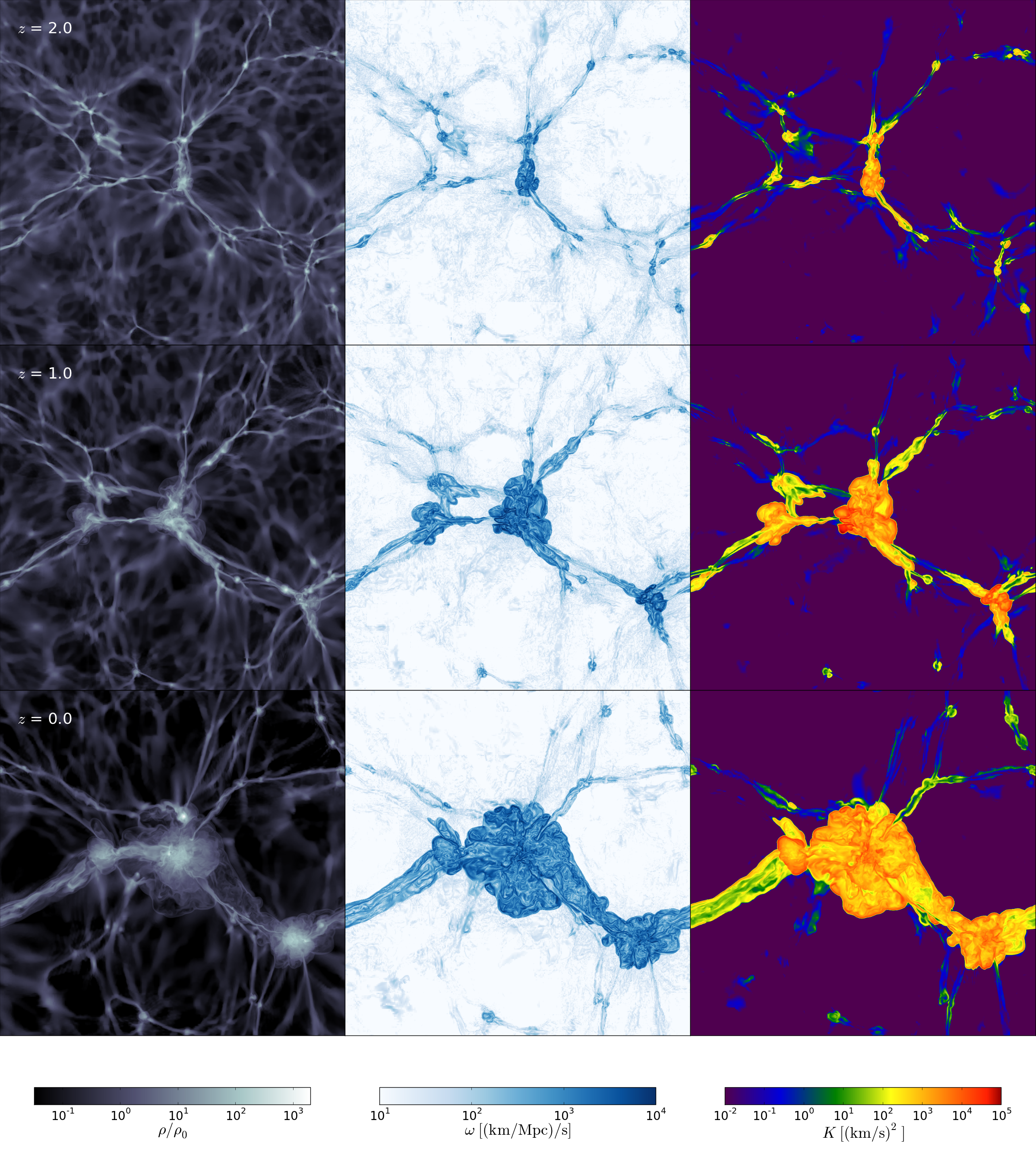}}
    \caption{Slices of the baryonic mass density (left), vorticity modulus (middle), and specific SGS turbulence energy
    	(right) at different redshifts for a simulation of the Santa Barbara cluster \cite{SchmAlm13}.
		The box size is $64\;\mathrm{Mpc}\;h^{-1}$.}
    \label{fig:sb_slices}
\end{figure}}

\epubtkImage{}{
  \begin{figure}[htbp]
  	\centerline{\includegraphics[width=\textwidth]{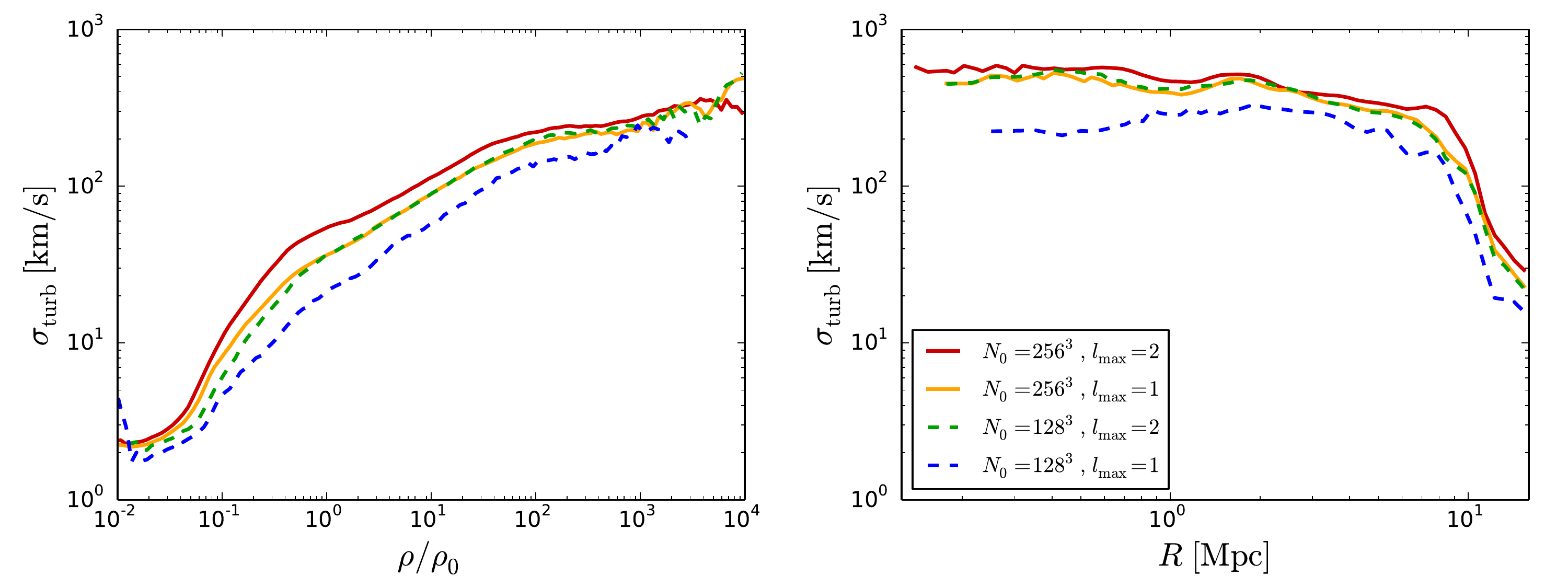}}
    \caption{Mean turbulent velocity dispersion defined by equation~(\ref{eq:sigma_turb})
    	vs.\ baryonic overdensity (left) and radius from the center (right) for simulations
		of the Santa Barbara cluster with different numerical resolutions \cite{SchmAlm13}.
		Starting from a uniform root-grid with $N_0$ grid cells, refinement by
		overdensity and vorticity modulus up to $l_{\rm max}$ levels is applied.}
    \label{fig:sb_profiles}
\end{figure}}

To ameliorate this problem as well as the dependence of $\sigma_{\rm turb}$ on the grid scale, 
the shear-improved SGS model outlined in 
Section~\ref{sec:kalman} has recently been applied to cosmological structure formation \cite{SchmAlm13}.
By running simulations of the Santa Barbara cluster \cite{HeitRick05} with the Nyx code \cite{AlmBell13}, 
it is demonstrated that the application of the standard SGS model for homogeneous turbulence can produce biases, 
particularly in regions of active turbulence production, such as the WHIM. In Figure~\ref{fig:sb_slices},
the growth of the cluster is illustrated by slices at different redshifts from a simulation
performed with the shear-improved SGS model. There clearly is a correlation between the high vorticity and 
SGS turbulence energy in the cluster and a sharp drop at the outer 
accretions shocks. Although a meaningful comparison to the results in \cite{IapViel13} cannot be made at this
point because the Santa Barbara cluster is based on a matter-dominated universe without cosmological constant
and the gas dynamics is adiabatic, some general conclusions regarding the nature of turbulence in clusters can
be drawn. In \cite{SchmAlm13}, the turbulent velocity dispersion is defined by
\begin{equation}
  \label{eq:sigma_turb}
   \sigma_{\rm turb} = \sqrt{U^{\prime\,2} + 2K}\,,
\end{equation}
where $\vect{U}^\prime$ is the fluctuating component of the velocity computed with the Kalman filter
(see Section~\ref{sec:kalman}). The average values of $\sigma_{\rm turb}$ for logarithmic bins of the baryonic
overdensity as well as the corresponding radial profiles 
are plotted for different numerical resolutions in Figure~\ref{fig:sb_profiles}.
Except for the lowest-resolution case, the radial profiles of $\sigma_{\rm turb}$ show
little sensitivity to numerical resolution. This is an important property of the turbulent velocity dispersion defined by equation~(\ref{eq:sigma_turb}). While the profiles are nearly flat for the ICM,
there is a sharp drop around $10\;\mathrm{Mpc}$, which is about the radius of the outer accretion shocks in
Figure~\ref{fig:sb_slices}. Since the gravitational pull of the cluster causes a large non-turbulent bulk flow toward
the center, the total flow velocity beyond the accretion shocks is much higher than $\sigma_{\rm turb}$. 
As a function of the overdensity $\rho/\rho_0$, $\sigma_{\rm turb}$ gradually increases towards the 
center of the cluster. Although the simulations in \cite{IapViel13} differ in important aspects, a roughly similar trend can be seen for the Doppler broadening parameter $b_{\rm t}$ in Figure~\ref{fig:doppler_broad}.

The turbulent kinetic energy $\frac{1}{2}\rho\sigma_{\rm turb}^2$ and the energy flux $\Sigma$ through the turbulent
cascade (see Section~\ref{sc:cosmo}) define the dynamical time scale 
\begin{equation}
  \label{eq:tau_dyn}
  \tau = \frac{\rho\sigma_{\rm turb}^2}{2\Sigma}\,.
\end{equation}
The radial profiles of $1/\tau$ plotted in Figure~\ref{fig:sb_profiles_prod} show that the
dynamical time scale in the ICM is several Gyr. 
The pronounced peak at $10\;\mathrm{Mpc}$ radius is a further indication of turbulence production by the
accretions shocks. Analogous to equation~(\ref{eq:tau_dyn}), the dissipation time scale can be defined as
\begin{equation}
  \label{eq:tau_eps}
  \tau_\epsilon = \frac{\sigma_{\rm turb}^2}{2\epsilon}\,.
\end{equation}
For statistically stationary turbulence, the balance between turbulence production and dissipation, $\Sigma\sim\rho\epsilon$, implies $\tau\sim\tau_\epsilon$. This can indeed be seen in Figure~\ref{fig:sb_profiles_prod} for the central region of the cluster. In the vicinity of the accretion shocks and outside the cluster, however,
$\tau\ll\tau_{\epsilon}$. In this case, the flow is far from equilibrium. These results demonstrate how 
(nearly) scale-invariant quantities computed with the shear-improved SGS model can be utilized to investigate statistical properties of turbulence. 

\epubtkImage{}{
  \begin{figure}[htbp]
    \centerline{\includegraphics[width=\textwidth]{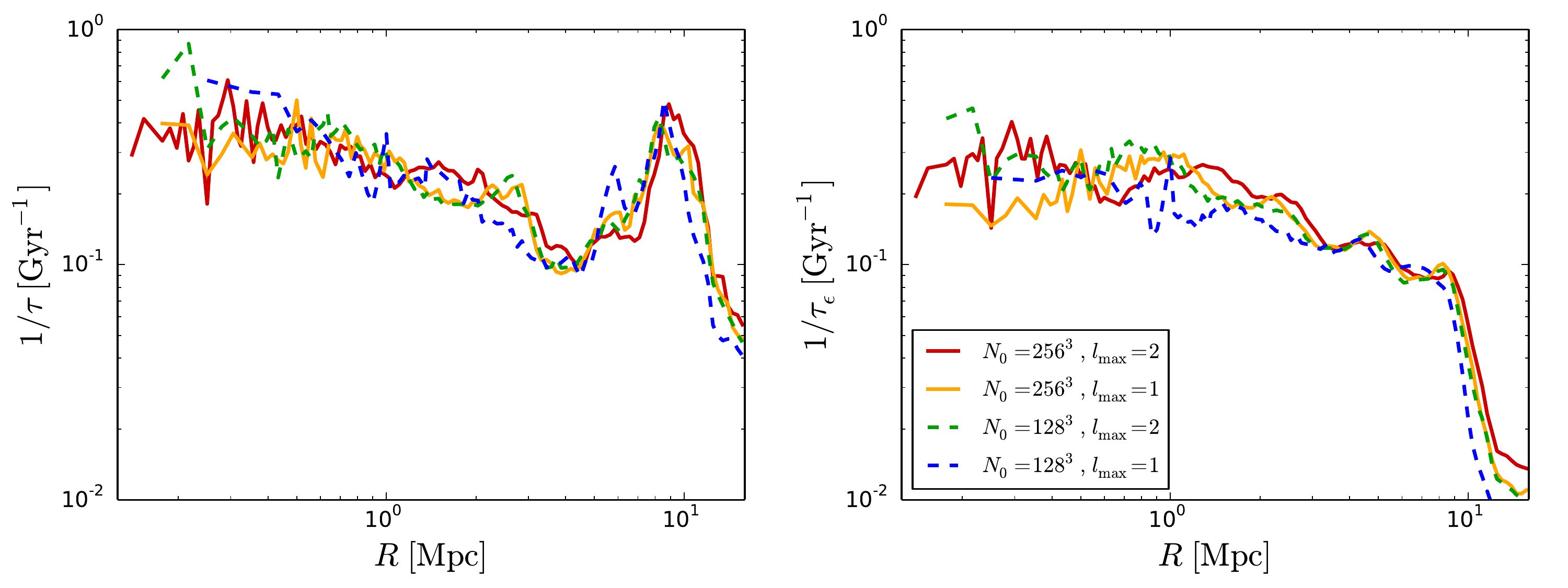}}
    \caption{Radial profiles of the inverse dynamical (left) and dissipation (right) time scales
    	for the same simulations as in Figure~\ref{fig:sb_profiles} \cite{SchmAlm13}.}
    \label{fig:sb_profiles_prod}
\end{figure}}
 
\newpage


\bibliography{turbulence}

\end{document}